\documentclass[usegraphicx,usenatbib,useAMS,twocolumn]{mn2e}

\usepackage{epsfig}
\usepackage{graphicx}
\usepackage{amsmath,bm}
\usepackage{color}
\usepackage{amssymb}
\usepackage{multirow}

\title[RSD measurement from the LOS-dependent power spectrum of DR12 BOSS galaxies]{The clustering of galaxies in the SDSS-III Baryon Oscillation Spectroscopic Survey: RSD measurement from the LOS-dependent power spectrum of DR12 BOSS galaxies}

\author[H. Gil-Mar\'in et al.]{H\'ector Gil-Mar\'in$^{1}$\thanks{hector.gil@port.ac.uk},  Will J. Percival$^{1}$, Joel R. Brownstein$^{2}$, Chia-Hsun Chuang$^{3}$, \and Jan Niklas Grieb$^{4,5}$, Shirley Ho$^{6}$,  Francisco-Shu Kitaura$^{7}$, Claudia Maraston$^1$, \and   Francisco Prada$^{3,8,9}$, Sergio Rodr{\'i}guez-Torres$^{3,10,8}$, Ashley J. Ross$^{11}$,\and Lado Samushia$^{12,13}$, David J. Schlegel$^{14}$ , Daniel Thomas$^1$, Jeremy L. Tinker$^{15}$ \and \& Gong-Bo Zhao$^{1,16}$ \\
 $^{1}$ Institute of Cosmology \& Gravitation, University of Portsmouth, Dennis Sciama Building, Portsmouth PO1 3FX, UK\\
  $^{2}$ Department of Physics and Astronomy, University of Utah, 115 S 1400 E, Salt Lake City, UT 84112, USA\\
    $^3$ Instituto de F\'{\i}sica Te\'orica, (UAM/CSIC), Universidad Aut\'onoma de Madrid,  Cantoblanco, E-28049 Madrid, Spain \\    
$^{4}$Universit\"ats-Sternwarte M\"unchen, Ludwig-Maximilians-Universit\"at M\"unchen, Scheinerstra\ss{}e 1, 81679 M\"unchen, Germany\\
$^{5}$Max-Planck-Institut f\"ur extraterrestrische Physik, Postfach 1312, Giessenbachstr., 85741 Garching, Germany\\
    $^{6}$ McWilliams Center, Carnegie Mellon University, Pittsburgh, PA 15213, USA \\
      $^{7}$ Leibniz-Institut f{\"u}r Astrophysik (AIP), An der Sternwarte 16, D-14482 Potsdam, Germany \\
      $^8$ Departamento de F\'{\i}sica Te\'orica, Universidad Aut\'onoma de Madrid, Cantoblanco, 28049, Madrid, Spain \\
$^9$ Instituto de Astrof\'{\i}sica de Andaluc\'{\i}a (CSIC), Glorieta de la Astronom\'{\i}a, E-18080 Granada, Spain \\
$^{10}$ Campus of International Excellence UAM+CSIC, Cantoblanco, E-28049 Madrid, Spain \\
  $^{11}$ Center for Cosmology and AstroParticle Physics, The Ohio State University, Columbus, OH 43210, USA\\
$^{12}$ Department of Physics, Kansas State University, 116, Cardwell Hall, Manhattan, KS, 66506, USA \\
$^{13}$ National Abastumani Astrophysical Observatory, Ilia State University, 2A Kazbegi Ave., GE-1060 Tbilisi, Georgia\\
$^{14}$ Lawrence Berkeley National Lab 1 Cyclotron Road, Berkeley, CA 94720 USA \\
$^{15}$ Center for Cosmology and Particle Physics, Department of Physics, New York University, New York, NY 10003, USA \\
$^{16}$National Astronomy Observatories, Chinese Academy of Science, Beijing, 100012, P. R. China }
\def\gs{\mathrel{\raise1.16pt\hbox{$>$}\kern-7.0pt
\lower3.06pt\hbox{{$\scriptstyle \sim$}}}}         
\def\ls{\mathrel{\raise1.16pt\hbox{$<$}\kern-7.0pt 
\lower3.06pt\hbox{{$\scriptstyle \sim$}}}}         

\begin{document}
\maketitle

\begin{abstract}

We measure and analyse the clustering of the Baryon Oscillation Spectroscopic Survey (BOSS) relative to the line-of-sight (LOS), for LOWZ and CMASS galaxy samples drawn from the final Data Release 12 (DR12). The LOWZ sample contains 361\,762 galaxies with an effective redshift of $z_{\rm lowz}=0.32$, and the CMASS sample 777\,202 galaxies with an effective redshift of $z_{\rm cmass}=0.57$. From the power spectrum monopole and quadrupole moments around the LOS, we measure the growth of structure parameter $f$ times the amplitude of dark matter density fluctuations $\sigma_8$ by modeling the Redshift-Space Distortion signal. When the geometrical Alcock-Paczynski effect is also constrained from the same data, we find joint constraints on $f\sigma_8$, the product of the Hubble constant and the comoving sound horizon at the baryon drag epoch $H(z)r_s(z_d)$, and the angular distance parameter divided by the sound horizon $D_A(z)/r_s(z_d)$. We find $f(z_{\rm lowz})\sigma_8(z_{\rm lowz})=0.394\pm0.062$, $D_A(z_{\rm lowz})/r_s(z_d)=6.35\pm0.19$, $H(z_{\rm lowz})r_s(z_d)=(11.41\pm 0.56)\,{10^3\rm km}s^{-1}$ for the LOWZ sample, and $f(z_{\rm cmass})\sigma_8(z_{\rm cmass})=0.444\pm0.038$, $D_A(z_{\rm cmass})/r_s(z_d)=9.42\pm0.15$, $H(z_{\rm cmass})r_s(z_d)=(13.92 \pm 0.44)\, {10^3\rm km}s^{-1}$ for the CMASS sample. We find general agreement with previous BOSS DR11 measurements.  Assuming the Hubble parameter and angular distance parameter are fixed at fiducial $\Lambda$CDM values, we find $f(z_{\rm lowz})\sigma_8(z_{\rm lowz})=0.485\pm0.044$ and $f(z_{\rm cmass})\sigma_8(z_{\rm cmass})=0.436\pm0.022$ for the LOWZ and CMASS samples, respectively.

\end{abstract}


\section{Introduction}\label{sec:intro}

The large-scale distribution of matter, as observed through galaxy clustering, encodes significant cosmological information. Much of this can be extracted from the shape and amplitude of the galaxy power spectrum multipole moments. The focus of this paper is the observed anisotropic redshift-space distortions (RSD; \citealt{Kaiser}) caused by peculiar velocities, which contain information about how gravity behaves at large scales and about the total matter content of the Universe. As these distortions depend on the growth of structure, they offer an independent and complementary technique to measure the matter content and to test gravity, compared to studies of the cosmic expansion history.

In this paper we measure the galaxy power spectrum monopole and quadrupole moments calculated from the galaxy sample of the Sloan Digital Sky Survey III \citep{EISetal:2011} Baryon Oscillation Spectroscopic Survey (BOSS; \citealt{Bolton12,Dawsonetal:2012,Smee13}) data release 12 (DR12; \citealt{dr12}). The galaxy catalogues drawn from the final data release from BOSS, DR12, cover the largest cosmic volume ever observed, with an effective volume $V_{\rm eff}=7.4\,{\rm Gpc}^3$ \citep{Reidetal:2015}. The number of independent modes contained allows us to observe the RSD with the highest ever significance, and use them to set tight constraints on the growth of structure $f$ times the amplitude of primordial dark matter power spectrum $\sigma_8$ at the effective redshifts of $z=0.32$ and $z=0.57$. Additionally we are able to constrain $D_A(z_{\rm eff})/r_s(z_d)$ and $H(z_{\rm eff})r_s(z_d)$ by performing the Alcock-Paczynski test (AP; \citealt{AP}). 

This paper forms part of an initial set of DR12 papers produced by the BOSS galaxy clustering team. In \citep[][companion paper]{cuesta} we measure BAO from the correlation function moments measuring the cosmological expansion history. A complimentary approach is provided in  \citep[][companion paper]{gil-marin15b}, which looks at the Baryon Acoustic Oscillation (BAO) in power spectrum monopole and quadrupole moments, with both papers applying an algorithm to reconstruct the primordial density distribution. The BOSS galaxy targeting and catalogue creation algorithms are presented in \citet{Reidetal:2015}, which includes an extensive analysis of the catalogues themselves, and the methods employed to correct for observational effects.

RSD measurements have a long history following the seminal theoretical paper of \citet{Kaiser}, and the development of large galaxy surveys. Key milestones include measurements from the PSCz galaxy survey \citep{tadros99}, 2-degree Field Galaxy Redshift Survey \citep{peacock01,hawkins03,percival04}, WiggleZ \citep{Blakeetal:2012}, 6-degree Field Galaxy Survey \citep{Beutleretal:2012} and the SDSS-II LRGs \citep{Okaetal:2014}. Previous measurements made from BOSS including DR9, when the survey was approximately one third complete \citep{Samushiaetal:2012}, and most recently from DR10/11 \citep{Chuangetal:2013,Beutleretal:2013,Samushiaetal:2013,Sanchezetal:2013,Reidetal:2013,Alametal:2015}. 

The RSD measurements made from the BOSS DR11 sample, which are discussed further in \S\ref{sec:analysis}, exhibit a scatter that is not negligible with respect to the statistical errors. This is caused by adopting different approaches to model quasi-linear and non-linear behaviour in the measured clustering, beyond the linear RSD signal. As the signal-to-noise increases as we move to small scales, RSD measurements are very sensitive to the behaviour of the model on the smallest scales fitted. It is therefore important that the model is general enough to encompass all of the unknown behaviour, whilst also being able to accurately model the data itself. Consequently, a large effort on developing a model that is able to account for: i) mode-coupling; ii) galaxy bias; iii) RSD ;  up to quasi- and non-linear scales is required. Some improvements have recently been achieved on i) beyond the linear regime for the real space power spectrum using perturbation theory schemes. There are several models that attempt to do this task: cosmological standard perturbation theory (see \citealt{Bernardeauetal:2002} and references therein), Lagrangian perturbation theories \citep{hivon95,LPT,Carlsonetal:2009,Okamuraetal:2011,Valageas_Nishimichi:2011}, time renormalization \citep{Pietroni:2008,anselmi}, Eulerian resumed perturbation theories \citep{Crocce_Scoccimarro:2006,multipoint,wang_szalay,anna,regpt,bernardeau12} and closure theory \citep{closure}.  
Recent improvements in the galaxy bias model describe accurately how the galaxies trace dark matter, including non-linear and non-local terms \citep{Nishimichietal:2011,McDonaldRoy:2009,Saitoetal:2014} in addition to primordial non-Gaussian terms \citep{Biagetti14}. The final point iii)  is to accurately model the mapping from real to redshift space statistics. Different approaches to this problem includes: the TNS model \citep{Taruyaetal:2010}, the  Distribution function approach model  \citep{Vlahetal:2012,Okumuraetal:2012} and the Gaussian streaming model \citep{Reid_White:2011}.  

In this paper we use a resumed perturbation theory approach (RPT) in order to  describe the dark matter non-linear power spectra components as it is described in \cite{HGMetal:2012}. We combine this approach with the non-linear and non-local galaxy bias model of \cite{McDonaldRoy:2009}, as it was done in previous analyses \citep{Beutleretal:2013,hector_bispectrum1}. Finally, we account for the RSD using the redshift space mapping presented in \cite{Taruyaetal:2010,Nishimichietal:2011}. This modelling has demonstrated to describe accurately the RSD for both dark matter particles and dark matter haloes \citep{HGMetal:2012,hector_bispectrum0}. Note that in previous approaches, \cite{Beutleretal:2013} use  a different model of RPT to describe the dark matter components and a different $k$-range is considered for obtaining the best-fit parameters. Also in this paper we treat $f$ and $\sigma_8$ as a free independent parameters during the fitting process, and is only at the end of the analysis which we constrain $f\sigma_8$. On the other hand \cite{Beutleretal:2013} constrain $f\sigma_8$ fixing the value of $\sigma_8$ on the non-linear components of the model, which do not depend on the particular combination $f\times\sigma_8$.

This paper is organised as follows. In \S\ref{sec:data} we present the description of the LOWZ and CMASS DR12 data samples and the resources used for computing the covariance matrices and for testing the theoretical models. In \S\ref{sec:estimator_ps} we present the estimator used for measuring the power spectrum multipoles. In \S\ref{sec:results}  the results including the best-fit parameters and their errors are presented. \S\ref{sec:modelling} contains information about the theoretical models used to describe the galaxy power spectrum multipoles in redshift space. In this section, we also include information about how to model the AP distortions and the effect of the survey window in the measurements. In \S\ref{sec:parameter_estimation} we present the details about the parameter estimation, including how the covariance matrices are extracted and how the best-fit parameters and their errors have been computed. In \S\ref{sec:analysis} we present a final analysis of the results presented in \S\ref{sec:results} and how their compare with other galaxy surveys and other BOSS analyses, as well as the effect that changing the cosmological model has on $f\sigma_8$. Finally, in \S\ref{sec:conclusions} we present the conclusions of this paper.

\section{Data and Mocks}\label{sec:data}

\subsection{The SDSS III BOSS data}\label{sec:bossdata}

As part of the Sloan Digital Sky Survey III \citep{EISetal:2011} the Baryon Oscillations Spectroscopic Survey (BOSS) \citep{Dawsonetal:2012} measured spectroscopic redshifts for more than 1 million galaxies and over 200\,000 quasars. The galaxies were selected from multi-colour SDSS imaging \citep{Fukugitaetal:1996,Gunnetal:1998,Smithetal:2002,Gunnetal:2006,Doietal:2010} focussing on the redshift range of $0.15\leq z \leq0.70$. The galaxy survey used two primary target algorithms, selecting samples called LOWZ, with 361\,762 galaxies in the final data release DR12 \citep{dr12} between $0.15\leq z \leq0.43$ and CMASS, with 777\,202 galaxies in DR12 between $0.43\leq z \leq 0.70$. The full targeting algorithms used and the method for calculating the galaxy and random catalogues are presented in \citet{Reidetal:2015}, which also shows that the samples jointly cover a large cosmic volume $V_{\rm eff}=7.4\,{\rm Gpc}^3$ with a number density of galaxies as a function of observed redshift,  that ensures that the shot noise does not dominate at BAO scales. Full details of the catalogues are provided in \citet{Reidetal:2015}, and we do not replicate this here.

In order to correct for several observational artifacts in the catalogues, the CMASS and LOWZ samples incorporate weights: a redshift failure weight, $w_{\rm rf}$, a fibre collision weight, $w_{\rm fc}$, and a systematic weight, $w_{\rm sys}$ (CMASS only), which combines a seeing condition weight and a stellar weight \citep{Rossetal:2012,Andersonetal:2013,Reidetal:2015}. Hence, each galaxy target contributes to our estimate of the target galaxy density field by
\begin{equation}
\label{eq:wc}w_c=w_{\rm sys}(w_{\rm rf}+w_{\rm fc}-1).
\end{equation}
The redshift failure weights account for galaxies that have been observed, but whose redshifts have not been measured: nearby galaxies, which are approximated as being ``equivalent'' are up-weighted to remove any bias in the resulting field. The fibre collision weight similarly corrects for galaxies that could not be observed as there was another target within $62''$, a physical limitation of the spectrograph (see \cite{Rossetal:2012} for details). The systematic weight accounts for fluctuations in the target density caused by changes in the observational efficiency. It is only present for the CMASS sample, which relies on deeper imaging data, and such a weight is not required for the brighter LOWZ sample.

Additionally, we use the standard weight to balance regions of high and low density \citep{FKP,Beutleretal:2013},
\begin{equation}
\label{eq:wfkp}w_{\rm FKP}({\bf r})=\frac{w_{\rm sys}({\bf r})}{w_{\rm sys}({\bf r})+w_c({\bf r})n({\bf r})P_{\rm bao}},
\end{equation}
where $n$ is the mean number density of galaxies and $P_{\rm bao}$ is the amplitude of the galaxy power spectrum at the scale where the error is minimised. We assume $P_{\rm bao}=10\,000\,{\rm Mpc}^3h^{-3}$, which corresponds to the amplitude of the power spectrum at scales $k\sim0.10\,h{\rm Mpc}^{-1}$ \citep{Reidetal:2015}.

\subsection{The mock survey catalogues}\label{sec:mocks}

Galaxy mock catalogues have become an essential tool in the analysis of precision cosmological data provided by galaxy surveys. They provide a fundamental test of large-scale structure analyses and help to determine errors on measurements. As much of the large-scale physics can be captured using approximate methods, we do not necessarily need to base mock catalogues on full N-body cosmological simulations: structure formation models can be calibrated with a small number of N-body simulations, and the parameter space studied using a more efficient scheme. In this paper we use mocks cerated by two different approaches: MultiDark-Patchy BOSS DR12 mocks\footnote{http://data.sdss3.org/datamodel/index-files.html} (hereafter \textsc{MD-Patchy})  \citep[][companion paper]{Kitauraetal:2015} and Quick-Particle-Mesh mocks (hereafter \textsc{qpm}) \citep{QPMmocks}. Both schemes incorporate observational effects including the survey selection window, veto mask and fiber collisions.

\textsc{MD-Patchy} mocks rely on Augmented Lagrangian Perturbation Theory (ALPT) formalism \citep{Kitaura_Hess:2013}, which is based on splitting the displacement field into a long and a short-range component. The long-range component is computed by second order Lagrangian Perturbation Theory (2LPT), whereas the short-range component is modeled using the spherical collapse approximation. The \textsc{MD-Patchy} mocks use 10 combined snapshots at $z=0.1885$, 0.2702, 0.3153, 0.3581, 0.3922,  0.4656, 0.5053, 0.5328, 0.5763, 0.6383. The underling cosmology for these mocks has been chosen to be $(\Omega_\Lambda,\, \Omega_m,\,\Omega_b,\,\sigma_8,\,h,\,n_s)=(0.692885,\,0.307115,\,0.048,\,0.8288,\,0.6777,\,0.96)$, being very close to the best-fit values of the last release of {\it Planck15} \citep{Planck_cosmology15}.  {  For this cosmological model the sound horizon at the baryon-drag redshift is $r_s(z_d)=147.66\,{\rm Mpc}$. }

The \textsc{qpm} mocks are based on low-resolution particle mesh simulations that accurately reproduce the large-scale dark matter density field, in combination with the halo occupation distribution technique (HOD) to populate the resolved haloes with galaxies. For the \textsc{qpm} mocks, the  snapshots are at the effective redshift of, $z_{\rm eff}=0.55$ for CMASS and $z_{\rm eff}=0.40$ for LOWZ.  The underling cosmology for these mocks has been chosen to be $(\Omega_\Lambda,\, \Omega_m,\,\Omega_b,\,\sigma_8,\,h,\,n_s)=(0.71,\,0.29,\,0.0458,\,0.80,\,0.7,\,0.97)$. {  For this cosmological model the sound horizon at the baryon-drag redshift is $r_s(z_d)=147.10\,{\rm Mpc}$. }

\subsection{Fiducial Cosmology}\label{sec:fiducial_cosmo}

We have opted to analyse both mocks and data using the same cosmological model. The fiducial value assumed for this is $\Omega^{\rm fid}_m=0.31$, which is in agreement with the last {\it Planck15} release. As a consequence we will analyse the mocks using a value of $\Omega_m$ that is different than their true values. When converting redshift into comoving distances, this will introduce an extra anisotropy to the one generated by the peculiar velocities. In our analysis this effect will be accounted by the AP scaling relations  presented in \S\ref{sec:AP}. The rest of cosmological parameters in the fiducial cosmology are ${\boldsymbol \Omega}^{\rm fid}\equiv(\Omega_\Lambda^{\rm fid},\, \Omega_m^{\rm fid},\,\Omega_b^{\rm fid},\,\sigma_8^{\rm fid},\,h^{\rm fid},\,n_s^{\rm fid})=(0.69,\,0.31,\,0.049,\,0.8475,\,0.6711,\,0.9624)$. {  For this cosmology the sound horizon at the baryon-drag redshift is $r_s(z_d)=148.11\,{\rm Mpc}$. }

\section{Measuring power spectrum moments}\label{sec:estimator_ps}

In order to compute the galaxy power spectrum we start by defining the Feldman-Kaiser-Peacock function \citep{FKP},
\begin{equation}
F({\bf r})=\frac{w_{\rm FKP}({\bf r})}{I_2^{1/2}} [w_c({\bf r})n({\bf r}) - \alpha n_s({\bf r})],
\label{eq:FKP_factor}
\end{equation}
where $n$ and $n_s$ are, respectively, the observed number density of galaxies and the number density of a synthetic catalog Poisson sampled with the same mask and selection function as the survey with no other cosmological correlations. The functions $w_c$ and $w_{\rm FKP}$ were defined in Eqs \ref{eq:wc} and \ref{eq:wfkp} respectively. The factor $\alpha$ is the ratio between the weighted number of observed galaxies over the random catalogue, $\alpha\equiv \sum_i^{N_{\rm gal}} w_c / N_s$, where $N_s$ denotes the number of objects in the synthetic catalog and $N_{\rm gal}$ the number of galaxies in the real catalog. The factor $I_2$ normalises the amplitude of the observed power in accordance with its definition in a universe with no survey selection,
\begin{equation}
\label{eq:I2def}I_2\equiv \int d^3{\bf r}\,w_{\rm FKP}^2\langle nw_c\rangle^2({\bf r}).
\end{equation}
Following the Yamamoto estimator \citep{Yamamotoetal:2006}, we define the multipole power spectrum estimator as,
\begin{eqnarray}
\label{eq:P_yama}
\nonumber{\hat P^{(\ell)}_{\rm Yama} }(k)& =& \frac{(2\ell + 1)}{I_2} \int \frac{d\Omega_k}{4\pi}\, \left[ \int d{\bf r}_1\, F({\bf r}_1) e^{i{\bf k}\cdot{\bf r}_1}\right.\\
\nonumber &\times& \left.\int d{\bf r}_2\, F({\bf r}_2) e^{-i{\bf k}\cdot {\bf r}_2}\mathcal{L}_\ell({\hat {\bf k}}\cdot {\hat {\bf r}_2})-P^{(\ell)}_{\rm Poisson}({\bf k})\right],\\
\end{eqnarray}
where $d\Omega_k$ is the solid angle element, { $\mathcal{L}_\ell$ is the Legendre polynomial of order $\ell$},  $P^{(\ell)}_{\rm Poisson}$ is the Poisson shot noise term,
\begin{equation}
P^{(\ell)}_{\rm Poisson}({\bf k}) = (1+\alpha)\int d{\bf r}\, \bar{ n} ({\bf r})w^2({\bf r})\mathcal{L}_\ell (\hat{\bf k}\cdot\hat{\bf r}),
\label{shot_noise}
\end{equation}
where the integral has been performed as a sum over the galaxy catalogue. 
For multipoles of order $\ell>0$, $P^{(\ell)}_{\rm Poisson}\ll {\hat P^{(\ell)}}$, and consequently the shot noise correction is negligible. This estimator keeps the relevant LOS information by approximating the LOS of each pair of galaxies with the LOS of one of the two galaxies. This is a reliable approximation on the scales of interest, which clearly improves on assuming a single fixed LOS for the whole survey for $l>0$, but will eventually break down at very large scales and high order multipoles \citep{Yooetal:2015,Samushiaetal:2015}. 

The implementation of the Yamamoto estimator is performed using multiple FFTs (Fast Fourier Transform), each measuring the LOS-weighted clustering along different axes as presented in \cite{Bianchietal:2015}. Thus, the computation of the monopole and quadrupole can be written in terms of $N\log N$ processes (where $N$ is the number of grid-cells used to discretise the galaxy field), which is significantly faster than performing the sum over galaxies used in previous analyses \citep{Beutleretal:2013}.

We use a random catalogue of number density of $\bar{n}_s({\bf r})=\alpha^{-1} \bar{n}({\bf r})$ with $\alpha^{-1}\simeq50$. We place the LOWZ and CMASS galaxy samples on $1024^3$ grids, of box side $L_b=2300\,h^{-1}{\rm Mpc}$ for the LOWZ galaxies, and $L_b=3500\,h^{-1}{\rm Mpc}$  to fit the CMASS galaxies. This corresponds to a grid-cell resolution of $3.42\,h^{-1}{\rm Mpc}$ for the CMASS galaxies and $2.25\,h^{-1}{\rm Mpc}$ for the LOWZ galaxies. The fundamental wave-lengths are therefore $k_f=1.795\cdot10^{-3}\,h{\rm Mpc}^{-1}$ and $k_f=2.732\cdot10^{-3}\,h{\rm Mpc}^{-1}$ for the CMASS and LOWZ galaxies, respectively.  We have checked that for $k\leq0.3\,h{\rm Mpc}^{-1}$, doubling the number of grid-cells per side, from $1024$ to $2048$, produces a negligible change in the power spectrum, $\ll1\%$. This result indicates that using $1024^3$ grid-cells provides sufficient resolution at the scales of interest. We apply the Cloud-in-Cells scheme (CiC) to associate galaxies to grid-cells, and bin the power spectrum $k-$modes in 60 bins between the fundamental frequency $k_f$ and a maximum frequency of $k_{\rm M}=0.33\,h{\rm Mpc}$,  with width  $\Delta \log_{10} k=\left[ \log_{10}(k_{\rm M})-\log_{10}(k_f) \right]/60$.

We limit the scales fitted as follows: our procedure for determining the largest scale we use for the fitting process is based on limiting the impact of the systematic weights, and is presented in Appendix~\ref{appendix_b}. We limit scales to $k>0.02\,h{\rm Mpc}^{-1}$ for the monopole and $k>0.04\,h{\rm Mpc}^{-1}$ for the quadrupole. The smaller (larger) the minimum scale ($k$-value) included, the more $k$-modes are used and therefore the smaller the statistical errors of the estimated parameters. However, small scales are poorly modeled in comparison to large scales, such that we expect the systematic errors to grow as the minimum scale decreases. Therefore, we empirically find a compromise between these two effects such that the systematic offset induced by poorly modelled non-linear behaviour is smaller than the statistical error. To do so, we perform different best-fit analysis for different minimum scales and check that the best-fit parameters of interest does not change significantly (compared to the statistical errors) as a function of this minimum scale. 

\section{The Power Spectrum Multipoles}\label{sec:results}

\begin{figure*}
\centering
\includegraphics[scale=0.45]{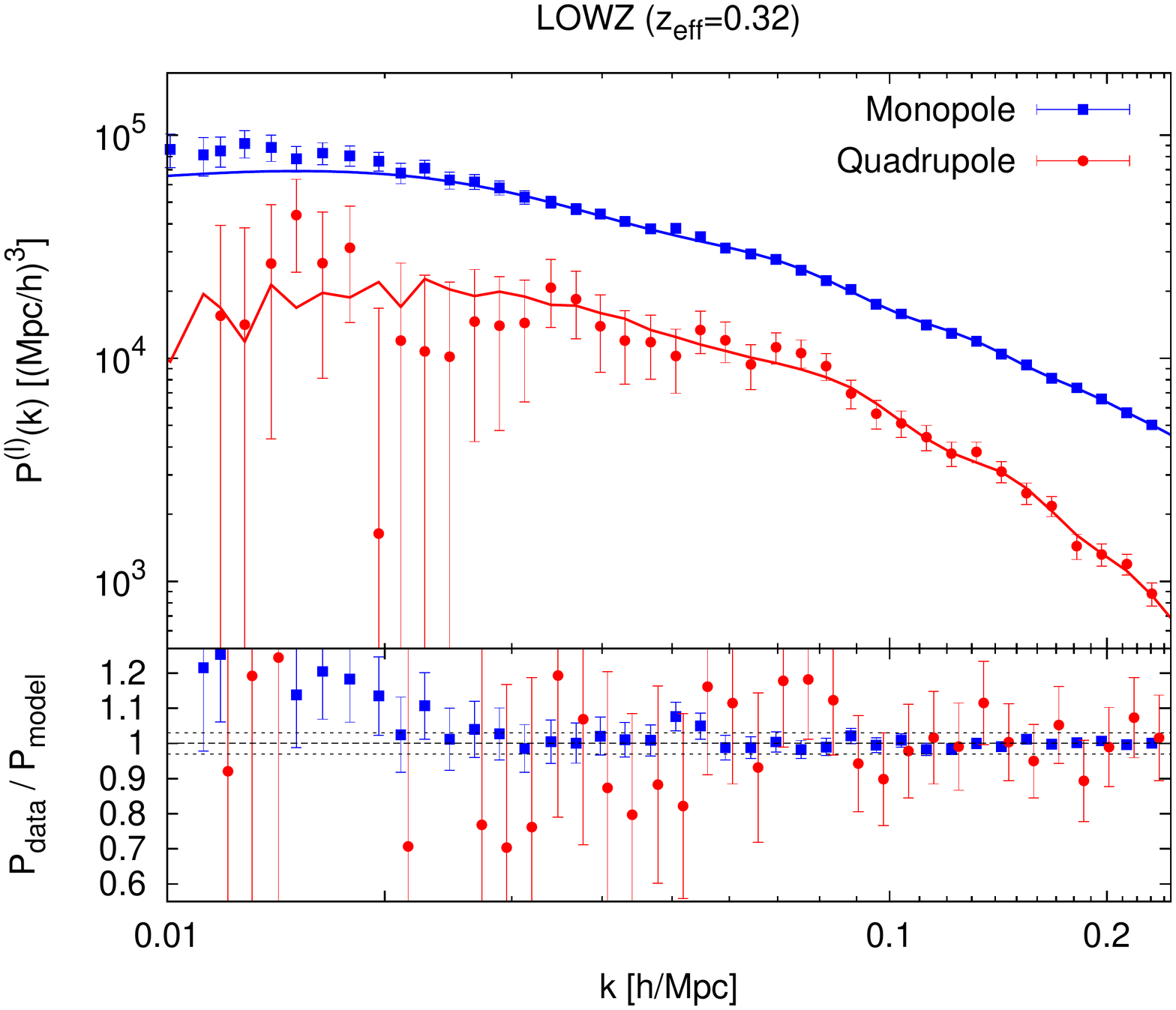}
\includegraphics[scale=0.45]{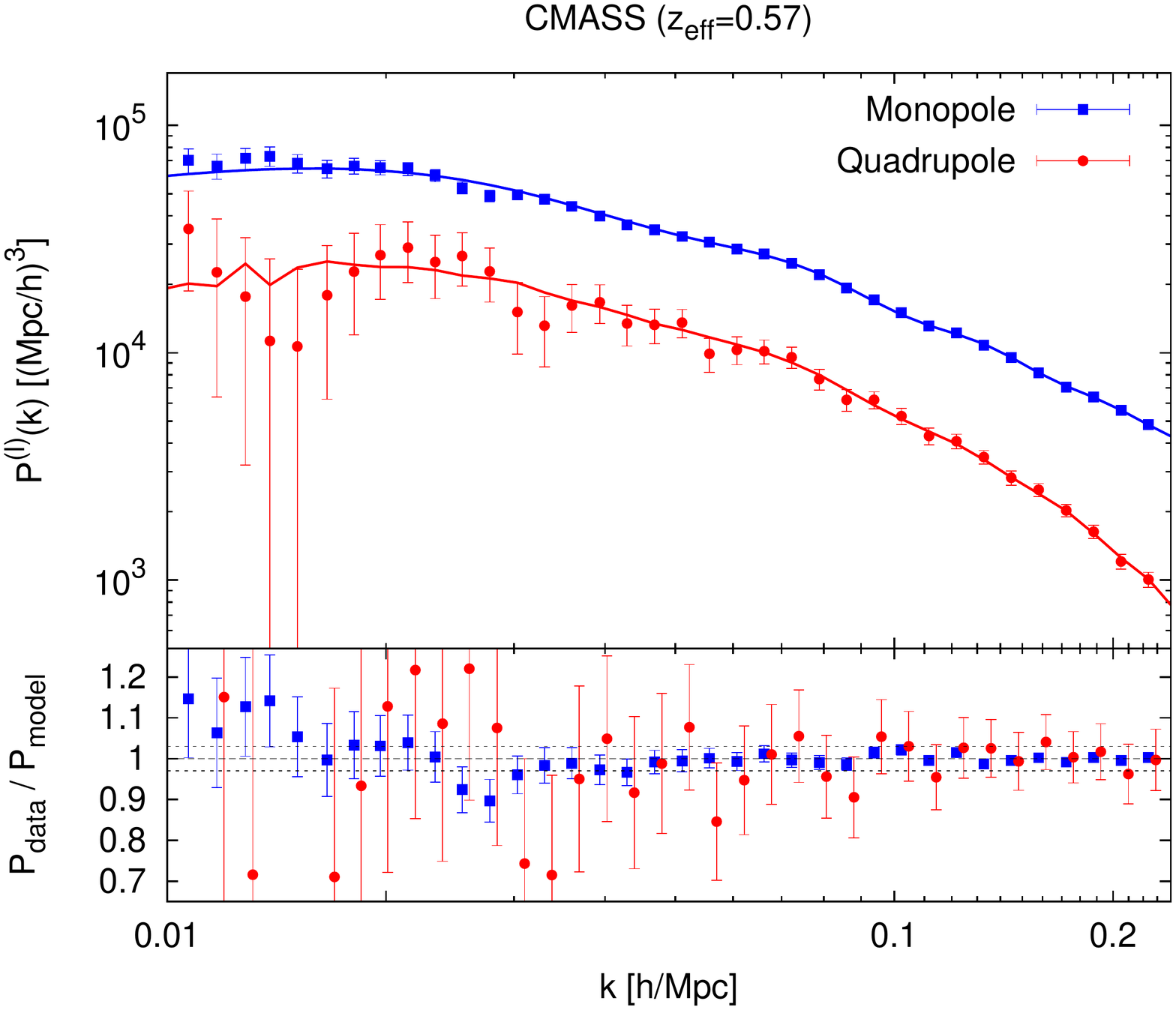}
\caption{The top sub-panels display the measured LOWZ- (top panel) and CMASS-DR12 (bottom panel), monopole (blue squares) and quadrupole (red circles) power spectra. For both cases, the measurements correspond to a combination of the northern and southern galaxy caps according to their effective areas as presented in Eq.\ref{eq:areas_combination}. The error-bars are the average values of the dispersion among realizations of the \textsc{qpm} and \textsc{MD-Patchy} mocks. The red and blue solid lines correspond to the best-fit model using the parameters listed in Table \ref{table_results1}, for $k_{\rm max}=0.24\,h{\rm Mpc}^{-1}$. For simplicity, we plot the average between models corresponding to the parameters obtained using \textsc{qpm} and \textsc{MD-Patchy} covariance matrices. The bottom sub-panels show the ratio between the power spectrum multipoles measurements and the best-fit model presented in the top sub-panel. The quadrupole symbols have been displaced horizontally for clarity.  The dotted black lines represent a $3\%$ deviation.}
\label{dataPS}
\end{figure*}

The top sub-panel of  Fig.~\ref{dataPS} presents the power spectrum monopole (blue squares) and quadrupole (red circles) for LOWZ and CMASS DR12 data measurements (top and bottom panels as labeled) from the combination of the NGC and SGC data.  This combination has performed by averaging the NGC and SGC power spectra weighting  by their effective area,
\begin{equation}
P^{(\ell)}=(P^{(\ell)}_{\rm NGC}A_{\rm NGC}+P^{(\ell)}_{\rm SGC}A_{\rm SGC})/(A_{\rm NGC}+A_{\rm SGC}),
\label{eq:areas_combination}
\end{equation}
where $A_{\rm NGC}$ and $A_{\rm SGC}$ are the effective areas of the NGC and SGC, respectively, whose values are $A_{\rm NGC}^{\rm lowz}=5836\,{\rm deg}^2$, $A_{\rm SGC}^{\rm lowz}=2501\,{\rm deg}^2$, $A_{\rm NGC}^{\rm cmass}=6851\,{\rm deg}^2$ and $A_{\rm SGC}^{\rm cmass}=2525\,{\rm deg}^2$\footnote{  The areas for both samples were   initially planned to be the equal, as both samples are coming from the same spectroscopic pointing. However, due to difficulties during the early phases of the project, the sky area of the LOWZ sample lags that of the full survey by approximately $1000\,{\rm deg}^2$ \citep{Andersonetal:2013}.}.
The best-fit model predictions are shown by the solid lines, taking the average of the models fitted using the covariance extracted from \textsc{qpm} and \textsc{MD-Patchy} mocks with parameters as reported in Table~\ref{table_results1} in \S\ref{sec:analysis}. Details about the models are presented in \S\ref{sec:modelling}, and covariance matrices in \S\ref{sec:covariance}. The blue solid line shows the model for the monopole and the red solid line for the quadrupole. Error-bars correspond to the {\it rms} of the mocks, averaging between those calculated using covariance matrices determined with \textsc{qpm} or \textsc{MD-Patchy} mocks. The comparison of the {\it rms} of the \textsc{qpm} and \textsc{MD-Patchy} mocks can be found in Fig.~\ref{plot:errormocks} and is discussed in \S\ref{sec:covariance}.

 In the lower sub-panels, we present the fractional differences between the data and the best-fit model. For both LOWZ and CMASS samples, the model is able to reproduce the monopole data points up to $k\simeq0.24\,h{\rm Mpc}^{-1}$, within $3\%$ accuracy (indicated by the black dotted horizontal lines). The model reproduces the measured quadrupole with an accuracy of $\sim10\%$ for the LOWZ sample and of $\sim5\%$ for the CMASS sample. However, the signal-to-noise ratio of the observed quadrupole is not sufficiently high to determine if the observed fluctuations are statistical or systematics of the model. For LOWZ and CMASS samples,  the fitting process ignores the  large scales ($k\leq0.02\,h{\rm Mpc}^{-1}$ for the monopole and $k\leq0.04\,h{\rm Mpc}^{-1}$ for the quadrupole) because of the effects of star contamination as it has been discussed in \S\ref{sec:bossdata} (see Appendix \ref{appendix_b} for a further discussion on how these limits have been decided). The effects of the fiber collisions on the best-fit parameters of the models are discussed in Appendix \ref{sec:fc}.

\section{ Modelling the power spectrum multipoles}\label{sec:modelling}
In this section we present the model used to analyse the monopole and quadrupole power spectra in \S\ref{sec:results}. The modellisation is done in the following four steps.
\begin{enumerate}
\item In \S\ref{sec51} we present the  galaxy bias model that maps the dark matter theoretical predictions into galaxy statistical observables. 
\item In \S\ref{sec52} we present the model that relates the real space statistical moments with redshift-space ones. 
\item In \S\ref{sec:AP} we  incorporate AP parameters in order to allow changes due to inaccuracies when converting redshifts into distances by assuming a different value of $\Omega_m$ than the actual one. 
\item In \S\ref{sec:geometry} we describe how the window survey mask is applied in order to account for observational effects due to the geometry of the survey. 
\end{enumerate}
\subsection{The bias model}\label{sec51}
We assume an Eulerian non-linear and non-local bias model proposed by \cite{McDonaldRoy:2009} and previously used for analysing the power spectrum multipoles and bispectrum of DR11 CMASS BOSS galaxies \citep{Beutleretal:2013,hector_bispectrum1}. 
 A priori the non-local galaxy bias model depends on 4 free parameters: the linear bias $b_1$, the non-linear bias $b_2$ and non-local bias parameters $b_{s^2}$ and $b_{3\rm nl}$. In order to reduce the number of free bias parameters, we assume that the bias is local in Lagrangian space, which sets the values of $b_{s^2}$ and $b_{3\rm nl}$ given the linear bias coefficient, $b_1$.  This assumption has been validated using N-body simulations, and it provides consistent results between the power spectrum and bispectrum for the CMASS sample \citep{hector_bispectrum0,hector_bispectrum1}.
In case the condition of local Lagrangian bias were relaxed, the parameters $b_{s^2}$ and $b_{3\rm nl}$ would be treated as free parameters, increasing, consequently, the number of free parameters of the bias model. In such case, the model would account for a more general  galaxy biasing, but the error-bars on the parameters of interest would also increase. Since, the results of the power spectrum and bispectrum of mocks and N-body simulations suggest that the bias model of galaxies and haloes is consistent with local Lagrangian, we also assume it  for the data. 
\subsection{Modelling the redshift space distortions}\label{sec52}

The mapping from real space to redshift space quantities involves the power spectrum of the velocity divergence. We assume that there is no velocity bias between the underling dark matter field and the galaxy field at least on the relatively large scales of interest. We follow the same redshift space modelling that in previous analysis \citep{Beutleretal:2013,hector_bispectrum1}, described in  \cite{Taruyaetal:2010} and \cite{Nishimichietal:2011}, which provides a prediction for the redshift space power spectrum in terms of the matter-matter, velocity-velocity and matter-velocity non-linear power spectra. {   Expressions for the non-linear power spectra used in this paper, were presented in \cite{hector_bispectrum1}; the model for the non-linear matter quantities was obtained using resumed perturbation theory at 2-loop level as described in \cite{HGMetal:2012}. The necessary linear power spectrum input was computed using CAMB \citep{camb}.}

We account for the Fingers-of-God  (hereafter FoG), through the Lorentzian damping factor, as  described in \cite{hector_bispectrum1}. This factor has one free parameter refered as $\sigma_{\rm FoG}$. With this parameter we aim to describe the  non-linear damping due to the velocity dispersion of satellite galaxies inside host haloes. However we treat this factor as an effective parameter that encode our poor understanding of the non-linear RSD and we  marginalised over.

In this paper we consider that the shot noise contribution in the power spectrum monopole may be modified from that of a pure Poisson sampling. We parametrise this deviation through a free parameter, $A_{\rm noise}$, i.e., $P^{(0)}_{\rm noise}=(1-A_{\rm noise})P^{(0)}_{\rm Poisson}$, where the terms  $P^{(0)}_{\rm Poisson}$ is the Poisson predictions for the shot noise as is presented in Eq. \ref{shot_noise}.  For $A_{\rm noise}=0$ we recover the Poisson prediction, whereas when $A_{\rm noise}>0$ we obtain a sub-Poisson shot noise term and $A_{\rm noise}<0$ a super-Poisson noise term.

\subsection{The Alcock-Paczynski effect}\label{sec:AP}

The AP effect \citep{AP} is caused by converting redshift into distance using a different cosmology from the actual one, which introduces a spurious  anisotropy in the power spectrum that can be measured. 
{  Along the LOS, the observed signal is sensitive to the Hubble parameter through $\propto H^{-1}(z)$, when the clustering is measured on scales that are small compared with cosmological changes in the distance-redshift relationship.} On the other hand, in the  angular direction the distortions depend on the angular distance parameter, $D_A(z)$. When a fiducial model is assumed to convert redshifts into distances, the AP effect can be described by the dilation scales,
\begin{eqnarray}
\alpha_{\parallel}&\equiv&\frac{H^{\rm fid}(z) r_s^{\rm fid}(z_d)}{H(z) r_s(z_d)},\\
\alpha_{\perp}&\equiv&\frac{D_A(z)r_s^{\rm fid}(z_d)}{D_A^{\rm fid}(z)r_s(z_d)},
\end{eqnarray}
where $\alpha_\parallel$ and $\alpha_\perp$ are the parallel- and perpendicular-to-the-LOS dilation scales, respectively. Here, $H^{\rm fid}(z)$ and $D_A^{\rm fid}(z)$ are the fiducial values (those corresponding to the assumed cosmology to convert redshifts into distances) of the Hubble constant and the angular dimeter distance at a given redshift $z$, respectively. On the other hand, the fiducial sound horizon at the baryon-drag redshift is given by $r_s^{\rm fid}(z_d)$. The factors $\alpha_{\parallel}$ and $\alpha_\perp$ describe how the true wave-length modes, $k'_{\parallel}$ and $k'_\perp$, have been distorted into the observed ones, $k_\parallel$ and $k_\perp$: $k_\parallel=\alpha_\parallel k'_\parallel$ and  $k_\perp=\alpha_\perp k'_\perp$, by the effect of assuming a different cosmological model. {  

This component of observed anisotropy is modelled through the $\alpha_\parallel$ and $\alpha_\perp$ parameters (see eq. 60 of \citealt{Beutleretal:2013}). The geometric AP effect also affects the BAO scale, and the assumed distance-redshift relation has the potential to shift the BAO peak position differently in the monopole and quadrupole moments of the observed comoving clustering signal. Therefore, the AP parameters are simultaneously measured in our analysis, even though our focus is on measuring the RSD. Analyses that do not wish to measure the RSD signal and focus on the BAO scale can make use of reconstruction and therefore, in general, provide better measurements of the AP effect (e.g. \citealt{cuesta,gil-marin15b}). We will comment on the potential differences further in \S\ref{sec:boss}.

}

Assuming the fiducial cosmology described in \S\ref{sec:fiducial_cosmo}, the fiducial values for $H(z)$ and $D_A(z)$ are, $H^{\rm fid}(z_{\rm lowz})=79.49\,{\rm km}{s}^{-1}{\rm Mpc}^{-1}$, $D_A^{\rm fid}(z_{\rm lowz})=999.23\,{\rm Mpc}$ for the LOWZ sample at $z_{\rm lowz}=0.32$, and  $H^{\rm fid}(z_{\rm cmass})=92.25\,{\rm km}{s}^{-1}{\rm Mpc}^{-1}$, $D_A^{\rm fid}(z_{\rm cmass})=1398.43\,{\rm Mpc}$ for the CMASS sample at $z_{\rm cmass}=0.57$. The value for the fiducial sound horizon distance is $r_s^{\rm fid}(z_d)=148.11\, {\rm Mpc}$.

\subsection{ The survey geometry}\label{sec:geometry}
The estimator presented in \S\ref{sec:estimator_ps} only provides an unbiased prediction of the true underlying power spectrum without any survey geometry effects. At intermediate and large scales, the measurement is affected by the shape of the survey, especially for high order multipoles. Given a theoretical anisotropic power spectrum $P^{\rm theo.}({\bf k}')$, the observed power spectrum due to the effects of the survey is windowed through the following expression, 
\begin{equation}
\label{Pwin}P^{\rm win.}({\bf k})=\int \frac{d^3{\bf k}'}{(2\pi)^3}\,P^{\rm theo.}({\bf k}')|W({\bf k}-{\bf k}')|^2,
\end{equation}
where $W$ is defined as, 
\begin{equation}
W({\bf k})\equiv\frac{\alpha}{I_2^{1/2}}\int d^3{\bf r}\,\bar{n}_s({\bf r})e^{i{\bf k}\cdot{\bf r}}.
\end{equation}
We refer to $|W|^2$ as the window function, which satisfy the normalization condition, $\int {d^3{\bf k}'}/{(2\pi)^3}\,|W({\bf k}')|^2=1$, imposed by the definition of the factor $I_2$ in Eq.~\ref{eq:I2def}.
The functional provided by $P^{\rm win.}[P^{\rm theo.}]$ in Eq. \ref{Pwin} is a convolution. Therefore, the convolution theorem can be apply making use of FFT techniques,  which allows the computation in a minutes-time scale per model $P^{\rm theo.}$.

We assume that the monopole and quadrupole provide all the information about the full $\mu$-shape of the power spectrum. Thus we can write,
\begin{equation}
\label{eq:Ptheo}P^{\rm theo.}({\bf k})={P^{(0)}}^{\rm theo.}(k)+{P^{(2)}}^{\rm theo.}(k)\mathcal{L}_2(\mu).
\end{equation}
We can define a windowed power spectrum $\ell$-multipole as, 
\begin{equation}
\label{eq:Pwinell}{P^{(\ell)}}^{\rm win.}(k)=\frac{2\ell+1}{2}\int d\mu_{\bf k}\,\mathcal{L}_\ell(\mu_{\bf k}) P^{\rm win.}({\bf k}),
\end{equation}
where $\mu_{\bf k}\equiv\hat{r}_z\cdot\hat{\bf k}$.
From Eq.~\ref{eq:Pwinell} it is clear to see that both ${P^{(0)}}^{\rm win.}$ and ${P^{(2)}}^{\rm win.}$  have contributions from both ${P^{(0)}}^{\rm theo.}$ and ${P^{(2)}}^{\rm theo.}$. Eq. \ref{eq:Pwinell} provides a full description of the effect of the window in the monopole and quadrupole. However, for practical reasons, is convenient to use this equation to calibrate a  matrix that is able to relate the ${P^{(\ell)}}^{\rm win.}$ at a given $k$-bin from an arbitrary shape of ${P^{(\ell)}}^{\rm theo.}$. Using Eq.~\ref{eq:Pwinell} we write the matrix elements,
\begin{eqnarray}
 \nonumber \mathcal{W}_{ij}^{\ell \ell'}&\equiv&\left[  \frac{2\ell'+1}{2} \int d\mu_{\bf k} \mathcal{L}_{\ell'}(\mu_k)\int d^3{\bf k}'\, |W({\bf k}-{\bf k}')|^2 \right.\\
 \nonumber&\times&\left. {P^{(\ell)} }^{\rm theo.}({ k}') \mathcal{L}_{\ell}(\mu')\Theta_{\rm TH}(k_i-k')   \right]/{P^{(\ell')}}^{\rm theo.}(k_j),\\
\label{eq:Wij}
\end{eqnarray}
where $\Theta_{\rm TH}(k_i-k')$ is a top hat function around the $k_i$-bin: $\Theta_{\rm TH}(k_i-k')$ is 1 when $k'$ belongs to $k_i$-bin and 0 otherwise.  In order to form a window-matrix which is able to mimic the behaviour described by Eq.~\ref{eq:Pwinell} we have used 1000 $k$-bins  between $k_f$ and $k=0.5\,h{\rm Mpc}^{-1}$ as an input of $k_j$ and 60 output $k_i$-bins which coincides with the $k$-values where we measure the dataset.  
Using the calibrated values of the window matrix described by Eq.~\ref{eq:Wij} we write the windowed monopole and quadrupole power spectra as, 
\begin{eqnarray}
\nonumber {P^{(0)}}^{\rm win.}(k_i)&=&\sum_j \mathcal{W}_{ij}^{00} {P^{(0)}}^{\rm theo.}(k_j) + \sum_j \mathcal{W}_{ij}^{02} {P^{(2)}}^{\rm theo.}(k_j),\\
\nonumber {P^{(2)}}^{\rm win.}(k_i)&=&\sum_j \mathcal{W}_{ij}^{20} {P^{(0)}}^{\rm theo.}(k_j) + \sum_j \mathcal{W}_{ij}^{22} {P^{(2)}}^{\rm theo.}(k_j).\\
\label{eq:Wijll}
\end{eqnarray}
Note that Eq.~\ref{eq:Pwinell} and \ref{eq:Wijll} describe the same survey window effect, but the latter one is much faster to be applied to minimization and \textsc{mcmc} algorithms.
In this paper we always use combined window that we obtain by weighting the individual windows of  NGC and SCG by their area. Since this is what we do with the power spectrum measurements, the combined window reproduce by definition the combined power spectrum.  

{  The matrix terms in Eq. \ref{eq:Wij} depend, in principle, on the choice of theoretical power spectrum,  ${P^{(\ell)} }^{\rm theo.}$. However this dependence is expected to be weak, as the term ${P^{(\ell)}}^{\rm theo.}$ appears in both denominator and numerator,  and the part of the signal in $W_{ij}^{\ell \ell'}$ coming from ${P^{(\ell)} }^{\rm theo.}$ is cancelled. In order to quantify this, we have built two window matrices with different power spectra, whose values are within $2\sigma$ of the data and no significant differences were observed.

}

\begin{figure*}
\centering
\includegraphics[scale=0.3]{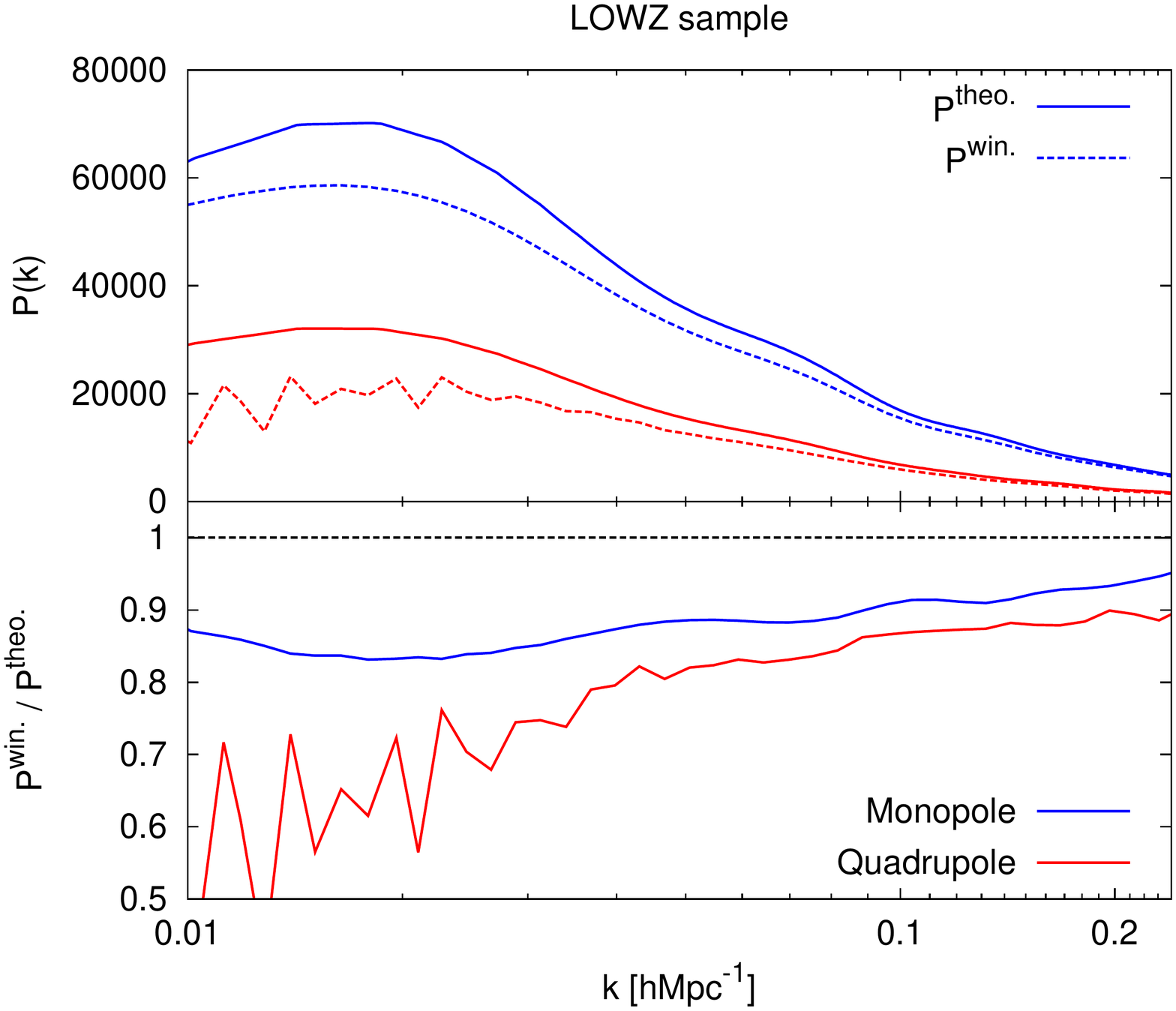}
\includegraphics[scale=0.3]{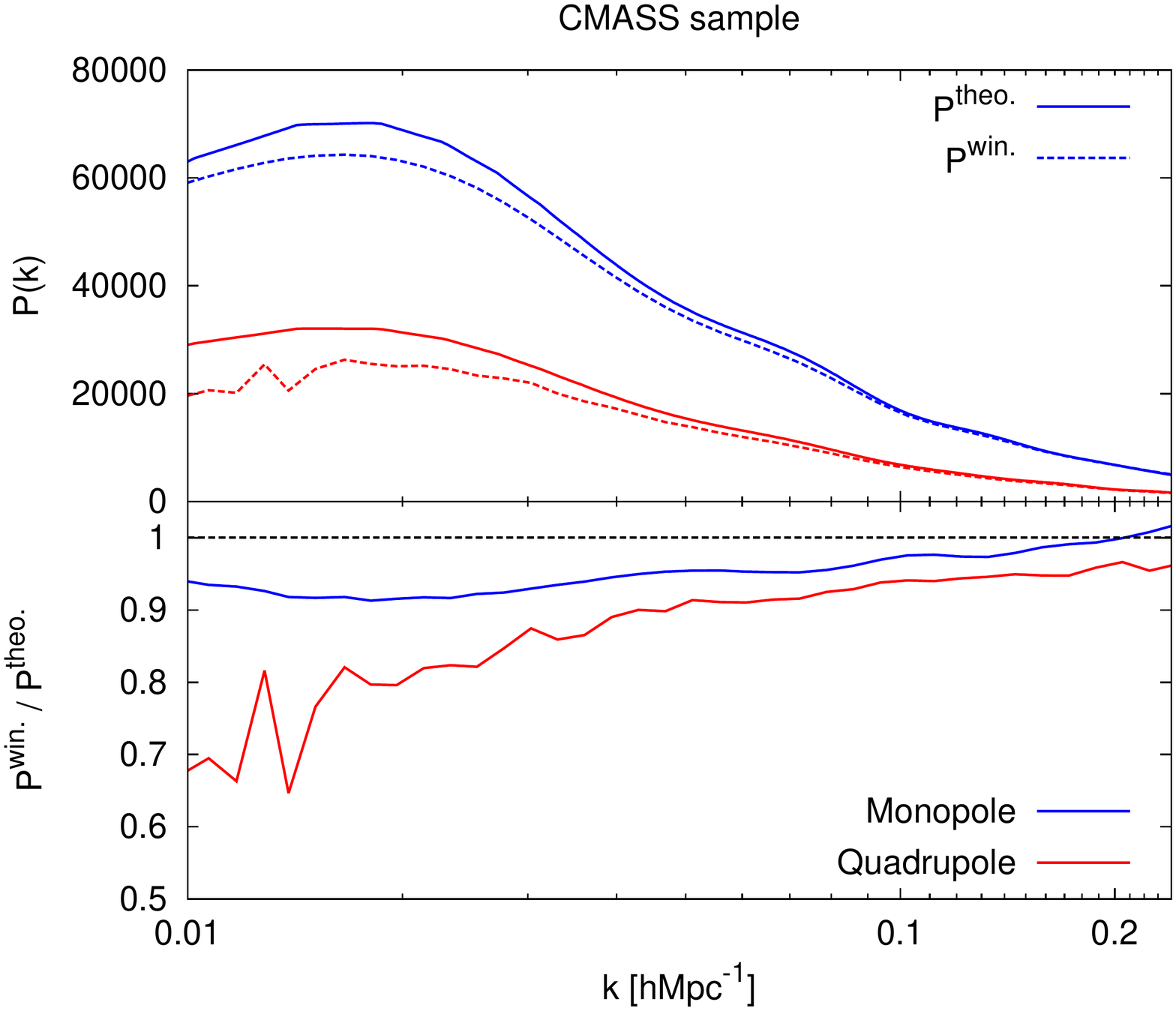}

\caption{Top sub-panels: Effect of the window function on the monopole (blue lines) and on the quadrupole (red lines) for the LOWZ-DR12 sample (left panel) and for the CMASS-DR12 sample (right panel). The solid lines correspond to a toy-model for ${P^{\rm theo.}}^{(0)}$ and ${P^{\rm theo.}}^{(2)}$ in Eq.~\ref{eq:Ptheo}. The dashed lines correspond to ${P^{(0)}}^{\rm win.}$ and ${P^{(2)}}^{\rm win.}$ in Eq.~\ref{eq:Wijll}. Lower sub-panels:  Relative deviation between ${P^{(\ell)}}^{\rm win.}$ and its convolution with the window mask. }
\label{plot:window}
\end{figure*}

Fig. \ref{plot:window} shows the effect of the LOWZ (left panel) and CMASS (right panel) DR12 samples on the power spectrum monopole and quadrupole. The upper panel shows a toy model input power spectrum monopole (blue solid lines) and quadrupole (red solid lines) and its output according to Eq.~\ref{eq:Wijll} (blue and red dashed lines). The lower panels show the ratio between the input and the output power spectrum multipole. We see that for CMASS the effect, both in the monopole and in the quadrupole, is smaller than for the LOWZ. This is due to the size of the sample. The larger the sample is, the smaller is the effect of the mask. At $k\simeq0.2\,h{\rm Mpc}^{-1}$, the effect of the window is near 0\% for the monopole in CMASS, $\sim5\%$ for the quadrupole in CMASS, $\sim7\%$ for the monopole in LOWZ and $\sim10\%$ for the quadrupole in LOWZ. {   At large scales, the windowed power spectrum predicted for the quadrupole has oscillatory behaviour, a consequence of the lack of $\mu$-modes in the $k$-bins at these scales: since the $k$-binning is logarithmic, the number of fundamental $k$-modes contained at large scales is much smaller than that at small scales. As a consequence, at these large scales the constraint $\int d\mu \mathcal{L}_\ell(\mu)\mathcal{L}_{\ell'}(\mu)=\delta_{\ell \ell'}$  is not satisfied for $\ell\,, \ell'>0$ with sufficient precision. The behaviour can be corrected by renormalising, both data and window by the sum of squared Legendre polynomials over the $\mu$-modes: $\mu$-modes $\sum_{\mu-{\rm modes}} \mathcal{L}^2_{\ell}(\mu)$. However, since for this paper we only use quadrupole data for $k>0.04\,h{\rm Mpc}^{-1}$, where the $\mu$-modes contained in the $k$-bins are large enough, the correction is not significant enough to be worth including. In any case, since the correction would have to be applied to both data and window, the relative ratio between theory and data would be unaffected by this correction.

}

In this paper we do not correct the power spectrum multipoles for the integral constrain, which only produce a significant effect at scales comparable to the size of the survey (see e.g. \citealt{Peackok_Nicholson91}), which in our case is $k_{\rm f}\sim0.002\,h{\rm Mpc}^{-1}$. Since the largest scales we consider are 10 and 20 times smaller for the monopole and quadrupole, respectively, in our analysis the integral constrain is a subdominant component compared to other effects such as systematic weights.  

\section{Parameter estimation}\label{sec:parameter_estimation}
In this section we describe how the parameters of interest and their errors, including the AP parameters and $f\sigma_8$ are estimated. We also present a systematic test on the power spectrum model presented in \S\ref{sec:modelling} using the galaxy mocks.
\subsection{Covariance Matrices}\label{sec:covariance_matrices}\label{sec:covariance}
The covariance matrix of the monopole and quadrupole is computed using the different realizations of the two sets of gtalaxy mocks described in \S\ref{sec:mocks}. We take into account the covariance of the monopole and quadrupole $k$-bins and also the cross-covariance between these two. Each element of the covariance matrix is calculated from the mocks, 
\begin{eqnarray}
\nonumber C_{i,j}^{\ell \ell'}=\frac{1}{N_m-1}\sum_{m=1}^{N_m}&& [P_m^{(\ell)}(k_i)-\langle P^{(\ell)}(k_i) \rangle]\\
&\times& [P_m^{(\ell')}(k_j)-\langle P^{(\ell')}(k_j) \rangle],
\end{eqnarray}
where $\langle P^{(\ell)}(k) \rangle \equiv \sum_m^{N_m} P_m^{(\ell)}(k) /N_m$ is the mean of the $\ell$-multipoles among realizations, and $N_m$ is the number of independent realizations.  The full covariance matrix for the monopole and quadrupole can be written in terms of the matrices $C^{\ell \ell'}$,  for $\ell=0,\,2$, as, 
\begin{equation}
C =
 \begin{pmatrix}
  C^{00} & C^{02} \\
  C^{20} & C^{22}
 \end{pmatrix}.
 \end{equation}
For the \textsc{qpm} mocks the number of independent realization is $N_m=1000$, whereas for \textsc{MD-Patchy} is $N_m=2048$. In both cases the number of elements is much larger than the total number of bins, $n_b$, which for this work is $n_b=120$.

Since the covariance matrix $C$ is estimated from a set of mocks, its inverse $C^{-1}$ is biased due to the limited number of realizations. We account for this effect by applying the correction proposed by \cite{Hartlap07}. In addition to this scaling, we have to propagate the error in the covariance matrix to the error on the estimated parameters. We do this by scaling the variance for each parameter by the factor of eq. 18 of \cite{Percival13}. However, we observe that the correction due to this effect is subdominant, namely $\leq2\%$.

\begin{figure*}
\centering
\includegraphics[trim = 20mm -5mm 41mm 5mm, clip=false,scale=0.35]{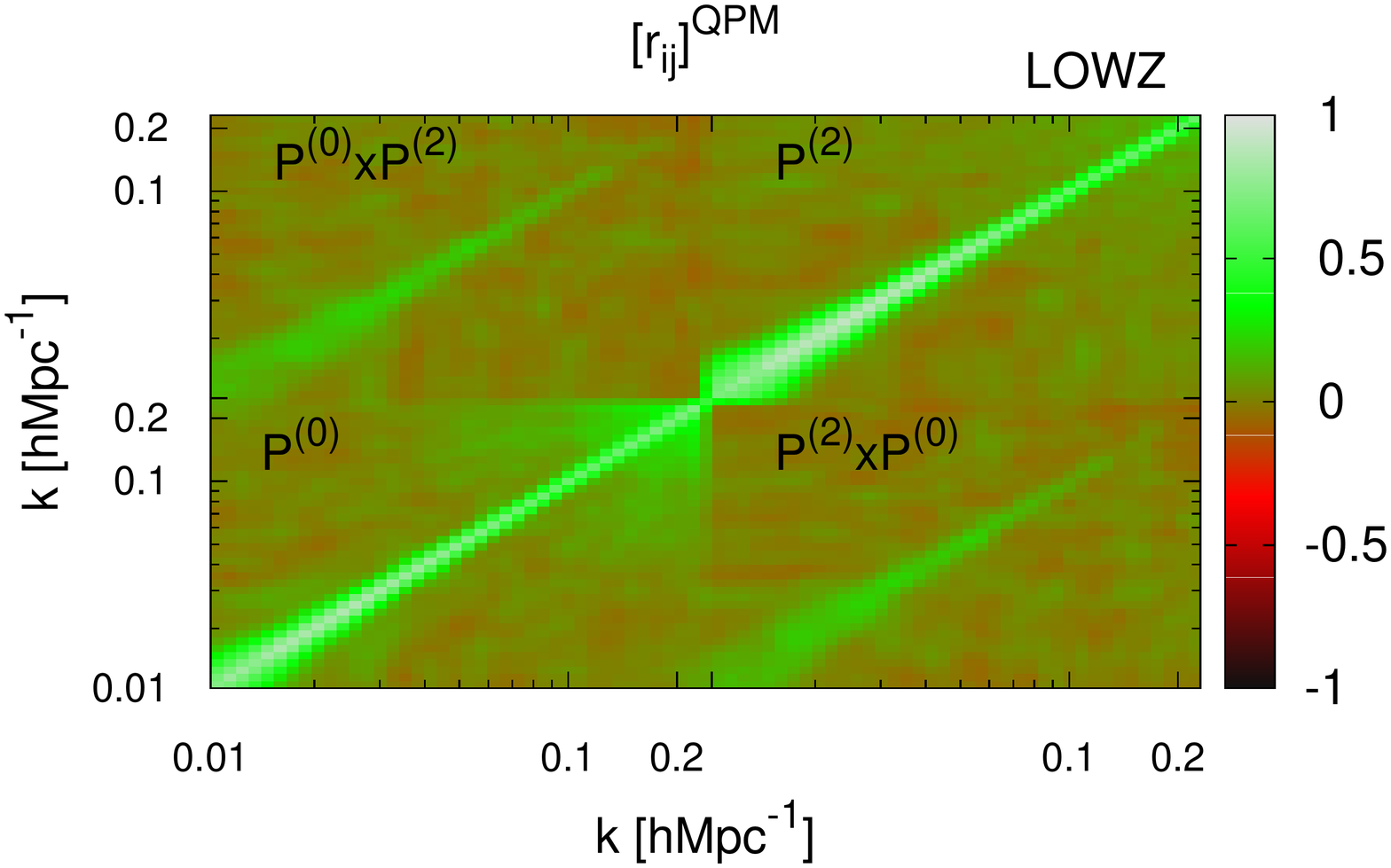}
\includegraphics[trim = 0mm -5mm 0mm 5mm, clip=false,scale=0.35]{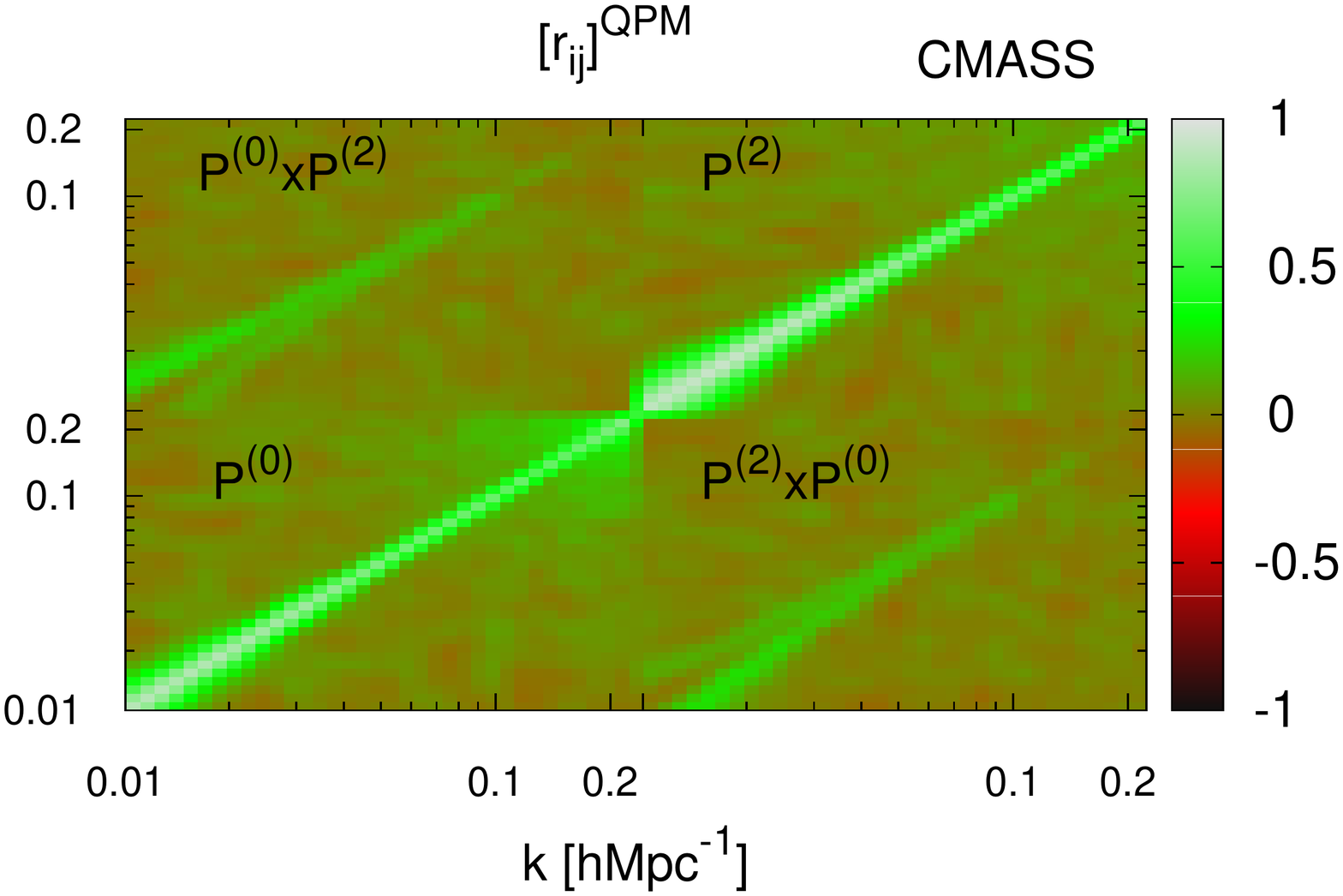}

\includegraphics[trim = 20mm 0mm 41mm 60mm, clip=false,scale=0.35]{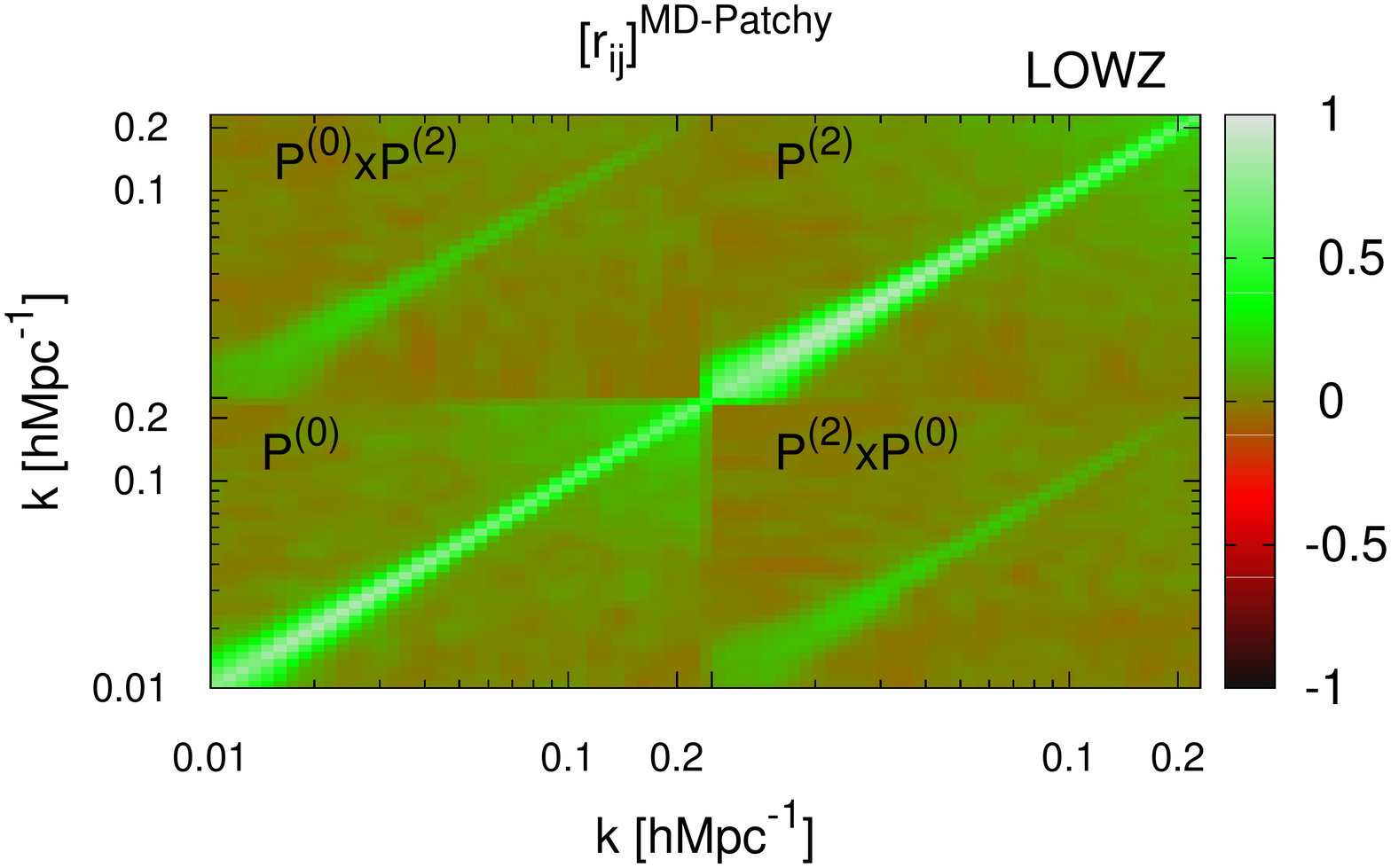}
\includegraphics[trim = 0mm 0mm 0mm 60mm, clip=fase,scale=0.35]{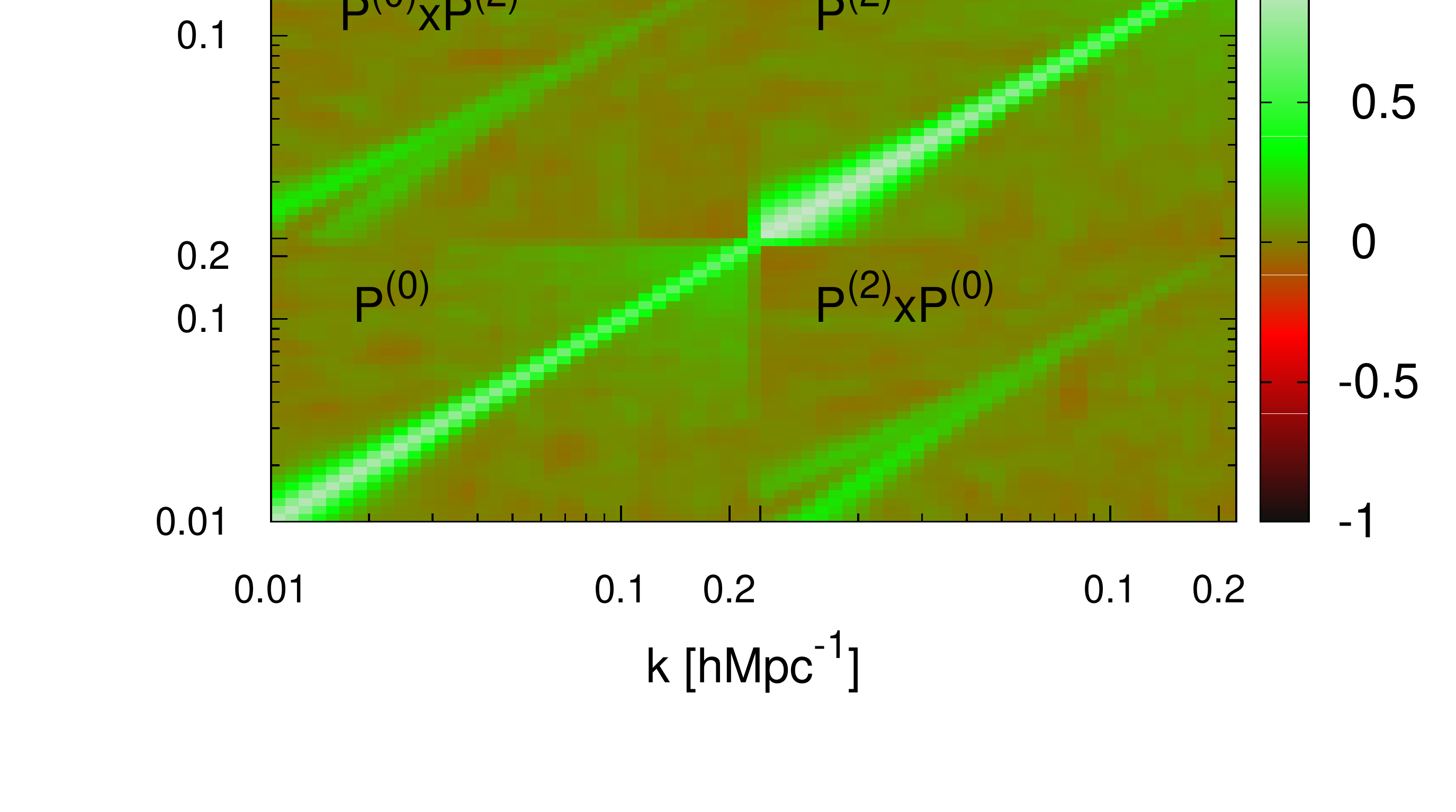}

\includegraphics[trim = 30mm -5mm 10mm 55mm, clip=false,scale=0.35]{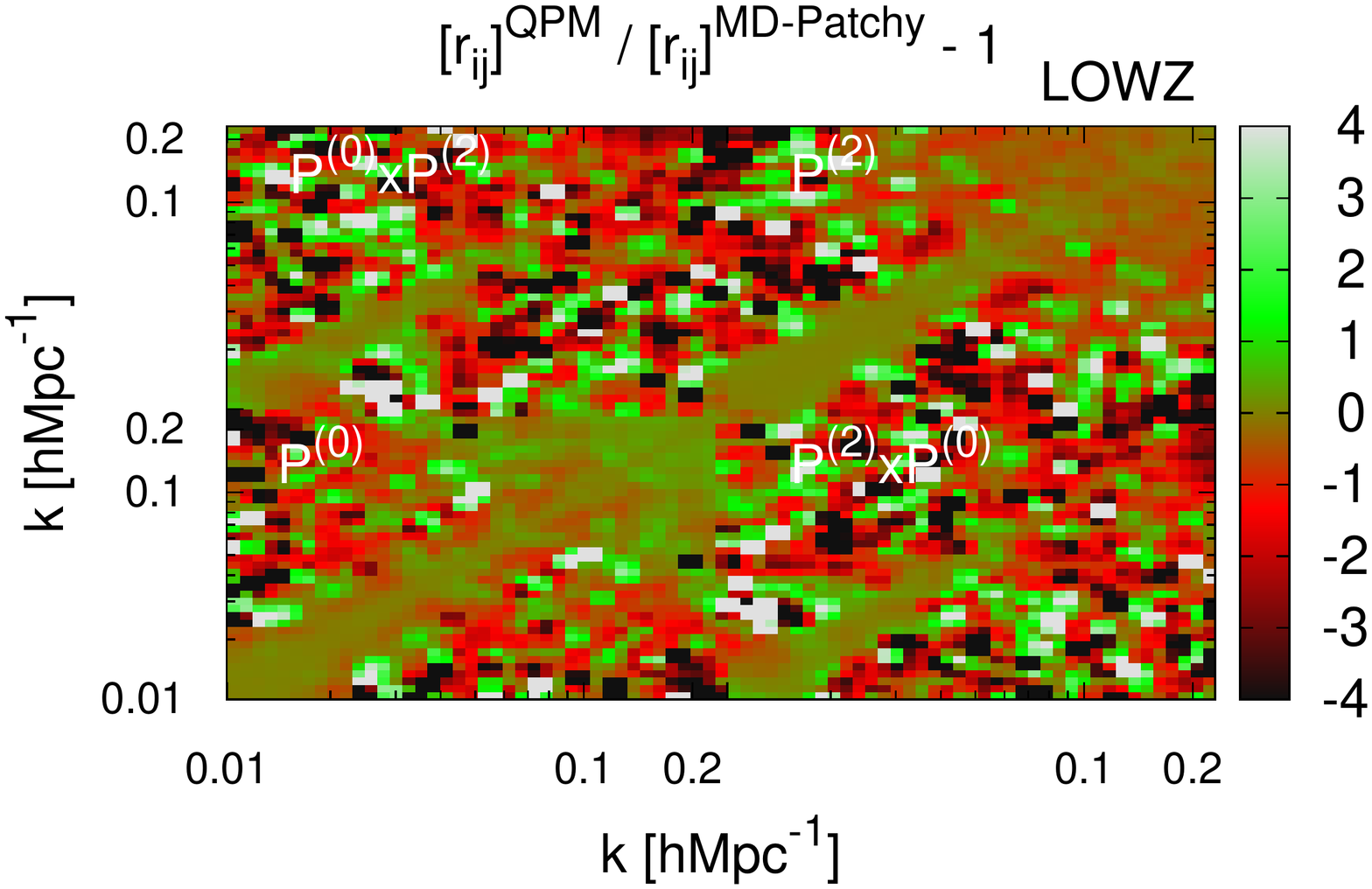}
\includegraphics[trim = 30mm -5mm 10mm 55mm, clip=false,scale=0.35]{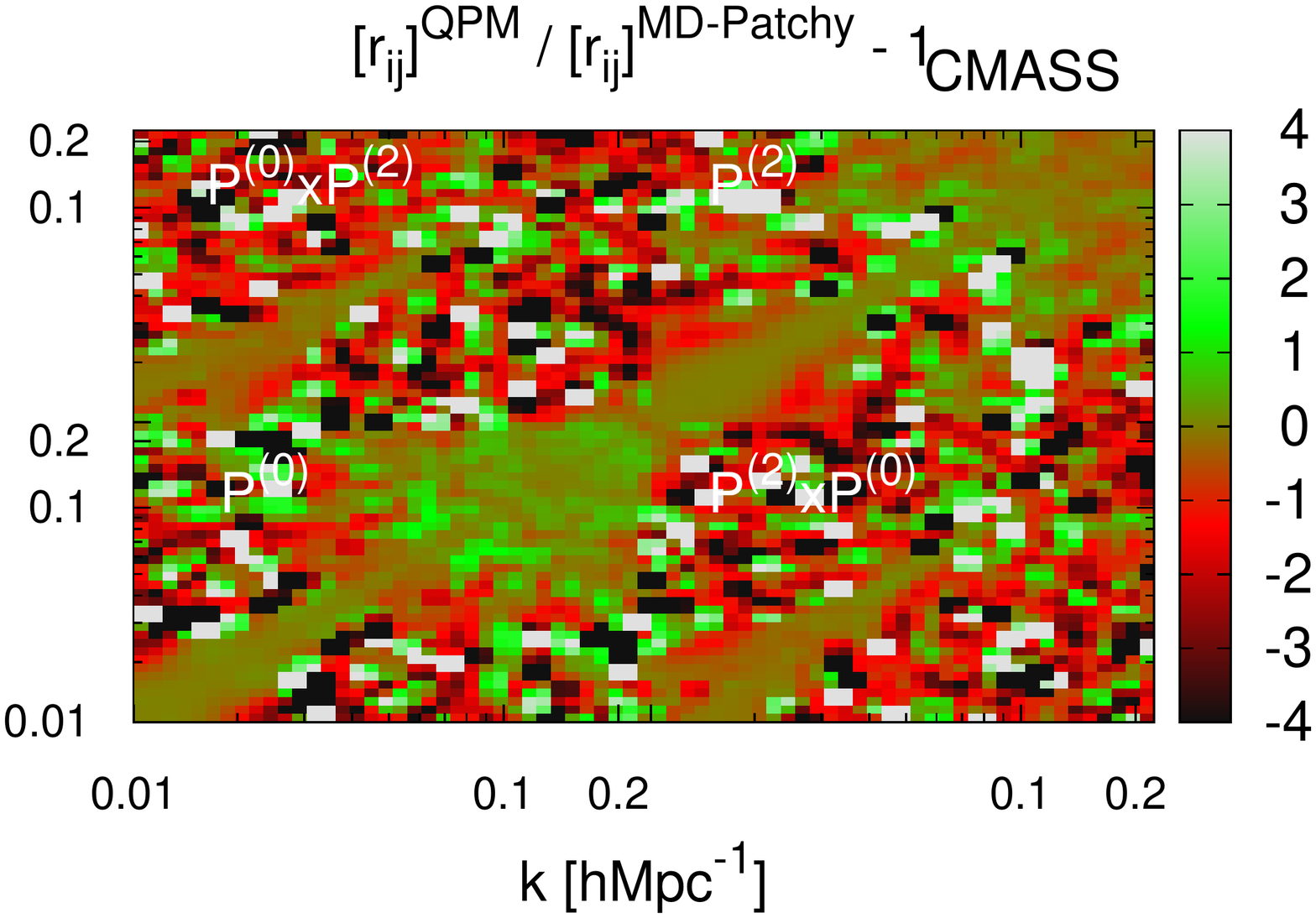}

\caption{Correlation coefficients of the monopole-quadrupole covariance matrix from LOWZ-DR12 sample (left panels) and from the CMASS-DR12 sample (right panels), extracted from 1000 realizations of the \textsc{qpm} mocks and from 2048 realizations of the \textsc{MD-Patchy} mocks. The top panels show the results for \textsc{qpm} mocks, the middle panels for the \textsc{MD-Patchy} mocks, and the bottom panels their ratio.}
\label{Cij}
\end{figure*}

The middle and top panels of Fig. \ref{Cij}  display the correlation coefficient matrices, $r_{i,j}\equiv C_{i,j}/[C_{i,i}C_{j,j}]^{1/2}$ for LOWZ and CMASS samples using the \textsc{qpm} mocks and the \textsc{MD-Patchy} mocks as labeled. In all cases NGC and SGC  have been combined into a single sample, as described in \S\ref{sec:results}.
We observe that the off-diagonal terms of the  auto-covariance (the covariance between monopole-monopole and quadrupole-quadrupole) are significantly correlated  at large scales because of the effect of the survey geometry, for both monopole and quadrupole. The off-diagonal terms of the cross-covariance between monopole and quadrupole present significantly smaller  correlation. 

 As we go to smaller scales the auto- and cross-covariance off-diagonal terms are reduced for both LOWZ and CMASS samples because the effect of the survey window is less important. For $k\geq0.2$ the auto-covariance off-diagonal elements start growing again because of the effect of mode coupling, which becomes more important as we go to smaller scales. At the same scales, the off-diagonal terms of cross-covariance stay very close to 0. This suggests that the mode coupling induces a strong correlation between close $k$-modes in the monopole and quadrupole, but not a correlation between these two statistics until $k=0.3\,h{\rm Mpc}^{-1}$. 
 
 The bottom panels of Fig. \ref{Cij} show the ratio between the \textsc{qpm} and \textsc{MD-Patchy} covariances for LOWZ and CMASS samples, as labeled, in order to stress their differences. We observe that most of the off-diagonal signal is very noisy, although some differences can be seen  for those off-diagonal terms close to the diagonal. For both LOWZ and CMASS samples, both   \textsc{qpm} and \textsc{MD-Patchy} mocks predict the same degree of correlation on large scales. As we go to smaller scales some differences arise. At scales where $k\geq0.1,\,h{\rm Mpc}^{-1}$, the off-diagonal terms of the monopole auto-correlation matrix, tend to be more correlated in the \textsc{qpm} mocks than in the \textsc{MD-Patchy} mocks, while the auto-correlation matrix estimates for the quadrupole are similar.

\begin{figure}
\centering
\includegraphics[scale=0.3]{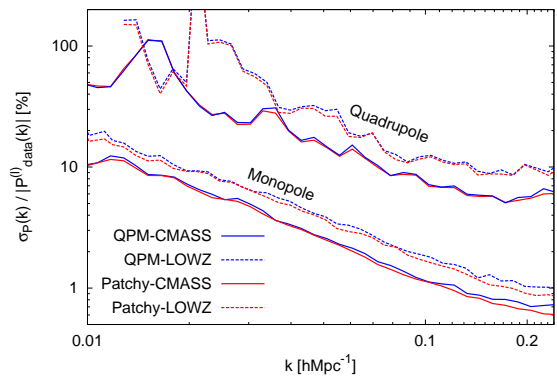}
\caption{Percentile diagonal errors corresponding to the $k$-bins used in \S\ref{sec:results} in which the monopole and quadrupole have been measured for the mocks. Solid lines display the CMASS-DR12 statistics and dashed lines the LOWZ-DR12 ones. Red lines are the predictions inferred from the \textsc{MD-Patchy} mocks, whereas the blue lines are according to \textsc{qpm} mocks. }
\label{plot:errormocks}
\end{figure}

In Fig. \ref{plot:errormocks} we show the percentile diagonal error for power spectrum multipoles, corresponding to the  $k$-bins used in \S\ref{sec:results}, relative the  power spectrum multipole amplitude of the data.  At large scales both \textsc{qpm} and \textsc{MD-Patchy} prediction agree well for both the LOWZ and CMASS samples for both monopole and quadrupole. At  small scales \textsc{qpm} mocks predict {\it higher} mode coupling in the monopole than \textsc{MD-Patchy}  and therefore the errors of the power spectrum monopole saturate {\it before} that of the \textsc{MD-Patchy} predictions. The precision of the monopole exceeds $1\%$ for $k\leq0.13\,h{\rm Mpc}^{-1}$. 

For the quadrupole the relative error predicted by \textsc{qpm} and \textsc{MD-Patchy} mocks is very similar in all the range of scales studied. 

Overall, the covariances extracted from \textsc{qpm} and \textsc{MD-Patchy} mocks are similar and the main discrepancies are for the monopole at small scales.  We are not able to discriminate  which of these to sets of mocks is best. 

\subsection{Best-fit and error estimation}\label{sec:errror_estimation}

We model the amplitude and the shape of the power spectrum monopole and quadrupole through a set of 8 free parameters ${\bf \Psi}=\{b_1, b_2, f, \sigma_8, A_{\rm noise}, \sigma_{\rm FoG}, \alpha_\parallel, \alpha_\perp\}$, which we briefly describe below. 
\begin{enumerate}
\item The galaxy bias is modelled using two bias parameters, $b_1$ and $b_2$ as described in \S\ref{sec51}. The value for the non-local bias parameters is set by the value of $b_1$ under the assumption of local Lagrangian bias.
\item The logarithmic growth factor $f$. This parameter can be predicted for a specific cosmological model (when the $\Omega_m$ value is known) if we assume a theory of gravity. In this paper we consider this as free parameter in order to test potential deviations from GR, or, if assuming GR, for not using a prior on the $\Omega_m$ value. 
\item The AP parameters, $\alpha_\parallel$ and $\alpha_\perp$. As described in \S\ref{sec:AP}, through varying $\alpha_\parallel$ and $\alpha_\perp$ we are able to parametrise the anisotropy generated in the power spectrum multipoles by assuming an incorrect value of $\Omega_m$. 
\item The amplitude of primordial dark matter power spectrum, $\sigma_8$. 
\item The amplitude of shot noise, $A_{\rm noise}$, as described in \S\ref{sec52}
\item The Fingers-of-God parameter, $\sigma_{\rm FoG}$, introduced in \S\ref{sec52}.
\end{enumerate}
Note that although we allow $f$ and $\sigma_8$  to vary  independently,  these two parameters are highly degenerate when constrained only using the power spectrum multipoles\footnote{$f$ and $\sigma_8$ are fully degenerated only in the large-scale limit, when Kaiser formula is still valid. At smaller scales the non-linear corrections break this degeneration. However, the signal-to-noise of the power spectrum at these scales is not sufficiently high to make reasonable constrains on $f$ and $\sigma_8$ alone, since the degeneration is poorly broken. Because of this we refer to these two parameters as highly degenerate but not fully degenerate.}. We will only quote the measured combined quantity $f\sigma_8$. Similarly, we will only report $b_1\sigma_8$, and $b_2\sigma_8$. 

The other cosmological parameters, such as $\Omega_m$, the spectral index $n_s$ and the Hubble parameter $h$ are fixed at  the fiducial values described in \S\ref{sec:fiducial_cosmo} during the fitting process. In \S\ref{sec:dependence_cosmo} we will vary the value of $\Omega_m$ to analyse the effect of such a change can produce in $f\sigma_8$. In this paper we always perform the parameter fitting process to the combined NGC+SGC sample, both for the mocks and for the data. 

In order to preform the parameter estimation, we  assume the monopole and quadrupole are drawn from a multivariate Gaussian distribution and use,
\begin{equation}
\chi^2({\bf \Psi})=[\Delta P({\bf \Psi})] [{\widetilde{C}^{-1}}][\Delta P({\bf \Psi})]^t
\end{equation}
where, $\Delta P({\bf \Psi})$ is the vector whose elements contain the difference between the data and the model for the power spectrum monopole and quadrupole, and $\widetilde{C^{-1}}$ is the inverse covariance matrix. 
By minimizing the $\chi^2$ function respect to ${\bf \Psi}$ we obtain  the best-fit set of parameters. The errors associated to each parameter are computed by exploring the likelihood surface when a specific parameter is fixed and the other parameters can vary freely. {  The likelihood surface is explored using a \textsc{simplex} minimization algorithm. In order to ensure that the minima found are global and not local, we run the algorithm multiple time with different starting points and different starting variation ranges, and check for convergence. }

\subsection{Tests for the galaxy mocks}\label{sec:results_mocks}

\begin{figure*}
\centering
\includegraphics[scale=0.31]{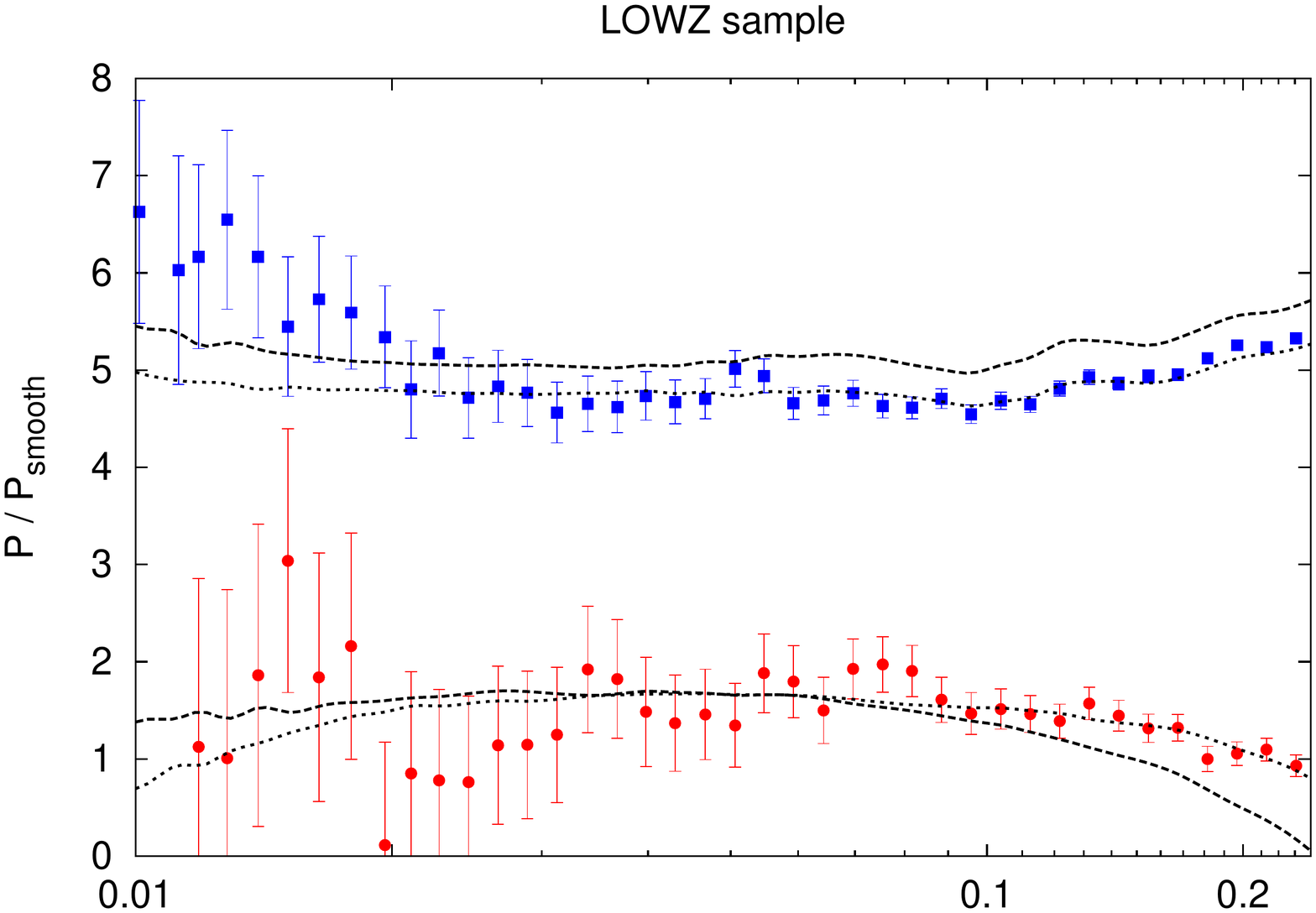}
\includegraphics[scale=0.31]{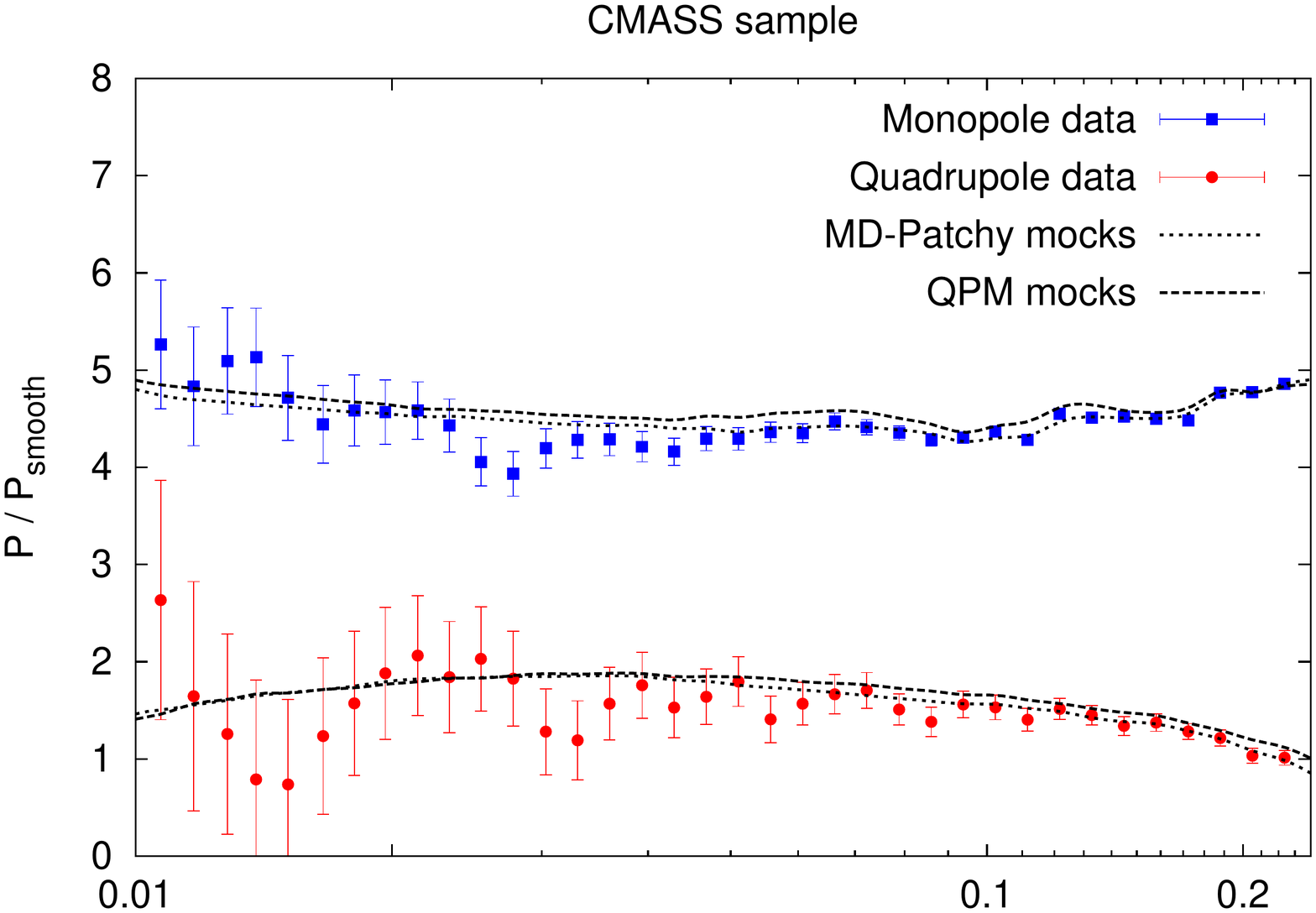}
\caption{Performance of the \textsc{qpm} (black dashed lines) and \textsc{MD-Patchy} (black dotted lines) mocks compared to the data, blue squares for the monopole and red circles for the quadrupole. The left and right panels display the results for the LOWZ and CMASS samples, respectively. For clarity, the amplitude of the power spectrum multipoles have been normalised by a smoothed linear power spectrum, $P_{\rm smooth}$. }
\label{plot:mockscomparison}
\end{figure*}

In this section we test for potential systematics of the model presented in \S\ref{sec:modelling} using the \textsc{MD-Patchy} galaxy mocks. We start by comparing the power spectrum multipoles for the mean of the 1000 and 2048 realizations of the \textsc{qpm} and \textsc{MD-Patchy} mocks, respectively, with the measurements of the DR12 data. This is shown in Fig. \ref{plot:mockscomparison},  where  the power spectrum monopole and quadrupole are divided by a smooth linear power spectrum. The left and right panels show the measurements corresponding to LOWZ and CMASS samples, respectively. Blue symbols represent the measurements for the data corresponding to the monopole, whereas the red symbols represent the quadrupole. The black dashed and dotted lines show the  mean values of the mocks  for \textsc{qpm} and \textsc{MD-Patchy}, respectively. 

For the LOWZ sample the \textsc{MD-Patchy} mocks describe accurately the data measurements for both monopole and quadrupole. The \textsc{qpm} mocks show some discrepancies with the data: \textsc{qpm} mocks systematically overestimate the monopole data by $\sim10\%$ at $k\geq0.02,h{\rm Mpc}^{-1}$ and fail to describe the quadrupole for $k\geq0.1\,h{\rm Mpc}^{-1}$. For the CMASS sample, \textsc{MD-Patchy}  mocks describe well the monopole and quadruple data. \textsc{qpm} mocks tend to overestimate both statistics by few percent.  The version of \textsc{qpm} mocks used in this paper was not designed to describe redshift space distortion features with few percent accuracy, as this version of the  \textsc{MD-Patchy} mocks was. Because of this, we will only use the \textsc{MD-Patchy} mocks multipole measurement to test the modelling of RSD. 

We combine the model described in \S\ref{sec:modelling} with the measured mock power spectrum monopole and quadrupole, averaged over the 2048 realization of the \textsc{MD-Patchy} mocks, in order to recover the $f\sigma_8$ parameter. Since we know the input cosmological parameters for the mock simulations, we can compare the obtained value with the expected one, and thus test which is the precision of the model when recovering $f\sigma_8$. Note that the measurements of the monopole and quadrupole of the mocks have been performed using the fiducial cosmology, which is different to the cosmology of the mocks as described in \S\ref{sec:mocks}. 

\begin{figure*}
\includegraphics[scale=0.3]{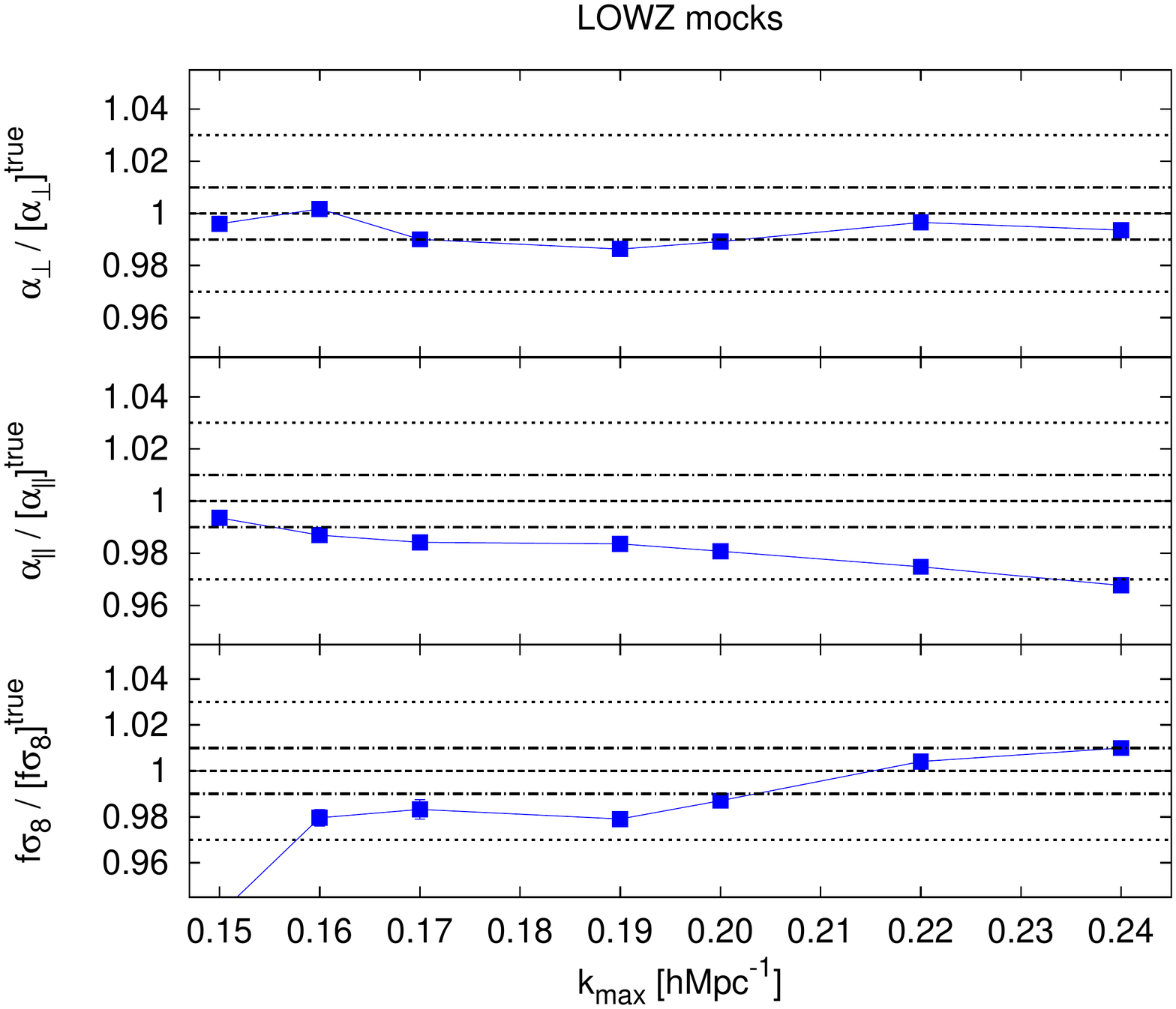} 
\includegraphics[scale=0.3]{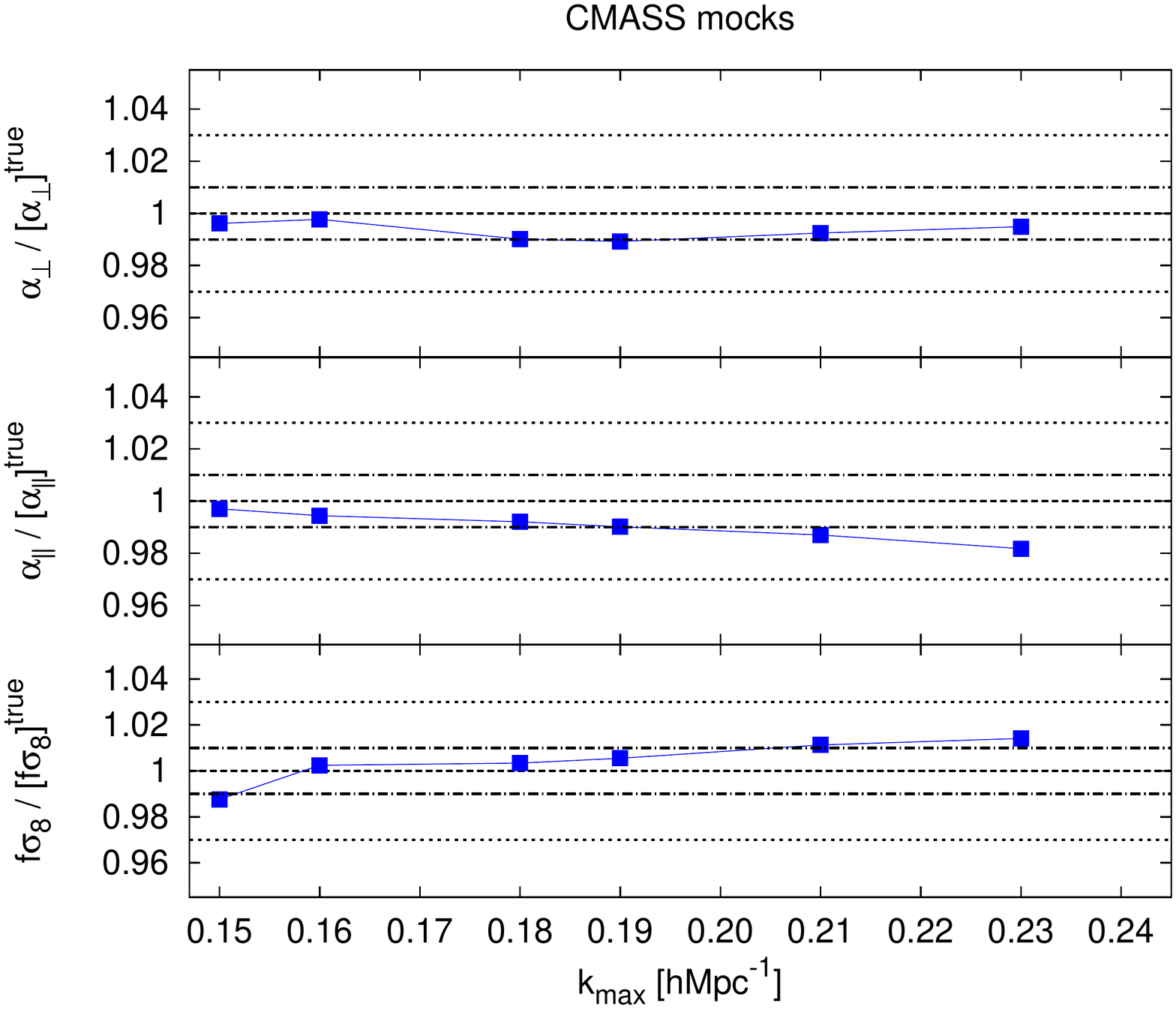} 
\caption{Best-fit $\alpha_\perp$, $\alpha_\parallel$ and $f\sigma_8$ parameters as a function of the minimum scale (maximum $k$) used for describing the LOWZ- (left panels) and CMASS (right panels) monopole and quadrupole power spectra predicted by the average of  2048 realizations of the \textsc{MD-Patchy} mocks (blue squares linked by solid lines). In order to mimic the analysis of the data,  the largest scales used for the fit are $k_{\rm min}=0.02\,h{\rm Mpc}^{-1}$ for the monopole and $k_{\rm min}=0.04\,h{\rm Mpc}^{-1}$ for the quadrupole. All values normalised by the corresponding true expected value. Horizontal dashed, dot-dashed and dotted lines show the 0\%, 1\% and 3\% deviations, respectively, with respect to the corresponding fiducial value. }
\label{plot:fs8mocks}
\end{figure*}

Fig \ref{plot:fs8mocks} displays in blue symbols linked by solid blue lines the obtained $f\sigma_8$, $\alpha_\parallel$ and $\alpha_\perp$ parameters as a function of the minimum scaled used to fit the model to the measurements of the \textsc{MD-Patchy} mocks. The error-bars correspond to the data error-bars scaled by a volume factor of $\sqrt{2048}$ in order to account for the volume difference.  In order to mimic the data analysis, the large scales cuts $k_{\rm min}=0.02\,h{\rm Mpc}^{-1}$ and $0.04\,h{\rm Mpc}^{-1}$ for the monopole and quadrupole, respectively, have been applied. In addition to $\alpha_\parallel$, $\alpha_\perp$, $f$ and $\sigma_8$, we allow  $b_1$, $b_2$, $A_{\rm noise}$ and $\sigma_{\rm FoG}$ to vary, as we do for the data.  The expected values for $f\sigma_8$ are shown in black dashed lines. For reference, $1\%$ and  $3\%$ deviations from the expected $f\sigma_8$ are also shown in dot-dashed and dotted black lines, respectively. 

For the LOWZ sample we see that the recovered value of $f\sigma_8$ agrees to $\leq2\%$ accuracy with the true one for $k_{\rm max}\geq0.16\,h{\rm Mpc}^{-1}$. We observe that if the truncation scale is $k_{\rm max}=0.15\,h{\rm Mpc}^{-1}$, the model produces a systematically low value of $f\sigma_8$. However, this does not occur when the truncation scale is higher than this value. In particular, if the truncation scale is $k_{\rm max}\geq 0.20\,h{\rm Mpc}^{-1}$ the accuracy of the obtained $f\sigma_8$ is $\lesssim1\%$, which is around 10 times smaller than the expected statistical error for this sample.  We observe that $\alpha_\perp$ is constrained with $\lesssim 1\%$ accuracy, at all studied scales. $\alpha_\parallel$ presents $\simeq3\%$ deviation from its fiducial value at $k_{\rm max}=0.24\,h{\rm Mpc}^{-1}$, and decreases to $\leq1\%$ as the scale of truncation is increased.

 For the CMASS sample we see that the recovered value of $f\sigma_8$ agree with $\lesssim1\%$ accuracy for truncation scales within the whole range of scales studied. Given the statistical errors for the CMASS sample, this systematic error is around 8 times smaller. As for the LOWZ case, $\alpha_\perp$ is recovered within $\leq1\%$ accuracy and $\alpha_\parallel$ within $\leq2\%$, in all the range of studied scales. 
 
 Since the observed systematic errors for both LOWZ and CMASS samples are much smaller than the statistical errors obtained for the data, we do not correct $f\sigma_8$ by any systematic shift.

\begin{figure*}
\centering
\includegraphics[scale=0.3]{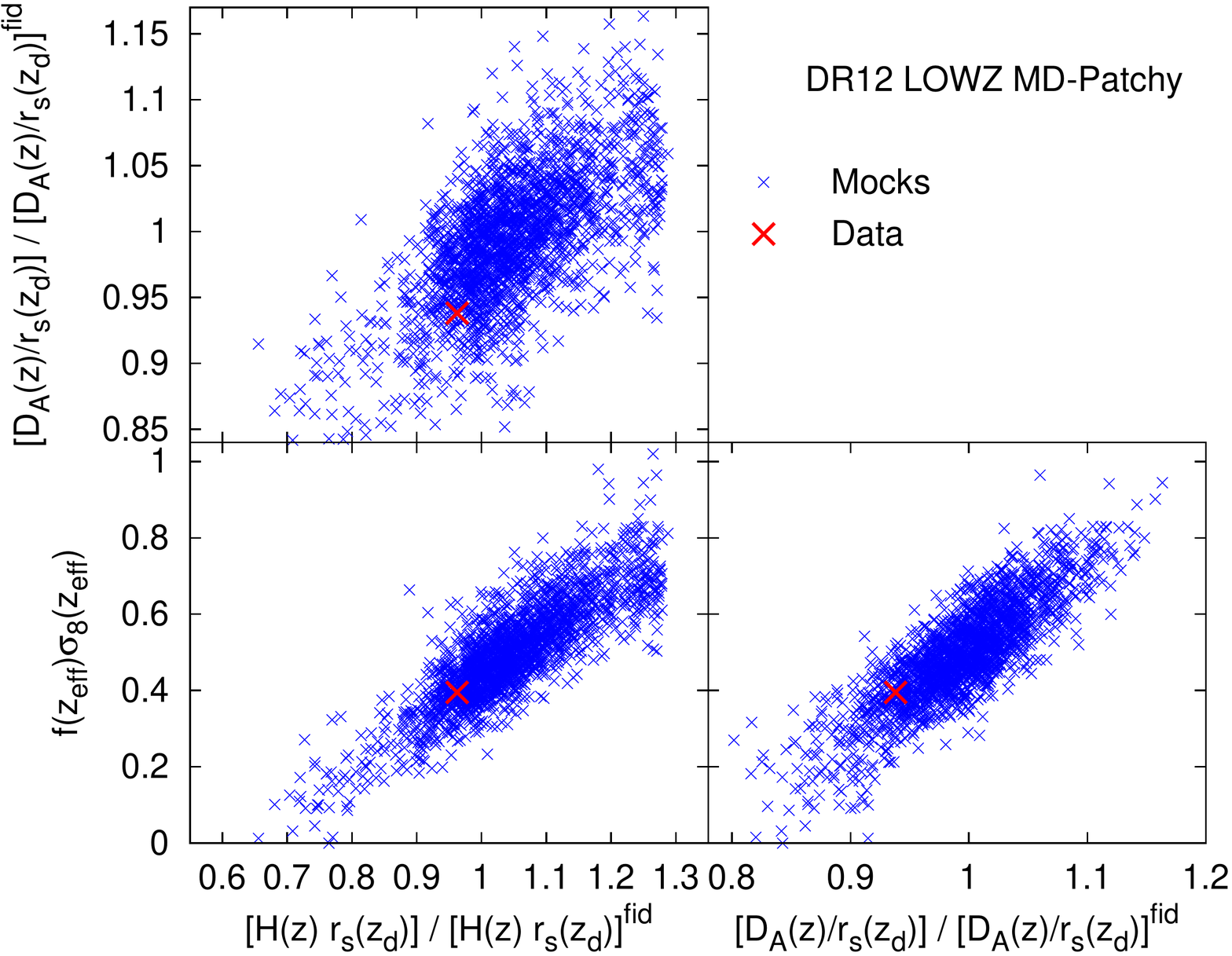}
\includegraphics[scale=0.3]{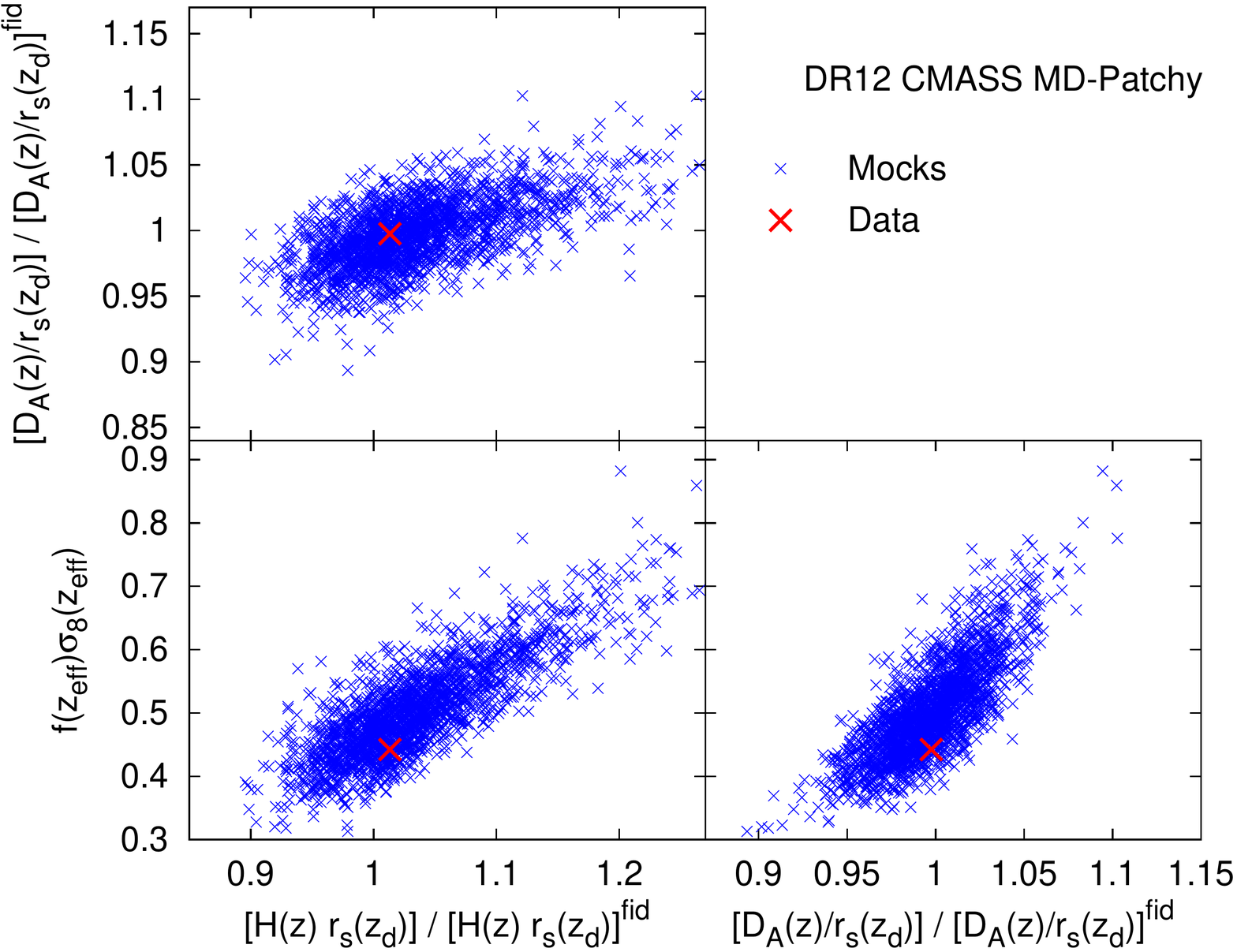}
\caption{Two dimensional distribution of  $f(z)\sigma_8(z)$, $H(z)r_s(z_d)$ and $D_A(z)/r_s(z_d)$, for the LOWZ  and CMASS samples, left and right panels, respectively. The blue points represent the best-fit solutions of 2048 independent \textsc{MD-Patchy} mocks realizations. The red crosses are the results for the data. The scatter of the mocks gives an idea of how these parameters are degenerate. Both mocks and data have been analysed using the \textsc{MD-Patchy} covariance. }
\label{plot:scatter}
\end{figure*}

Fig. \ref{plot:scatter} shows the distribution of parameters $f(z)\sigma_8(z)$, $H(z)r_s(z_d)$ and $D_A(z)/r_s(z_d)$ for the 2048 realizations of the \textsc{MD-Patchy} mocks in the LOWZ and CMASS samples. The blue crosses represent the best-fit parameter set for each realization of the mocks, whereas the red cross is for the data. 
The shapes of the clouds formed by the blue crosses provide information of how these parameters are degenerate. Thus, these plots show that the measurements of $f\sigma_8$, $H(z) r_s(z_d)$ and $D_A(z) / r_s(z_d)$ are strongly correlated. Assuming that the data results can be approximated by a multivariate Gaussian likelihood, in \S\ref{sec:multivariate_likelihood} we provide the covariance matrix among these 3 parameters for the measurements from the data.

\begin{figure*}
\centering
\includegraphics[scale=0.3]{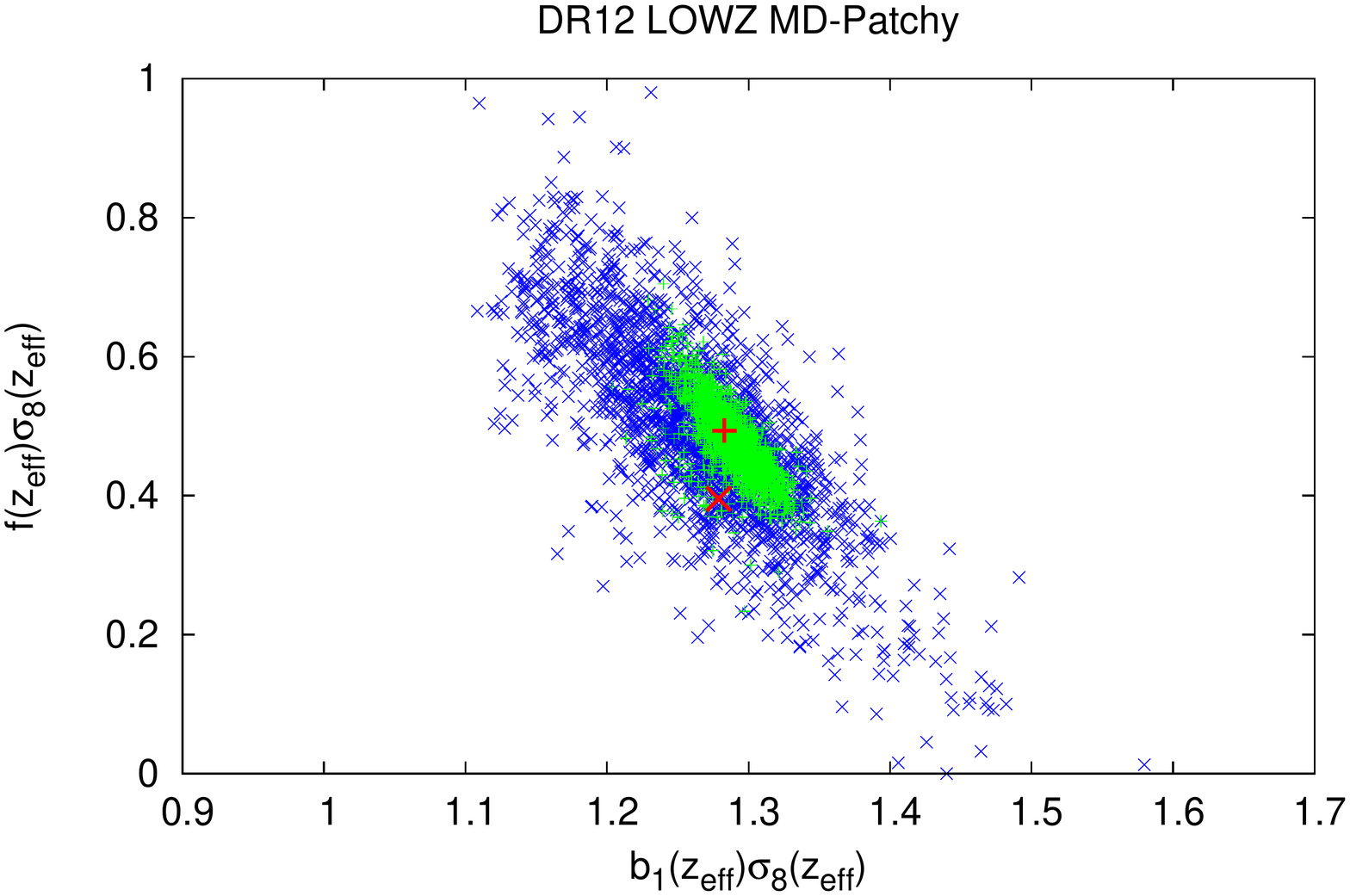}
\includegraphics[scale=0.3]{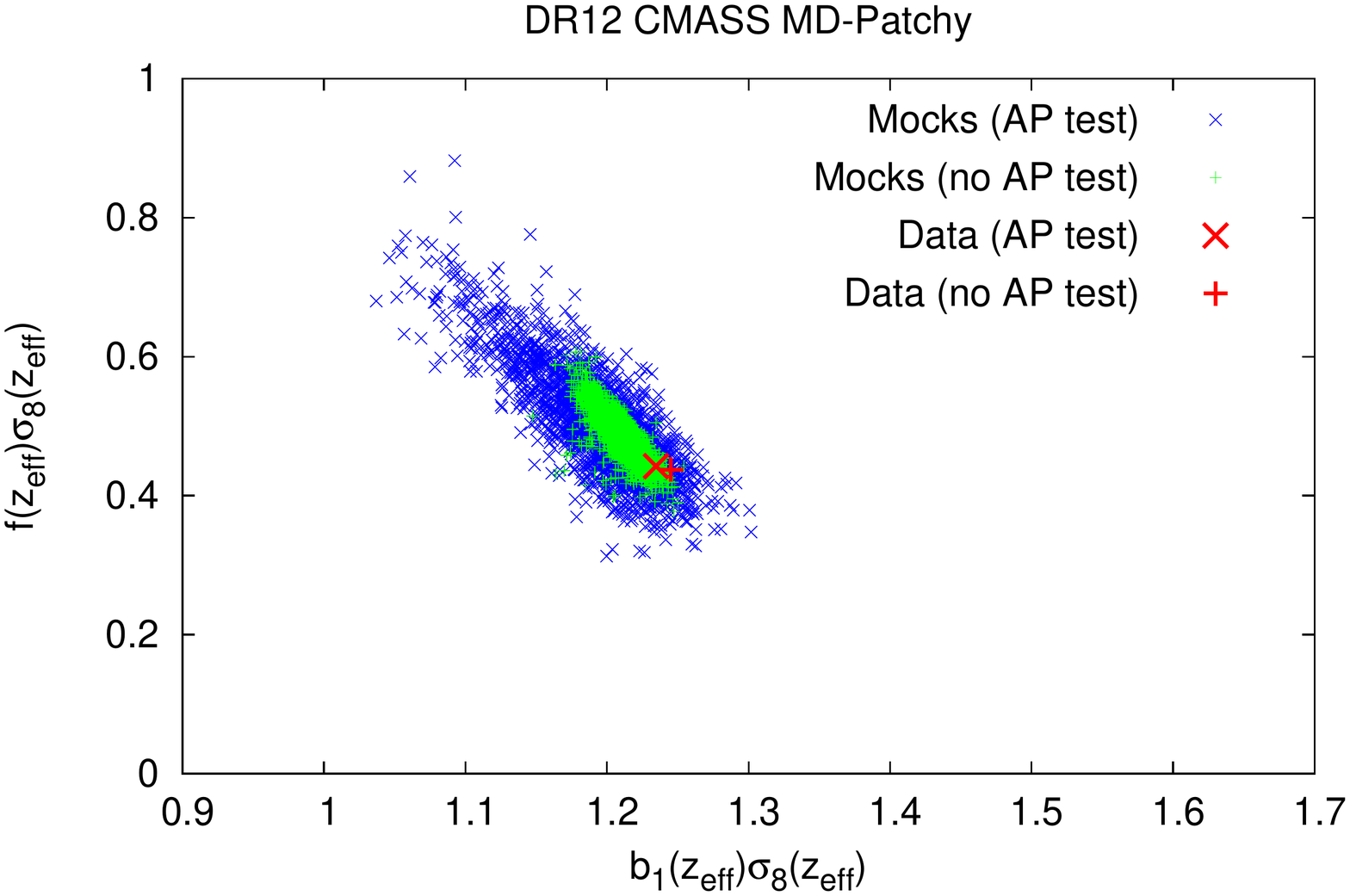}
\caption{Two-dimensional distribution for the parameter space corresponding to $b_1\sigma_8$ and $f\sigma_8$. Left and right panels display the results for the LOWZ and CMASS samples. Blue $\times$-symbols correspond to the best-fit solution for each realization (out of 2048) of the \textsc{MD-Patchy} mocks when all parameters are freely varied. The green $+$-symbols correspond to the best-fit solution when the AP parameters, $\alpha_\parallel$ and $\alpha_\perp$ are set their fiducial values, respectively. The red $\times$-symbol display the best-fit solution for the data sample when all the parameters are varied. The red $+$-symbol display the best-fit solution for the data sample when the AP parameters have been set to their fiducial values. In all the cases the covariance matrix have been inferred using the \textsc{MD-Patchy} mocks.}
\label{plot:fs8scatter}
\end{figure*}

Fig. \ref{plot:fs8scatter} shows the two-dimensional parameter space of $b_1\sigma_8$ and $f\sigma_8$ for the 2048 realizations of the  \textsc{MD-Patchy} mocks in the LOWZ and CMASS samples. 

We see that when the AP parameters are set to 1, the scatter between best-fit values from different mocks reduces significantly forming a small cloud (green crosses) which is contained by the blue one. 
The red crosses represent the actual values for the data when the \textsc{MD-Patchy} covariance is used. In both cases we see that the data points lies well within the clouds formed by the different realizations of the \textsc{MD-Patchy} mocks. 

\section{BOSS DR12 Measurements}\label{sec:analysis}

The values of the parameters of the model corresponding to the fits to the data presented in Fig. \ref{dataPS} are listed in Table \ref{table_results1}, where the minimum cut-off scale used for the fit is $k_{\rm max}=0.24\,h{\rm Mpc}^{-1}$. We observe that the difference between parameters (using \textsc{qpm} and \textsc{MD-Patchy} covariances) and their error-bars is not significant. This indicates that the impact of the different covariance matrices in the parameter estimation is sub-dominant compared to other effects, such as the scale where the fitting process is truncated, or systematics of the model itself.

\begin{figure*}
\includegraphics[scale=0.3]{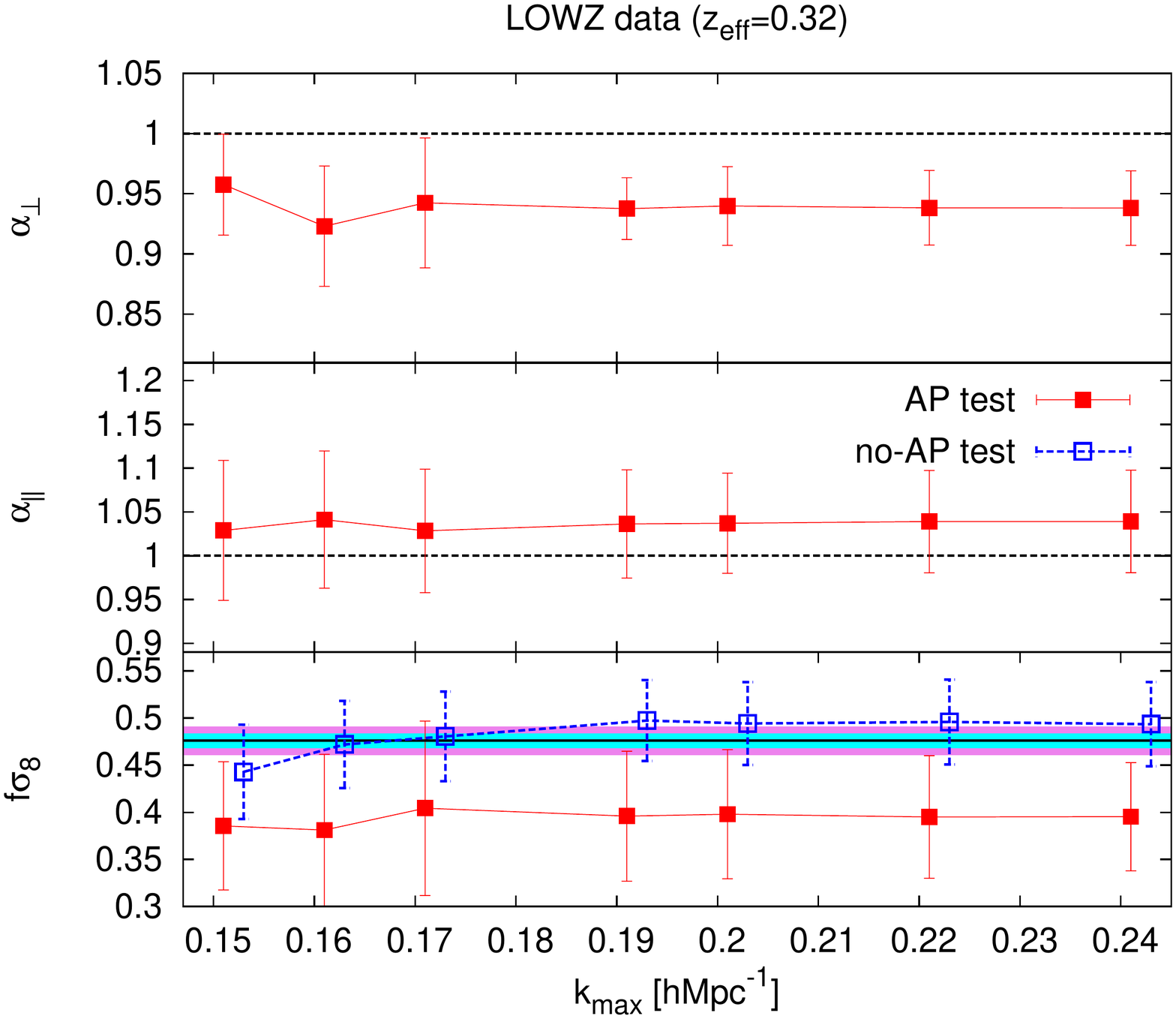} 
\includegraphics[scale=0.3]{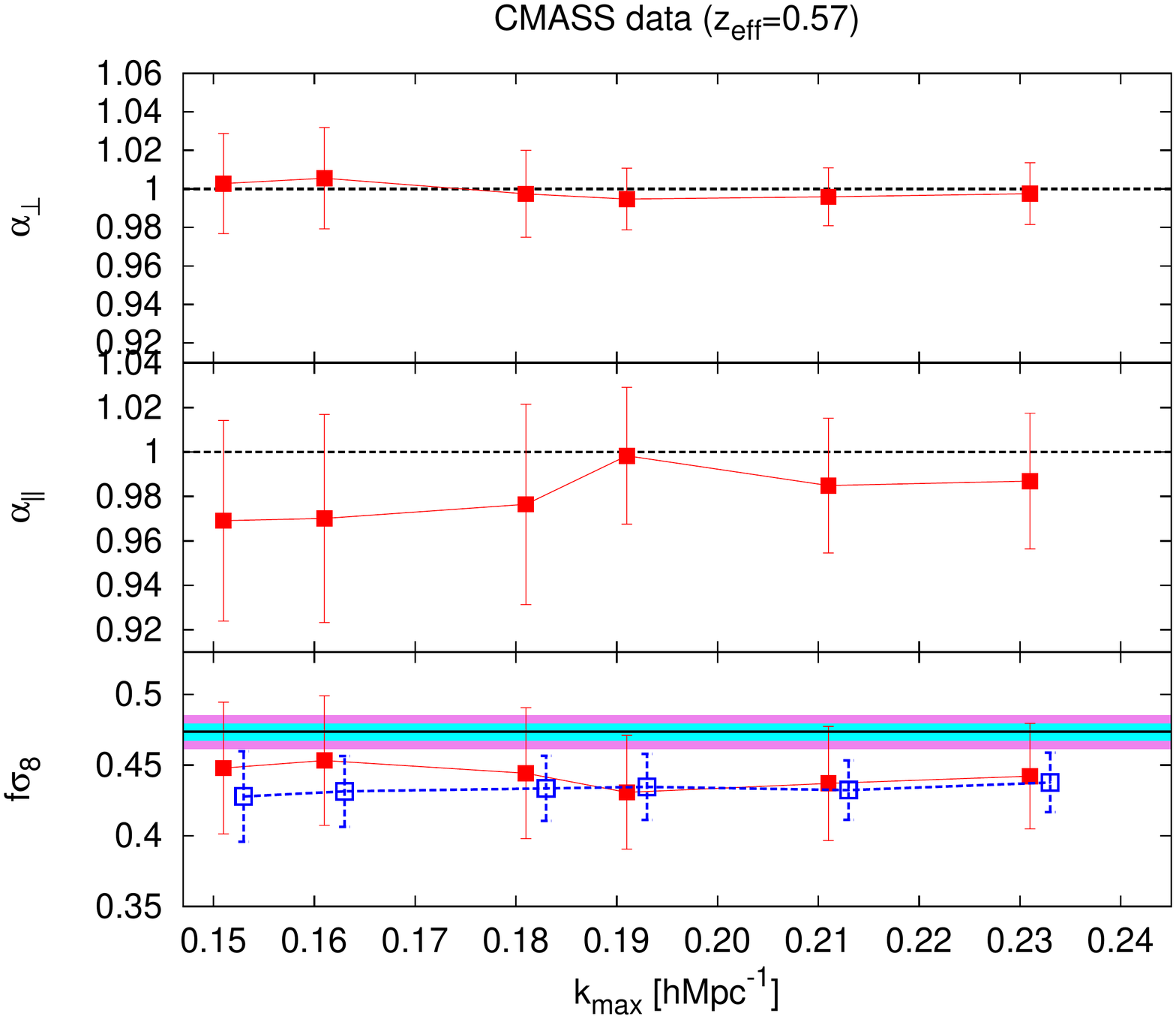} 
\caption{ Power spectrum monopole and quadrupole best-fit parameters as a function of the minimum scale considered, $k_{\rm max}$. The results for the LOWZ- (left panels) and CMASS-DR12 (right panel) BOSS data   are shown in red solid lines in combination with red filled symbols. In the $f\sigma_8$ panel we also show in dashed blue lines and empty blue symbols the best-fit values when the AP parameters have been set to their fiducial value. In this case, the symbols have been displaced horizontally for visualization reasons.  For all the cases the largest scales used for the fitting are $k_{\rm min}=0.02\,h{\rm Mpc}^{-1}$ for the monopole and $k_{\rm min}=0.04\,h{\rm Mpc}^{-1}$ for the quadrupole. In the $f\sigma_8$ sub-panels we show as a black solid line the predictions from {\it Planck15}.  The cyan and magenta bands represent the $1\sigma$ and $2\sigma$ error-bars, respectively, around the {\it Planck15} best-fit solution.}
\label{plot:kmax_data}
\end{figure*}

\begin{table*}
\begin{center}
\begin{tabular}{|c|c|c|c|c}
\hline
Sample (Cov.) & LOWZ (\textsc{qpm}) & LOWZ (\textsc{MD-Patchy}) & CMASS (\textsc{qpm}) & CMASS (\textsc{MD-Patchy}) \\
 \hline
 $f(z_{\rm eff})\sigma_8(z_{\rm eff})$ & $0.392\pm0.061$  & $0.395\pm0.064$  & $0.445\pm0.038$  & $0.442\pm0.037$  \\
 $H(z_{\rm eff})r_s(z_d)\, [10^3\rm{kms}^{-1}]$ & $11.48\pm0.55$ & $11.33\pm0.56$  & $13.99\pm0.44$ & $13.84\pm0.43$ \\
$D_A(z)/r_s(z_d)$ & $6.38\pm0.18$  & $6.33\pm0.19$  & $9.43\pm0.15$  & $9.42\pm0.15$ \\
 \hline
  $\alpha_\parallel(z_{\rm eff})$ & $1.025\pm0.052$ & $1.039\pm0.054$  & $0.977\pm0.030$  & $0.987\pm0.030$ \\
  $\alpha_\perp(z_{\rm eff})$ & $0.945\pm 0.027$  & $0.938\pm0.028$  & $0.999\pm 0.016$  & $ 0.998\pm0.016$ \\
  $b_1\sigma_8(z_{\rm eff})$ & $1.283\pm0.032$ & $1.279\pm0.037$ & $1.218\pm0.022$  & $1.225\pm0.020$ \\
  $b_2\sigma_8(z_{\rm eff})$ & $-0.19\pm0.64$  & $-0.38\pm0.36$  & $0.67\pm0.74$ & $0.40\pm0.66$ \\
  $A_{\rm noise}$ & $-0.30\pm0.22$ & $ -0.43\pm0.21$ & $-0.041\pm0.078$  & $-0.057\pm0.093$ \\
  $\sigma_{\rm FoG}(z_{\rm eff})[{\rm Mpc}h^{-1}]$ & $3.94\pm0.56$ & $4.23\pm0.56$  & $3.35\pm0.32$  & $3.42\pm0.31$ \\
  \hline
  $\chi^2/ {\rm d.o.f}$ & 29.62/(53-8) & 31.48/(53-8) & 26.168/(48-8) & 33.661/(48-8) \\
  \hline
\end{tabular}
\end{center}
\caption{Best-fit parameters obtained from fitting the monopole and quadrupole BOSS DR12 data using the theoretical approach described in \S\ref{sec:modelling}. The two first columns are the parameters obtained from fitting the LOWZ-DR12 data, whereas the third and forth columns are obtained from fitting the CMASS-DR12 data. The first and third columns are the parameters obtained when the covariance matrix is inferred from \textsc{qpm} mocks, whereas the second and forth columns parameters are obtained when the covariance matrix is inferred from \textsc{MD-Patchy} mocks. For all cases, the minimum scale used is $k_{\rm max}=0.24\,h{\rm  Mpc}^{-1}$ and the largest scales used are $k_{\rm min}=0.02\,h{\rm  Mpc}^{-1}$ for the monopole and $k_{\rm min}=0.04\,h{\rm  Mpc}^{-1}$ for the quadrupole. The error-bars represent $1\sigma$ deviations and have been inferred from the analysis of the likelihood function as it is described in \S\ref{sec:errror_estimation}. The cosmological parameters, $f\sigma_8$, $H(z)r_s(z_d)$ and $D_A(z)/r_s(z_d)$ are correlated, so we encourage using the multivariate Gaussian likelihood presented in \S\ref{sec:multivariate_likelihood}. }
\label{table_results1}
\end{table*}

Table \ref{table_results2} displays the best-fit results at $k_{\rm max}=0.24\,h{\rm Mpc}^{-1}$, with $H(z)r_s(z_d)$ and $D_A(z)/r_s(z_d)$ set to the fiducial cosmology prediction. As in Table \ref{table_results1}, the differences between the parameters and their errors obtained using either \textsc{qpm} and \textsc{MD-Patchy} mocks are not significant. 

\begin{table*}
\begin{center}
\begin{tabular}{|c|c|c|c|c}
\hline
Sample (Cov.) & LOWZ (\textsc{qpm}) & LOWZ (\textsc{MD-Patchy}) & CMASS (\textsc{qpm}) & CMASS (\textsc{MD-Patchy}) \\
 \hline
 $f(z_{\rm eff})\sigma_8(z_{\rm eff})$ & $0.476\pm0.043$  & $0.494\pm0.045$ & $0.434\pm0.023$  & $0.438\pm0.021$ \\
  $b_1\sigma_8(z_{\rm eff})$ & $1.290\pm0.016$  & $1.283\pm0.017$  & $1.236\pm0.012$& $1.236\pm0.012$ \\
  $b_2\sigma_8(z_{\rm eff})$ & $-0.23\pm0.95$ & $-0.37\pm0.60$  & $0.38\pm0.60$ & $0.25\pm0.56$\\
  $A_{\rm noise}$ & $-0.32\pm0.22$ & $-0.41\pm0.21$  &  $-0.044\pm0.089$  & $-0.064\pm0.086$ \\
  $\sigma_{\rm FoG}[{\rm Mpc}h^{-1}]$ & $ 4.04\pm0.61$  & $4.27\pm0.59$  & $3.41\pm0.31$  & $ 3.45\pm0.30$ \\
  \hline
    $\chi^2/ {\rm d.o.f}$ & 33.45/(53-6) & 36.08/(53-6) & 26.96/(48-6) & 33.99/(48-6) \\
\hline
\end{tabular}
\end{center}
\caption{Best-fit parameters obtained from fitting the monopole and quadrupole BOSS DR12 data as it has been described in Table \ref{table_results2}, where in this case the AP parameters, $D_A(z)/r_s(z_d)$ and $H(z)r_s(z_d)$ have been set to their fiducial value, $H(z_{\rm lowz})r(z_d)=11.773\,{\rm kms}^{-1}$, $D_A(z_{\rm lowz})/r_s(z_d)=6.7466$, $H(z_{\rm cmass})r(z_d)=13.663\,\cdot10^3{\rm kms}^{-1}$, $D_A(z_{\rm lowz})/r_s(z_d)=9.4418$. }
\label{table_results2}
\end{table*}

In Fig. \ref{plot:kmax_data} we show how the best-fit values of  $f\sigma_8$ and the AP parameters, $\alpha_\parallel$ and $\alpha_\perp$, change with the minimum scale considered.

From the left panels of Fig. \ref{plot:kmax_data}  we observe that when varying  $k_{\rm max}$ from $0.15\,h{\rm Mpc}^{-1}$ to $0.24\,h{\rm Mpc}^{-1}$, the best-fit value of $f\sigma_8$ is very stable compared to the size of the error-bars. This behaviour applies when the AP parameters are varied and also when they are fixed to their fiducial value
When the AP parameters are varied the best-fit value of $f\sigma_8$ is consistently low by $\sim1\sigma$ respect to {\it Planck15}+GR prediction. We also observe that, whereas $\alpha_\parallel$ is consistent with unity for all values of $k_{\rm max}$ studied, $\alpha_\perp$ presents a low value of $2\sigma-3\sigma$ respect to the fiducial cosmology prediction. When the AP parameters, $\alpha_\parallel$ and $\alpha_\perp$, are set 1, the LOWZ-sample $f\sigma_8$ parameter, is shifted $\sim1\sigma$ upwards and is more in accordance with the {\it Planck15} cosmology prediction.  We observe that this effect is independent of the model truncation scale. In \S\ref{sec:dependence_cosmo} we will reanalyze the data assuming different fiducial models for $\Omega_m$ and we will discuss how this tension changes. So far we can say that the LOWZ sample data has a preference for low values of $f\sigma_8$ and $\alpha_\perp$ (which is translated into low values of $D_A(z)/r_s(z_d)$, according to the definitions in \S\ref{sec:AP}), which is within 1 and $2\sigma$ of the fiducial values when GR is assumed. Both parameters are  correlated as we will show in \S\ref{sec:multivariate_likelihood}, so when $\alpha_\perp$ is fixed to its fiducial value, $f\sigma_8$ is automatically shifted towards a solution which, in this case, is consistent with {\it Planck15}+GR within $1\sigma$. 

From the right panels of Fig. \ref{plot:kmax_data} we see that the values of AP parameters are both consistent with their fiducial values within $1\sigma$ for all $k_{\rm max}$, suggesting that there is no strong tension between the fiducial cosmology and the actual. Also, the inferred $f\sigma_8$ value when the full AP test is performed, is consistent with {\it Planck15}+GR within $1\sigma$.  By setting the AP parameters to their fiducial values, we do not observe any significant change on the $f\sigma_8$ parameter, besides  a reduction of the error-bars, which increase the tension with {\it Planck15}+GR by up to $2\sigma$.

In Table \ref{table_finalresults} we present the final measurements of this paper. They have been obtained by averaging the results presented in \S\ref{sec:results} (Table \ref{table_results1} and \ref{table_results2}) using the \textsc{qpm} and \textsc{MD-Patchy} covariances.  We show the results for both the LOWZ and CMASS sample where the NGC and SGC have been combined into a single measurement.
The main result of this paper is $f(0.57)\sigma_8(0.57)=0.448\pm0.038$ for the CMASS and $f(0.32)\sigma_8(0.32)=0.402\pm0.060$ for the LOWZ sample when the AP test is performed; and $f(0.57)\sigma_8(0.57)=0.438\pm0.022$ for the CMASS and $f(0.32)\sigma_8(0.32)=0.497\pm0.0436$ for the LOWZ sample then the AP parameters have been tuned to their fiducial value . In the following subsections we analyse how these results are affected by a change in the cosmological models (\S\ref{sec:dependence_cosmo}), how these results compares with other values of $f(z)\sigma_8(z)$ extracted from papers based on the Data Release 11 of the BOSS sample (\S\ref{sec:boss}), and how these results compare to other values of $f(z)\sigma_8(z)$ obtained from other redshift surveys (\S\ref{sec:surveys}). We also present the multivariate likelihood surface at $f\sigma_8$, $H(z)r_s(z_d)$ and $D_A(z)/r_s(z_d)$ (\S\ref{sec:multivariate_likelihood}). 

\begin{table*}
\begin{center}
\begin{tabular}{|c|c|c|c|c}
\hline
 & LOWZ & CMASS & LOWZ (no-AP) & CMASS (no-AP) \\
 \hline
$f(z_{\rm eff})\sigma_8(z_{\rm eff})$ & $0.394\pm0.062$  & $0.444\pm0.038$  & $0.485\pm0.044$  & $0.436\pm0.022$ \\
$H(z_{\rm eff})r_s(z_d)\, [10^3\rm{kms}^{-1}]$ & $11.41\pm0.56$ & $13.92\pm0.44$   & 11.773 & 13.663 \\
$D_A(z)/r_s(z_d)$ & $6.35\pm0.19$  & $9.42\pm0.15$  & 6.7466  & 9.4418 \\
\hline
$\alpha_\parallel(z_{\rm eff})$ & $1.032\pm0.053$ &  $0.982\pm0.031$ & 1 & 1\\
 $\alpha_\perp(z_{\rm eff})$ & $0.942\pm 0.027$ & $0.998\pm 0.016$ & 1 & 1 \\
 $b_1\sigma_8(z_{\rm eff})$ &  $1.281\pm0.035$ &$1.222\pm0.021$  & $1.287\pm0.017$ & $1.236\pm0.012$\\
$b_2\sigma_8(z_{\rm eff})$ &  $-0.29\pm0.50$ & $0.53\pm0.70$ & $-0.30\pm0.78$ & $0.32\pm0.58$\\
 $A_{\rm noise}$ &  $-0.36\pm0.22$ & $-0.049\pm0.086$ & $-0.36\pm0.22$ & $-0.054\pm0.077$\\
 $\sigma_{\rm FoG}[{\rm Mpc}h^{-1}]$ & $4.08\pm0.56$  & $3.39\pm0.32$ & $4.16\pm0.60$ & $3.43\pm0.31$\\
  \hline
\end{tabular}
\end{center}
\caption{Combined best-fit parameters for the LOWZ- and CMASS-DR12 samples. The first column display the LOWZ results; the second column display the CMASS results; in both cases the AP test is performed. The third and forth column display the results for LOWZ and CMASS, respectively, when no-AP test is performed. The combination has been performed taking the average of the best-fit values of parameters obtained when the covariance matrix is extracted from either \textsc{qpm} and \textsc{MD-Patchy} mocks, as it has been listed in Tables \ref{table_results1} (when AP test is performed) and \ref{table_results2} (when no-AP test is performed). The displayed errors are also taken as the average as the ones listed in Tables \ref{table_results1} and \ref{table_results2}. The minimum scale for the fit is $k_{\rm max}=0.24\,h{\rm Mpc}^{-1}$, and the large scale cuts are $k_{\rm min}=0.02\,h{\rm Mpc}^{-1}$ for the monopole and $k_{\rm min}=0.04\,h{\rm Mpc}^{-1}$ for the quadrupole.}
\label{table_finalresults}
\end{table*}%

\subsection{Dependence with cosmology}\label{sec:dependence_cosmo}
{ In this section we study how sensitive  the measurements of $\alpha_\perp$, $\alpha_\parallel$ and $f\sigma_8$} are respect to the assumed cosmology for converting redshifts into comoving distances. In order to test this effect we have measured the galaxy power spectrum multipoles of the data assuming two additional cosmological models, one with $\Omega_m=0.292$ and another with $\Omega_m=0.332$ (see table 4 of \citealt{hector_bispectrum1} for details about these two extra cosmologies, listed as H-Planck13 and L-Planck13). These two values of $\Omega_m$ are at $1\sigma$ tension of the best-fit value reported by \cite{Planck13}.  
\begin{figure*}
\centering
\includegraphics[scale=0.3]{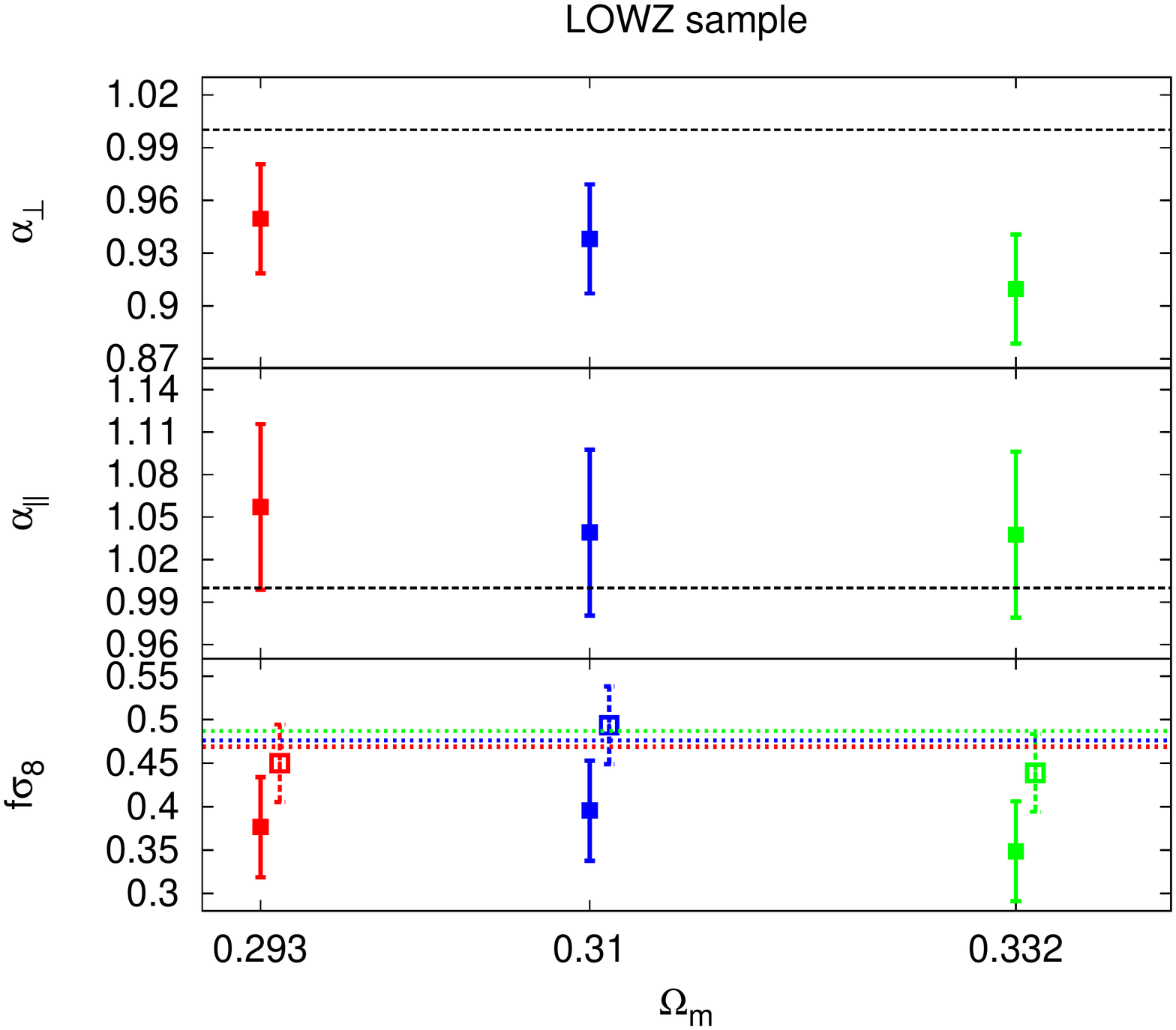}
\includegraphics[scale=0.3]{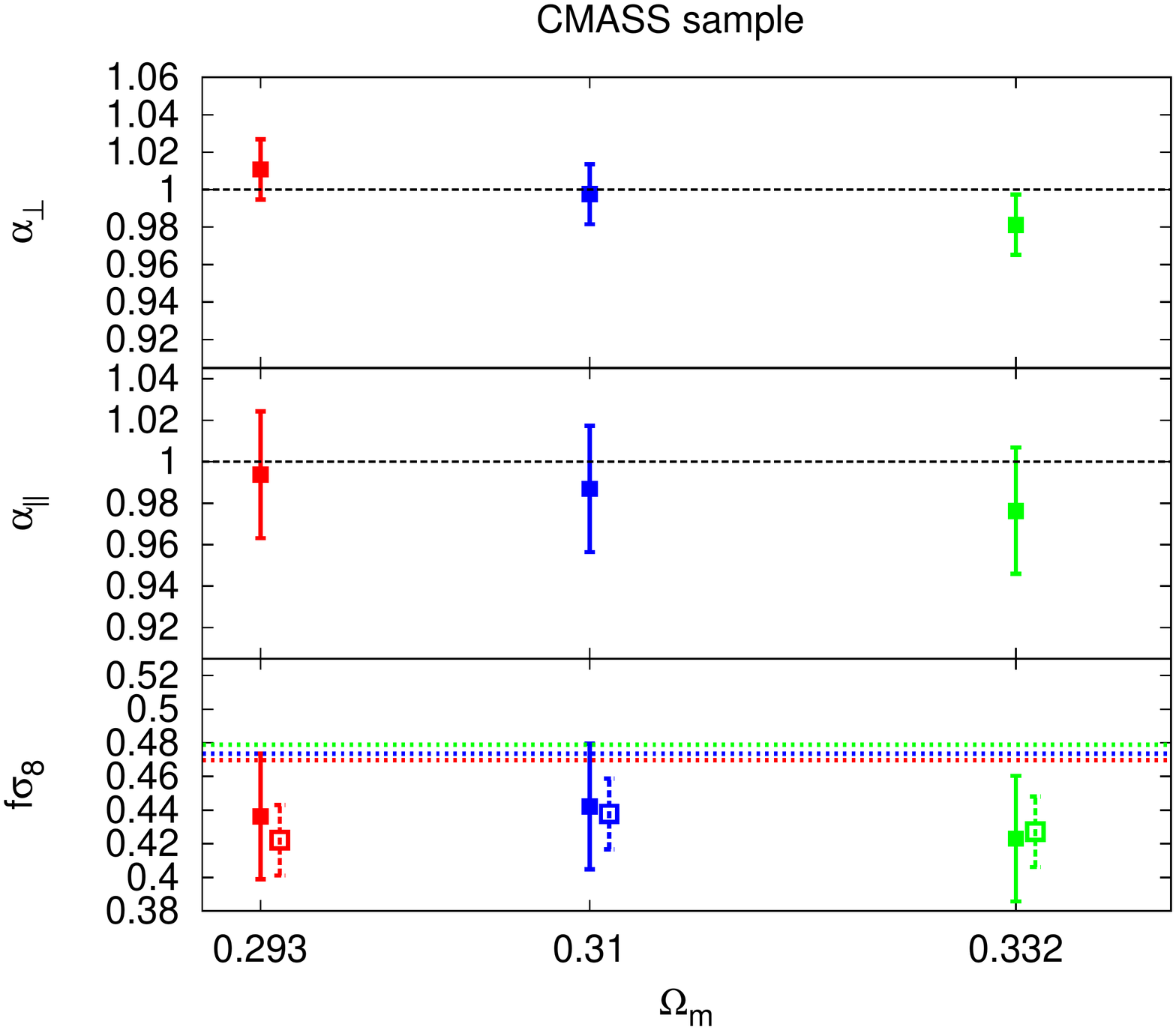}
\caption{The left and right panels show the best-fit $\alpha_\perp$ (top sub-panel), $\alpha_\parallel$ (middle sub-panel) and $f\sigma_8$ (bottom sub-panel) value for LOWZ and CMASS sample, respectively, as a function of the chosen  cosmological model  to convert redshift into distances. The different colours show 3 different models, $\Omega_m=0.293$ (red), $\Omega_m=0.31$ (blue) and $\Omega_m=0.332$ (green). The filled squares show the results when the AP parameters are varied, whereas for the empty squares the AP parameters have been set to their fiducial values. In the bottom sub-panel, the horizontal dashed lines show the $f\sigma_8$ prediction assuming GR ($f(z)=\Omega_m(z)^\gamma$, with $\gamma=0.545$) and $\sigma_8=0.815$. }
\label{plot:cosmology}
\end{figure*}
Our power spectrum models are as described in \S\ref{sec:modelling}, but based on the linear power spectrum of the assumed cosmological model (see fig. 6 in \citealt{hector_bispectrum1} for a comparison between these linear dark matter power spectra). In order to be consistent, we have re-analysed the \textsc{MD-Patchy} mocks assuming the new value of $\Omega_m$ for converting redshifts into comoving distances, in order to recompute a new covariance matrix, consistent with the new measurement of the monopole and quadrupole. 
{  In Fig. \ref{plot:cosmology} we show the values of $\alpha_\perp$, $\alpha_\parallel$ and $f\sigma_8$} obtained as a function of the chosen value for $\Omega_m$ for the LOWZ and CMASS sample in the left and right panels, respectively, using the corresponding \textsc{MD-Patchy} covariance. The red, blue and green points show the results of assuming $\Omega_m=0.292,\, 0.310,\, 0.332$, respectively. The filled symbols represent the results when the full AP test is performed, whereas the empty symbols represent the values  $f\sigma_8$ obtained when the AP parameters have been set to their fiducial value (in this case $\alpha_\parallel$ and $\alpha_\perp$ are set to 1). {  In the bottom sub-panel panel, the dotted horizontal coloured lines show the corresponding value of $f\sigma_8$ when GR is assumed and the value of $\sigma_8$ is set to 0.815 (which is the fiducial value for the {\it Planck15} cosmology). In the $\alpha_\perp$ and $\alpha_\parallel$ panels, the horizontal black dotted line represent the fiducial value $\alpha_\perp=\alpha_\parallel=1$. }

For the CMASS sample, the results do not strongly depend on the fiducial cosmology chosen. For both AP and non-AP cases, the changes in $f\sigma_8$ as a function of the cosmological model assumed are typically of $\sim0.5\sigma$. Furthermore,  when the  $\Omega_m=0.31$ model is assumed,  the tension between the observed $f\sigma_8$ and its {\it Planck15}+GR prediction minimises. Therefore, CMASS data suggests that when GR is assumed as a theory of gravity, $\Omega_m=0.31$ is the most likely model among those studied. {  Also, for this case the AP parameters are the closest to the fiducial values.}  

{ 
 For the LOWZ sample, when the AP test is performed we observe a $\sim2\sigma$ tension for $\Omega_m^{\rm fid}= 0.293$ and 0.310 on $f\sigma_8$. The tension is even higher when  $\Omega_m^{\rm fid}= 0.332$. Similar findings apply to the best-fit values of $\alpha_\perp$ parameter, whereas the best-fit value for $\alpha_\parallel$ is similar for the 3 studied models}.  When the AP test is turned off, and consequently $\alpha_\parallel$ and $\alpha_\perp$ are set to their fiducial values, we observe that the $f\sigma_8$ tension is reduced to well within $1\sigma$ for all models studied.  This suggests, that the cause of the $f\sigma_8$ tension observed in Fig. \ref{plot:kmax_data} cannot be explained by the choice of $\Omega^{\rm fid}_m$. Thus, according to LOWZ data, the observed $D_A(z)/r_s(z_d)$ parameter is in $\leq2\sigma$ tension with its fiducial value, which induces a $\leq2\sigma$ tension in the observed $f\sigma_8$ compared to its {\it Planck15}+GR prediction, due to the correlation between parameters. This tension is relaxed when $D_A(z)/r_s(z_d)$ is set to its fiducial value. In any case, the $\Omega_m=0.31$ model is the choice that presents less tension between the observed $f\sigma_8$ and its  {\it Planck15}+GR prediction. 

Overall, we conclude that both CMASS and LOWZ galaxies data is in agreement with the fiducial model $\Omega_m^{\rm fid}=0.31$ consistent with {\it Planck15} data: for the CMASS sample the tension is below $1\sigma$ and for the LOWZ sample within $2\sigma$. For both samples, $\Omega_m^{\rm fid}=0.31$ is the model that minimises the tension between the observed $f\sigma_8$ and the predicted by {\it Planck15}+GR.  

\subsection{Comparison of $f\sigma_8$ with other DR11 BOSS Data Releases analyses}\label{sec:boss}
\begin{figure*}
\centering 
\includegraphics[scale=0.3]{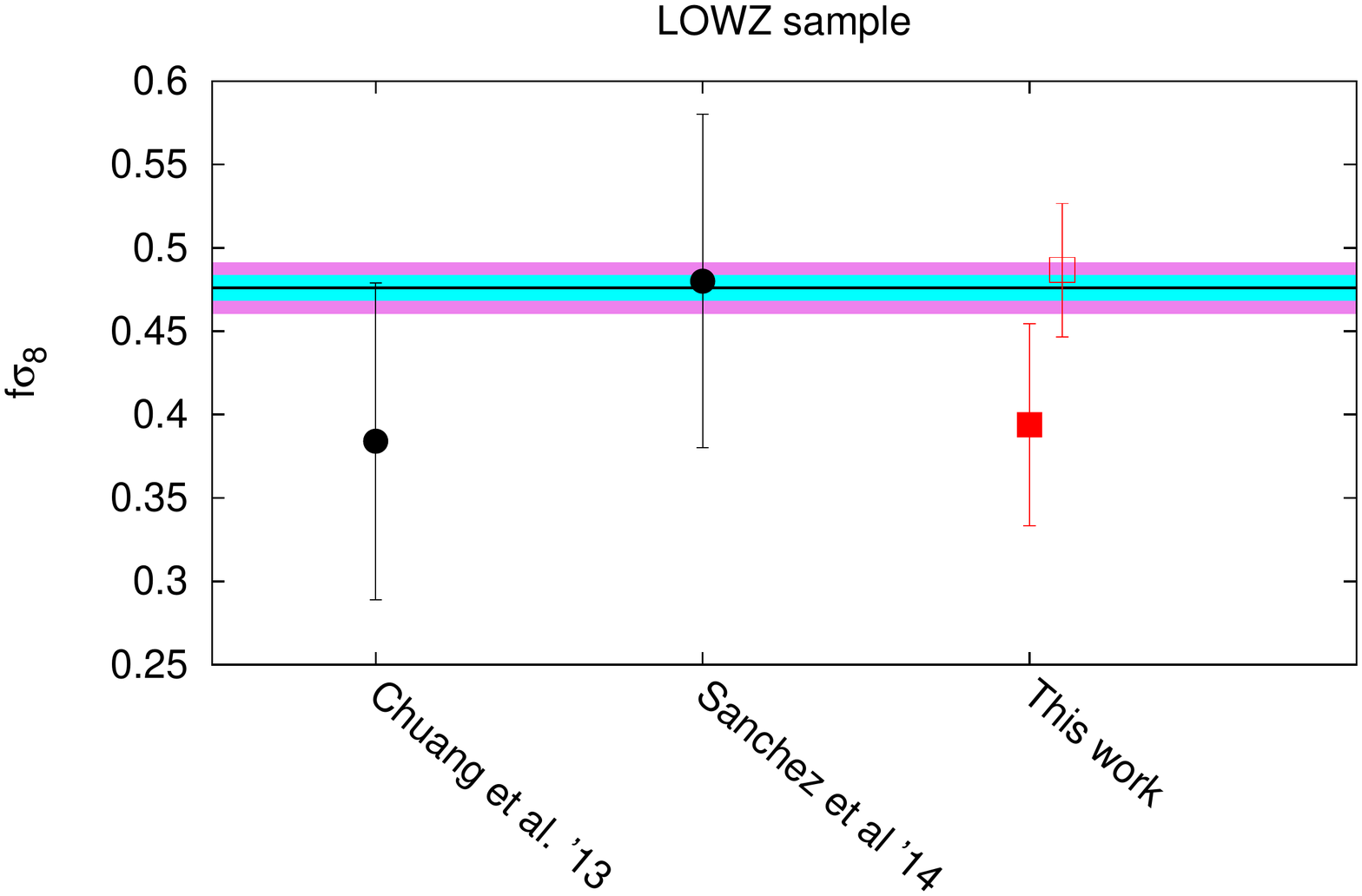}
\includegraphics[scale=0.3]{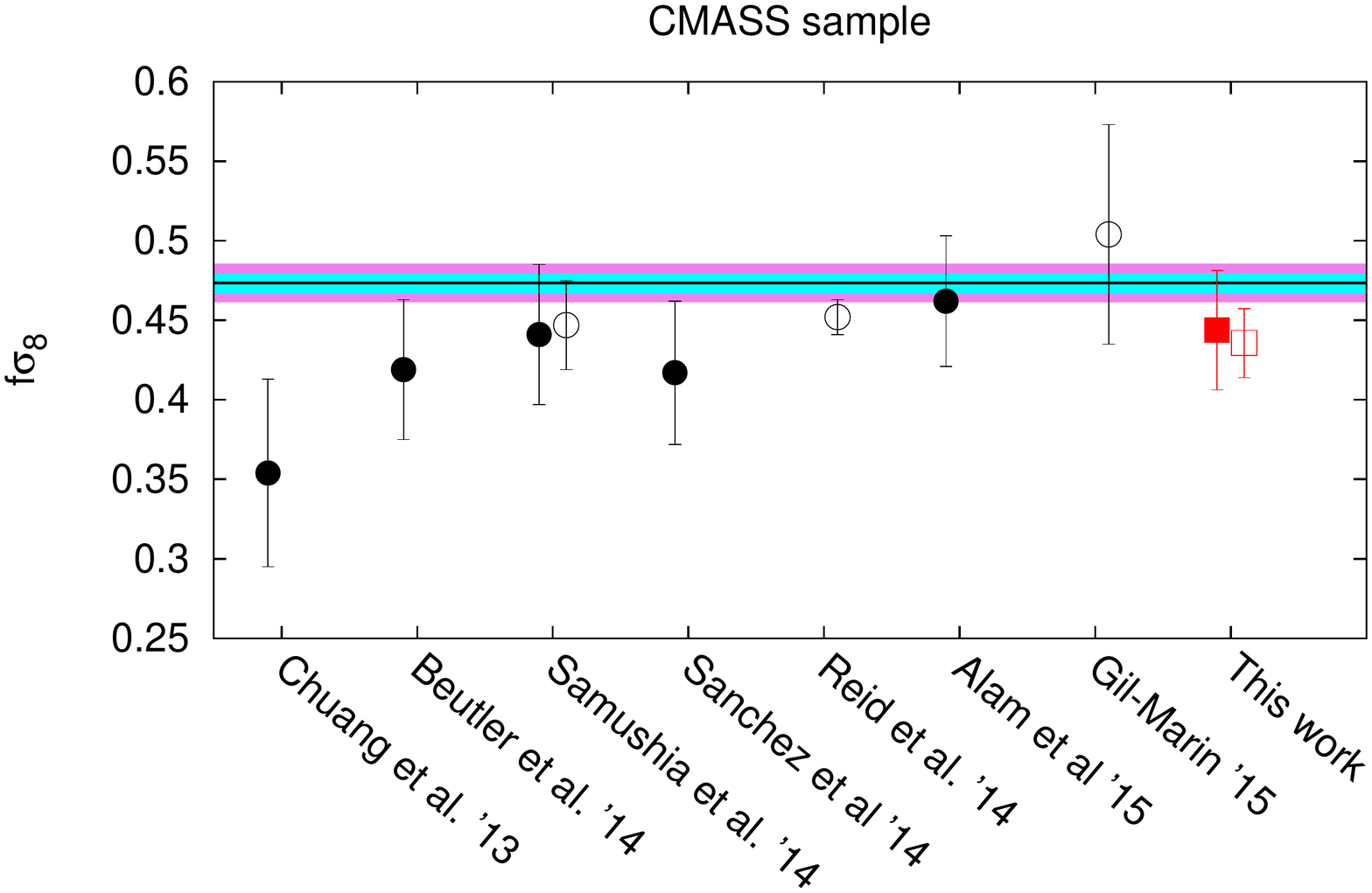}
\caption{The constrains on $f\sigma_8$ from LOWZ- (left panel) and CMASS-DR11 samples (right panel) are displayed in black circles \citep{Chuangetal:2013,Beutleretal:2013,Samushiaetal:2013,Sanchezetal:2013,Reidetal:2013,Alametal:2015,hector_bispectrum1}. In red squares the predictions LOWZ and CMASS-DR12 samples, respectively, as they are listed in Table \ref{table_finalresults}. Empty symbols are analysis with no-AP test, whereas filled symbols represents analysis where AP-test is performed.  The cyan and magenta bands show the $1$ and $2\sigma$-range, respectively, allowed by Planck TT+lowP+lensing in the base $\Lambda$CDM+GR model \citep{Planck_cosmology15}. }
\label{plot:boss}
\end{figure*}

In this section we compare our results on $f\sigma_8$ with other studies of  RSD based on the previous Data Release 11 (DR11) of the BOSS sample. DR11 contains $\sim10\%$ less galaxies than DR12, however, we expect that this change is not sufficiently large for producing signifiant changes (other than systematics) in the obtained best-fit of $f\sigma_8$. In Fig. \ref{plot:boss} we show the measurements of $f\sigma_8$ based on the following DR11 works: \citep{Chuangetal:2013,Beutleretal:2013,Samushiaetal:2013,Sanchezetal:2013,Reidetal:2013,Alametal:2015,hector_bispectrum1}.

\begin{enumerate}
\item \cite{Chuangetal:2013} analyse the DR11 LOWZ and CMASS two-point correlation function monopole and quadrupole in configuration space in the range of $56\,{\rm Mpc}h^{-1}\leq r \leq 200\, {\rm Mpc}h^{-1}$ for CMASS and $32\,{\rm Mpc}h^{-1}\leq r \leq 200\, {\rm Mpc}h^{-1}$ for LOWZ . Their work include $f\sigma_8$ measurements for both LOWZ and CMASS when the AP parameters are varied,  $f(0.57)\sigma_8(0.57)=0.354\pm0.059$ and $f(0.32)\sigma_8(0.32)=0.384\pm0.095$.

\item \citet{Beutleretal:2013} compute the DR11 CMASS  power spectrum monopole and quadrupole in $k$-space in the range of  $0.01\,h{\rm Mpc}^{-1}\leq k\leq 0.20\,h{\rm Mpc}^{-1}$. They report $f(0.57)\sigma_8(0.57)=0.419\pm0.044$ for the CMASS sample varying the AP parameters as well. 

\item \cite{Samushiaetal:2013} compute  the DR11 CMASS  two-point correlation function monopole and quadrupole in configuration space in the range of $24\,{\rm Mpc}h^{-1}\leq r \leq 152\, {\rm Mpc}h^{-1}$.  When they analyse the CMASS sample they obtain $f(0.57)\sigma_8(0.57)=0.441\pm0.044$  when the AP test is performed,  and $f(0.57)\sigma_8(0.57)=0.447\pm0.028$ when the AP parameters are tuned to their fiducial value. 

\item \cite{Sanchezetal:2013} analyse the two-point correlation function and the clustering wedges, parallel and perpendicular, for the DR11 LOWZ and CMASS sample using scales of $40\,{\rm Mpc}h^{-1}\leq r \leq 180\,{\rm Mpc}h^{-1}$  with AP test. They report $f(0.57)\sigma_8(0.57)=0.417\pm0.045$  and $f(0.32)\sigma_8(0.32)= 0.48\pm0.10$.

\item \cite{Reidetal:2013} perform a small scale analysis in the range $0.8\,{\rm Mpc}{h}^{-1}\leq r \leq 32\,{\rm Mpc}h^{-1}$ using halo occupation distribution and Planck cosmology. They report  $f(0.57)\sigma_8(0.57)=0.452\pm0.011$ for the CMASS sample. Note that the errors on this measurements are considerably smaller, but at the same time, they rely in significant modelling and cosmological assumptions, such as tuning the AP parameters to their fiducial value.

\item \cite{Alametal:2015} compute the DR11 CMASS two-point correlation function monopole and quadrupole in configuration space in the range of $30\,{\rm Mpc}h^{-1}\leq r \leq 126\, {\rm Mpc}h^{-1}$. Their analysis includes AP results and report $f(0.57)\sigma_8(0.57)=0.462\pm0.041$ for the CMASS sample.

\item \cite{hector_bispectrum1} compute the DR11 CMASS power spectrum monopole and bispectrum monopole in the range of $0.03\,h{\rm Mpc}^{-1}\leq k \leq 0.17\,h{\rm Mpc}^{-1}$.  They report a value of $f^{0.43}(0.57)\sigma_8(0.57)=0.582\pm0.084$ using RSD and no {  AP effect\footnote{{  The quantity constrained by the power spectrum monopole in combination with the bispectrum monopole is not $f$ times $\sigma_8$, but $f$ to the power 0.43 times $\sigma_8$. For a more detailed discussion about this topic see \cite{hector_bispectrum1}}}}. When this value is combined with $f^{\rm Planck13}=0.777$ ({  where {\it Planck13} is the fiducial cosmology used in that paper; see table 4 of \citealt{hector_bispectrum1})}, a $f\sigma_8$ value can be obtained, $f(0.57)\sigma_8(0.57)=0.504\pm0.069$. Note that this measurement is not based on the power spectrum quadrupole and relies on different cosmological assumptions. 
\end{enumerate}

The observed differences between the DR11 results are expected to be due to different systematics in the different models, scales and statistics considered to describe the shape of the power spectrum or two-point correlation function. With the exception of \cite{Chuangetal:2013} in CMASS, all the results are consistent within 1 and $2\sigma$. All the results that perform an AP analysis have a well  similar error-bars. 

For the CMASS sample, our $f\sigma_8$ measurement is consistent within $1\sigma$  with the DR11 reported values, with the exception of \cite{Chuangetal:2013}, {  which is $1.5\sigma$ below our findings and $\sim2\sigma$ below {\it Planck15} prediction}. In particular, our results are very close to those by \cite{Samushiaetal:2013} and \cite{Alametal:2015}, which are also consistent with {\it Planck15} data within $1\sigma$. When the AP parameters are tuned to their fiducial value, our result on $f\sigma_8$ is also very close to that of \cite{Samushiaetal:2013} and consistent within $1\sigma$ to the  \cite{Reidetal:2013} results. 

For the LOWZ sample our result lies between the best-fit values of \cite{Chuangetal:2013} and \cite{Sanchezetal:2013}. 

\subsection{Comparison of the geometrical AP parameters with other DR12 BOSS Data Releases analyses}\label{sec:boss}

 In this section we compare our results on the geometrical AP parameters, $H(z)r_s(z_d)$ and $D_A(z)/r_s(z_d)$ with other studies based on the analysis of the anisotropic 2-point function statistics of the Data Release 12 (DR12) of the BOSS sample. 

 In Table \ref{table:consensus} we compare the results from this paper (RSD-PS) with the BAO-based analysis of the anisotropic power spectrum (BAO-PS) by \cite{gil-marin15b} and the BAO analysis of the anisotropic correlation function (BAO-CF) by \cite{cuesta}. The BAO analyses provide results based on the same dataset used by the RSD analysis, which we refer to as pre-reconstruction analyses (Pre-BAO), and on the dataset obtained after applying the reconstruction algorithm, which we refer as post-reconstruction analyses (Post-BAO).

 The results of the pre-reconstruction analyses, Pre-BAO-PS, Pre-BAO-CF and RSD-PS, are very consistent with each other for both LOWZ and CMASS samples, and for both $H(z)r_s(z_d)$ and $D_A(z)/r_s(z_d)$. For LOWZ, $H(z)r_s(z_d)$ is almost exactly the same for Pre-BAO-PS, Pre-BAO-CF and RSD-PS, and $D_A(z)/r_s(z_d)$ Pre-BAO-PS and RSD-PS are $<0.5\sigma$ apart. For CMASS, $H(z)r_s(z_d)$, there is $<0.5\sigma$ between RSD-PS and Pre-BAO results, and the $D_A(z)/r_s(z_d)$ result is also very consistent, being the same for RSD-PS and Pre-BAO-PS, and very similar with Pre-BAO-CF as well ($<0.5\sigma$). In general, the small differences observed between Pre-BAO-CF and Pre-BAO-PS could come from i) observational systematics (such as photometric calibration, systematic and fibre collision weights) that may affect differently the measurement of the correlation function and the power spectrum, ii) from the fact that the correlation function and power spectrum (in finite scale-ranges) do not contain exactly the same information, or iii) simply that the statistical noise affects the measurements differently in the same way as a different binning of the data. The differences between Pre-BAO-PS and RSD-PS could have their origin in systematics of the model itself, or observational systematics that enter differently in the BAO and RSD modelling. In any case, these systematics are smaller than $0.5\sigma$.

  On the other hand, the results between post-reconstruction analyses and pre-reconstruction (including RSD-PS) are more significant, but always less than $2\sigma$. Reconstruction is known to reduce non-linear effects on the BAO signal, and some of the discrepancy could be due to this. This is discussed in more detail in both \cite{cuesta,gil-marin15b}. For this reason, we recommend using the post-reconstruction results of $D_A(z)/r_s(z_d)$ and $H(z)r_s(z_d)$, either from Post-BAO- CF, Post-BAO-PS, or the consensus value between both results presented in both papers, for constraining cosmological models. In case also the value of $f\sigma8$ is wanted to be used (in combination with the geometrical AP parameters) for constraining cosmological models, this paper is, at present, date the only BOSS DR12 paper that provides these measurements for the CMASS and LOWZ samples.

\begin{table}
\begin{center}
\begin{tabular}{c|c|c|c|}
Sample & Method-Statistic & $H(z)r_s(z_d)$ & $D_A(z)/r_s(z_d)$  \\
 \hline
 \hline
 \multirow{5}{*}{LOWZ} &   Post-BAO-PS & $11.64\pm0.62$ & $6.85\pm0.17$  \\  &  Post-BAO-CF & $11.65\pm0.84$ & $6.67\pm0.13$ \\ \cline{2-4}  & Pre-BAO-PS & $11.44\pm0.75$ & $6.48\pm0.27$  \\ &  Pre-BAO-CF & $11.44\pm0.74$ & $6.40\pm0.37$ \\ \cline{2-4}  & RSD-PS & $11.41\pm 0.56$ & $6.35 \pm 0.19$  \\
\hline
\multirow{5}{*}{CMASS} &   Post-BAO-PS & $14.56\pm0.38$ & $9.42\pm0.13$   \\  &  Post-BAO-CF & $14.75\pm0.54$ & $9.52\pm0.14$ \\ \cline{2-4} &  Pre-BAO-PS & $14.10\pm0.65$ & $9.42\pm0.22$ \\  & Pre-BAO-CF & $14.14\pm0.71$ & $9.51\pm0.19$ \\ \cline{2-4}  & RSD-PS & $13.92\pm 0.44$ & $9.42 \pm0.15$ \\
\end{tabular}
\caption{Geometrical AP parameters for LOWZ and CMASS samples: $H(z)r_s(z_d)$ (in $10^3\,{\rm km}s^{-1}$ units), $D_A(z)/r_s(z_d)$,  inferred from the BAO power spectrum analysis (BAO-PS) \citep{gil-marin15b}, BAO correlation function analysis (BAO-CF) \citep{cuesta}, and the RSD power spectrum analysis (RSD-PS) (this work). For the BAO analyses we present the pre- and post-reconstruction results, Pre-BAO and Post-BAO, respectively. Note that the RSD results are essentially a pre-reconstruction analysis, since the RSD signal is removed in the reconstruction process.  }
\label{table:consensus}
\end{center}
\end{table}

\subsection{Comparison with other galaxy surveys}\label{sec:surveys}

In this section we compare  our measurement on $f\sigma_8$ for the LOWZ and CMASS samples with the reported $f\sigma_8$ values of other galaxy surveys at different redshifts, and with {\it Planck15} predictions.

\begin{figure}
\centering 
\includegraphics[scale=0.3]{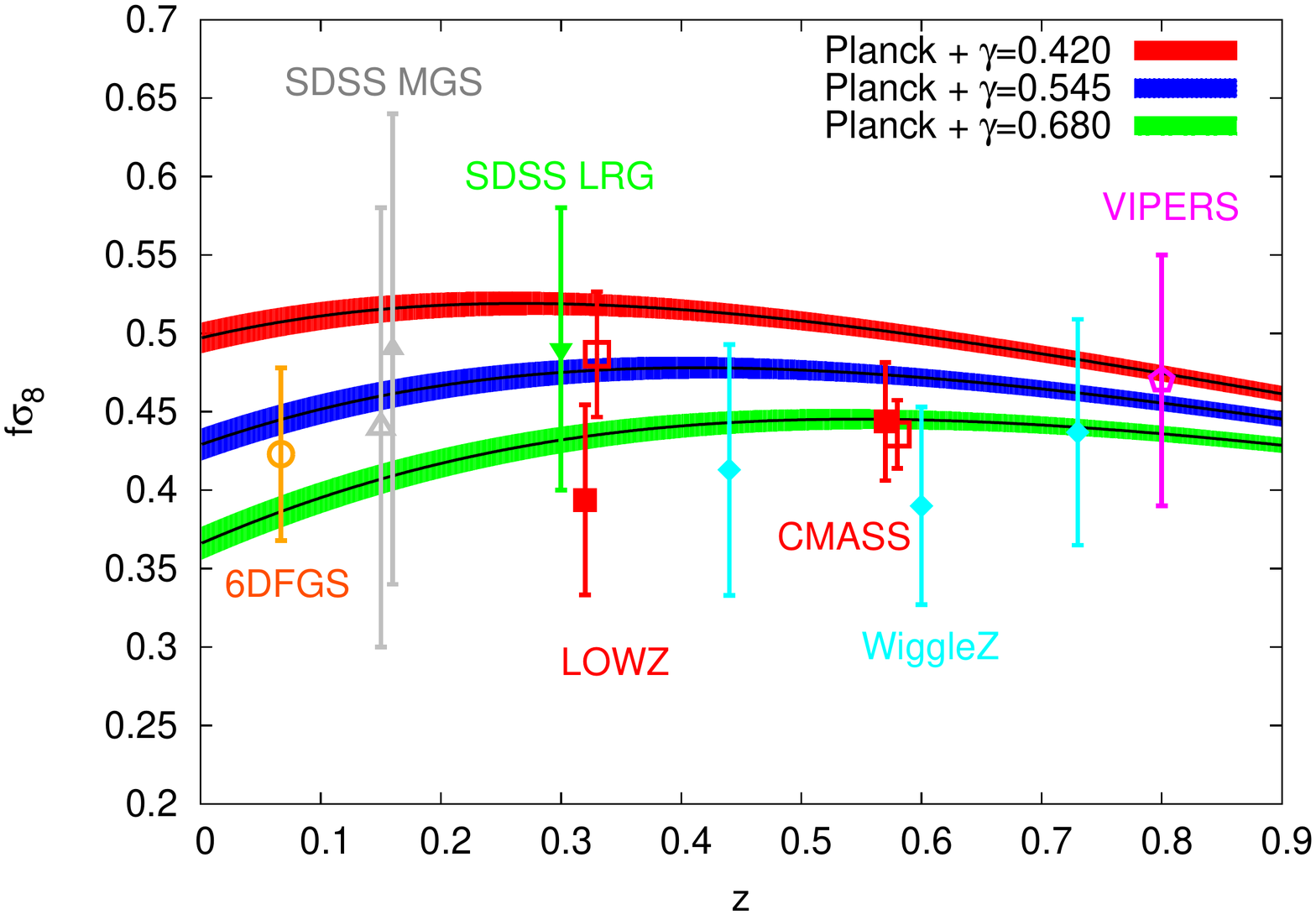}
\caption{Constraints on $f\sigma_8$ from several redshift surveys in the base of $\Lambda$CDM, with $f(z)=\Omega_m(z)^\gamma$: orange circle (6dFGRS by \citet{Beutleretal:2012}); gray triangle (SDSS Main Galaxy Sample by \citet{Howlettetal:2015}); green inverse triangle (SDSS Luminous Red Galaxies by \citet{Okaetal:2014}); cyan diamonds (WiggleZ by \citet{Blakeetal:2012}); and purple pentagon (VIPERS by \citet{Vipers}). In red squares the results from BOSS-DR12 for the LOWZ and CMASS sample according to Table \ref{table_finalresults}. Filled symbols represent the $f\sigma_8$ measurement when the RSD and the AP effect is considered, and filled symbols when only RSD effect is considered. For the empty red squares the $z$-position has been slightly displaced for clarity. The red, blue and green bands show the $1\sigma$-range allowed by Planck TT+lowP+lensing in the base $\Lambda$CDM model \citep{Planck_cosmology15} when $\gamma=0.420$, $\gamma=0.545$ (GR) and $\gamma=0.680$, respectively.}
\label{plot:surveys}
\end{figure}

Fig. \ref{plot:surveys} displays  the different RSD analyses of redshift galaxy surveys. The analyses with filled symbols solve simultaneously the RSD and AP effect, whereas those  represented with empty  symbols only consider RSD, keeping the cosmology fixed.  
\begin{enumerate}
\item 
The 6dFGRS survey analysis by \citet{Beutleretal:2012} reports a value of $f(0.067)\sigma_8(0.067)=0.423\pm0.055$. Their computation is based on measuring the redshift space correlation function in 2D. Since the effective redshift is very low, their measurement is insensitive to the AP effect. 
\item  
The Sloan Digital Sky Survey (SDSS) Data Release 7 Main Galaxy Sample analysis  by \cite{Howlettetal:2015} reports the measurement of the two-point correlation function. Using only RSD effect they obtain $f(0.15)\sigma_8(0.15)=0.44^{+0.16}_{-0.12}$. When the AP effect is included\footnote{In this result the AP effect is only partially included since $\alpha_\parallel$ and $\alpha_\perp$ are set to be equal but freely vary. When both parameters are set free the reported value is $f\sigma_8=0.63^{+0.24}_{-0.27}$.} they obtain $f(0.15)\sigma_8(0.15)=0.49^{+0.15}_{-0.14}$.
\item 
The analysis of the Luminous Red Galaxy (LRG) sample in the Data Release 7 of the Sloan Digital Sky Survey II by \cite{Okaetal:2014} measures the galaxy monopole, quadrupole and hexadecapole power spectrum multipoles. They use the RSD and AP effect to report $f(0.3)\sigma_8(0.3)= 0.49\pm0.09$.
\item 
The WiggleZ analysis by \cite{Blakeetal:2012}  measures the galaxy correlation function and power spectrum. Considering both RSD and AP effects, they report $f\sigma_8$ at 3 different redshifts, $f(0.44)\sigma_8(0.44)=0.413\pm0.080$, $f(0.6)\sigma_8(0.6)=0.390\pm0.063$ and $f(0.73)\sigma_8(0.73)=0.437\pm0.072$. 
\item The analysis of VIPERS by \cite{Vipers} measures the two-point correlation function. Using the RSD effect they report $f(0.8)\sigma_8(0.8)=0.47\pm0.08$. Given the volume of the survey, the results are very insensitive to the AP effect, which is not taken into account. 
\end{enumerate}

In Fig. \ref{plot:surveys} we also plot the model predictions for {\it Planck15} best-fit, $\Omega_m=0.308$, when different theories of gravity are assumed. We work on the assumption of $f=\Omega_m(z)^\gamma$. The prediction  for GR with a cosmological constant is $\gamma=0.545$, which is plotted in blue bands (the $1\sigma$ limits). We also plot the predictions for two additional values of $\gamma$, in red bands we plot $\gamma=0.420$ and in green bands $\gamma=0.680$.

{  Under the assumption of {\it Planck15}+{GR}, the $\gamma$ factor depends on the dark energy equation of state, $\omega$. For $\omega=-1$ we have the cosmological constant prediction, $\gamma=0.545$. We observe that in general all the results are in agreement with this prediction within $1\sigma$ and $2\sigma$ confidence levels. 
For dark energy models with $\omega>-1$, such as the parametrised Post-Friedman scalar field models, $\gamma>0.545$; whereas for $\omega<-1$, such as phantom dark energy models, $\gamma<0.545$. In the light of results of Fig. \ref{plot:surveys}, we observe that qualitatively, redshift galaxy surveys observations slightly favour models with $\omega>-1$, if $\Lambda {\rm CDM}+{\rm GR}$ is assumed, although the deviation with respect $\omega=-1$ is not very significant. On the other hand, if the GR condition is relaxed, $\gamma$ can also change. This is the case for example of the DGP model \citep{dgp}, whose prediction for $\gamma$ is 0.68, and it is slightly favoured with respect to GR. }

\subsection{Correlated Measurements}\label{sec:multivariate_likelihood}
In this subsection we present the multivariate Gaussian likelihoods calculated from the DR12 data from which marginalised results were presented in Table \ref{table_results1}. We focus on those parameters which are of cosmological interest, such as $f\sigma_8$, $H(z)r_s(z_d)$ and $D_A(z)/r_s(z_d)$. The errors of these parameters are  highly correlated, as one can infer from  Fig. \ref{plot:scatter}. Therefore, to jointly use these data, one needs to use  their covariance matrix.  

We start by defining the data vector containing the cosmology parameters of interest $f(z)\sigma_8(z)$, $H(z) r_s(z_d)$ (in $10^3 {\rm km}s^{-1}$ units) and $D_A(z)/r_s(z_d)$, 
\begin{equation}
D^{\rm data}(z) = 
 \begin{pmatrix}
  f(z)\sigma_8(z)  \\
  H(z) r_s(z_d)\, [10^3 {\rm km}s^{-1}] \\
  D_A(z)/r_s(z_d)
 \end{pmatrix}.
 \end{equation}
The best-fit values of the LOWZ and CMASS samples for these parameters are presented in Table \ref{table_results1}, given a covariance matrix, either \textsc{qpm} or \textsc{MD-Patchy}.

The covariance matrices of these parameters are,

\begin{equation}
C_{{\rm LOWZ}}^{\rm \textsc{MD-Patchy}} = 10^{-3}
 \begin{pmatrix}
4.1028 & 25.549 & 7.4600  \\
 - & 310.08 & 51.366 \\
 - & - & 34.912
 \end{pmatrix},
 \end{equation}
 
\begin{equation}
C_{{\rm LOWZ}}^{\rm \textsc{qpm}} = 10^{-3}
 \begin{pmatrix}
3.7082 & 22.721 & 7.2898  \\
 - & 301.46 & 50.403 \\
 - & - & 32.718
 \end{pmatrix},
 \end{equation}

\begin{equation}
C_{{\rm CMASS}}^{\rm \textsc{MD-Patchy}} = 10^{-3}
 \begin{pmatrix}
1.3424 & 10.597 & 3.7495 \\
 - & 179.80 & 34.180 \\
 - & - & 23.495
 \end{pmatrix},
 \end{equation}

\begin{equation}
C_{{\rm CMASS}}^{\rm \textsc{qpm}} = 10^{-3}
 \begin{pmatrix}
1.4475 & 11.244 & 4.0507 \\
 - & 188.69 & 36.234 \\
 - & - & 24.698
 \end{pmatrix},
 \end{equation}
for LOWZ and CMASS samples, using the \textsc{qpm} or \textsc{MD-Patchy} mocks to infer the best fit parameters from the power spectrum multipoles measurements, as labeled. These covariance matrices are symmetric by construction ($C_{ij}\equiv C_{ji}$) and consequently we only provide the results of half of the matrix.

From these matrices, the likelihood of any cosmological model is given by,
\begin{equation}
\mathcal{L}\propto \exp\left[ -(D^{\rm data} - D^{\rm model})^T C^{-1} (D^{\rm data} - D^{\rm model}) / 2  \right],
\end{equation}
where $D^{\rm model}$ is the vector with the model prediction for the same cosmological parameters as $D^{\rm data}$.

\begin{figure*}
\centering
\includegraphics[scale=0.31]{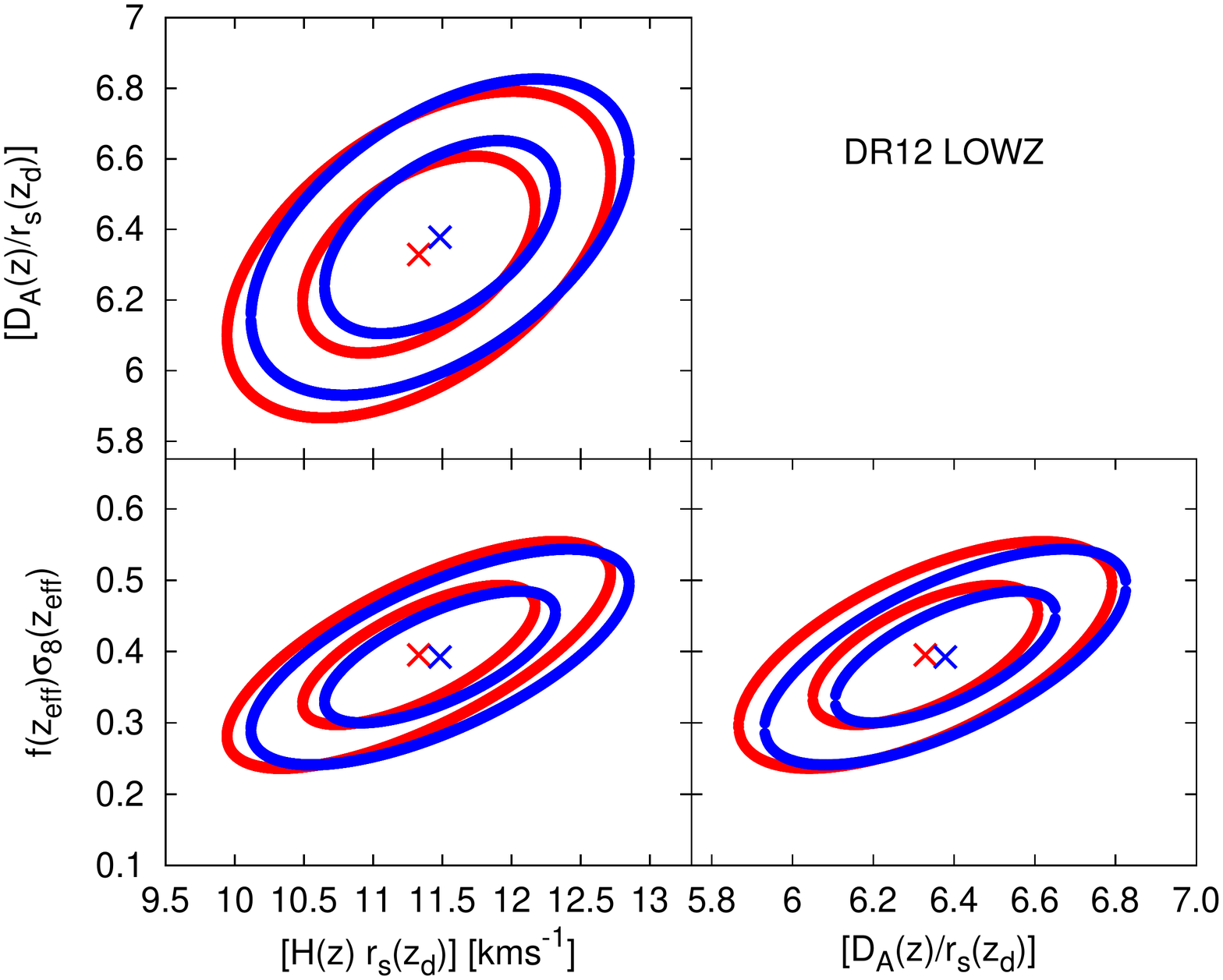}
\includegraphics[scale=0.31]{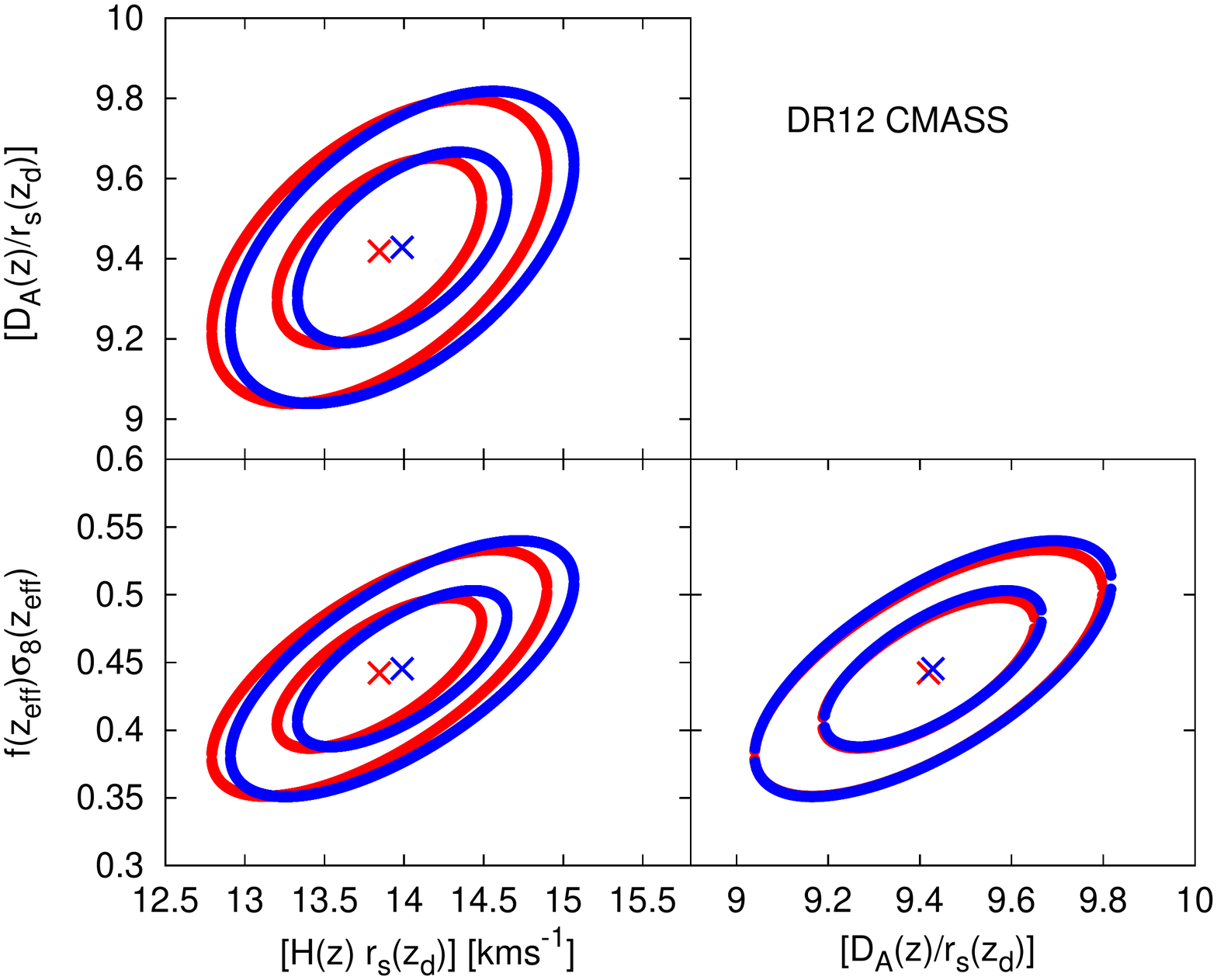}
\caption{Likelihood surfaces of the parameters $f\sigma_8$, $D_A(z)/r_s(z_d)$ and $H(z)r_s(z_d)$ extracted from the DR12 data, LOWZ sample (left panel) and CMASS sample (right panel), using the covariance matrices provided by \textsc{MD-Patchy} mocks (red ellipses) and \textsc{qpm} mocks (blue ellipses). Each ellipse correspond to $1\sigma$ ($\Delta\chi^2=2.30$) and $2\sigma$ ($\Delta\chi^2=6.17$) confidence levels, when the 3 parameters are marginalised. Each type of ellipse has been centered on the minimum solution presented in Table \ref{table_results1}.}
\label{plot:elipses}
\end{figure*}

In Fig. \ref{plot:elipses} we show the ellipses which represent the likelihood surface of $1\sigma$ ($\Delta\chi^2=2.30)$ and $2\sigma$ $(\Delta\chi^2=6.17)$, corresponding to the covariance matrices presented above, for LOWZ and CMASS samples, left and right panels, respectively, for \textsc{MD-Patchy} (red lines) and for \textsc{qpm} mocks (blue lines). Each ellipse is centered in the minimum solution presented in Table \ref{table_results1}. We observe that the differences between the covariances are small, and most of the shifts are  in the minimum where they are centered, and not in the shape and orientation of the ellipsoid itself. Therefore, we conclude that the difference in covariance obtained by using either \textsc{MD-Patchy} or \textsc{qpm} mocks is not significant. 

\section{Conclusions}\label{sec:conclusions}
In this paper we have presented a measurement of the isotropic and anisotropic power spectrum relative to the LOS of the LOWZ and CMASS DR12 galaxy samples of the Baryon Oscillations Spectroscopic Survey of the Sloan Digital Sky Survey III. We have analysed the  redshift space distortions in the power spectrum multipoles and the constraints imposed on the growth factor times the amplitude of linear power spectrum, $f\sigma_8$. We have also considered the geometrical Alcock-Paczynski effect, which allows us to set  constrains on the angular diameter distance parameter $D_A(z_{\rm eff})/r_s(z_d)$ and the Hubble parameter $H(z_{\rm eff})r_s(z_d)$. 
We study the BAO peak position on the monopole and quadrupole power spectrum in a companion paper which is released at the same time of the present paper \citep{gil-marin15b}. 

In order to extract cosmological information from the galaxy power spectrum multipole measurements we have used a non-local and non-linear bias model \citep{McDonaldRoy:2009} which depends on 4 parameters, $b_1$, $b_2$, $b_{s^2}$ and $b_{3\rm nl}$. Imposing that the bias is local in Lagrangian space, $b_{s^2}$ and $b_{3\rm nl}$ are set by the value of $b_1$, and therefore only two free parameters are left to marginalise over, $b_1$ and $b_2$. To model the RSD we use the TNS model \cite{Taruyaetal:2010,Nishimichietal:2011} which has been used in previous data releases of BOSS to describe the power spectrum multipoles, as well as the bispectrum monopole. The RSD model depends on the logarithmic rate of structure growth, $f$,  on the FoG damping parameter, $\sigma_{FoG}$ and on the value of the shot noise. Although $f$ is directly related to the assumed $\Omega_m$ given a theory of gravity, we have kept it free in order to test possible deviations from GR. In order to model the real space quantities in the TNS model, we have used  the resumed perturbation theory at 2-loop order presented in \cite{HGMetal:2012}. 
In our model we have also included the geometrical AP effect, through the dilation parameters $\alpha_\parallel$ and $\alpha_\perp$, which modifies the wave modes parallel and perpendicular to the LOS, respectively. These parameters are related to the angular diameter distance and the Hubble parameter, which we are also able to constrain. In our analysis we have fixed the shape of the linear power spectrum using the fiducial cosmology ${\boldsymbol \Omega}^{\rm fid}$, but we have marginalised over the amplitude $\sigma_8$. In total, our galaxy redshift space power spectrum model has 8 free parameters we marginalise over. 

We have tested possible systematics of our model using the \textsc{MD-Patchy} mocks simulations, which have been designed to reproduce up to $k\simeq0.3\,h{\rm Mpc}^{-1}$ the power spectrum and bispectrum of data and simulations \citep[][companion paper]{Kitauraetal:2015}. Using the mean from the 2048 simulations to gain signal, we find that we are able to recover $f\sigma_8$ with an accuracy of $<3\%$ for LOWZ and  $<2\%$ for CMASS with a $k_{\rm max}\leq 0.24\,h{\rm Mpc}^{-1}$. The systematics observed are significantly smaller than the statistical errors we measure for the DR12 BOSS data, and therefore we do not consider to correct the measurements, neither the corresponding errors. 

We have computed the full covariance matrix of the power spectrum multipoles using two different types of galaxy mocks, 1000 realizations of \textsc{qpm} mocks and 2048 realizations of \textsc{MD-Patchy} mocks, and we have performed two parallel analyses using these two covariance matrices. Since the differences in the best fit parameters and their corresponding errors have been found to be small, we have decided to take the average among these two approaches, to generate a unique set of results. 

We find that for the DR12 LOWZ sample $f(z_{\rm lowz})\sigma_8(z_{\rm lowz})=0.394\pm0.062$, $D_A(z_{\rm lowz})/r_s(z_d)=6.35\pm0.19$, $H(z_{\rm lowz})r_s(z_d)=(11.41\pm 0.56)\,{10^3\rm km}s^{-1}$, where $z_{\rm lowz}=0.32$. For DR12 CMASS we find $f(z_{\rm cmass})\sigma_8(z_{\rm cmass})=0.444\pm0.038$, $D_A(z_{\rm cmass})/r_s(z_d)=9.42\pm0.15$, $H(z_{\rm cmass})r_s(z_d)=(13.92 \pm 0.44)\, {10^3\rm km}s^{-1}$,{  where $z_{\rm cmass}=0.57$}. A covariance matrix for these measurements was also presented. These are the main results of this paper and are in general agreement with previous BOSS DR11 measurements. Furthermore, we have been able to reduce the previous error-bars on $f\sigma_8$, shrinking them down to $15\%$ for LOWZ sample and $8.5\%$ for CMASS sample, which are the most precise  measurements of this parameter at these redshifts at the moment, when the full AP effect is considered.  Additionally, if we assume that  the Hubble parameter and angular distance parameter are fixed at fiducial $\Lambda$CDM values, we find $f(z_{\rm lowz})\sigma_8(z_{\rm lowz})=0.485\pm0.044$ and $f(z_{\rm cmass})\sigma_8(z_{\rm cmass})=0.436\pm0.022$ for the LOWZ and CMASS samples, respectively. In this case the error bars represent a $9.1\%$ for LOWZ and  $5.0\%$ for CMASS.

Moreover, we have analysed the data with two additional cosmological models, $\Omega_m=0.292$ and $\Omega_m=0.332$, which in this case we have changed the shape of the linear power spectrum of our model, accordingly. Overall, we observe that both CMASS and LOWZ galaxies data is in agreement with the fiducial model $\Omega_m^{\rm fid}=0.31$ consistent with {\it Planck15} data: for the CMASS sample the tension is below $1\sigma$ and for the LOWZ sample within $2\sigma$. For both samples, when the cosmological model is changed, the value of $f\sigma_8$ only changes by $\leq0.5\sigma$.  

The constraints on $f(z_{\rm eff})\sigma_8z_{\rm eff}$, along with $H(z_{\rm eff})r_d(z)$ and $D_a(z_{\rm eff})r_d(z_{\rm eff})$, will be useful in a joint analysis with other cosmological data sets (in particular CMB data) for setting stringent constraints on neutrino mass, dark energy, gravity, curvature as well as number of neutrino species.

\section*{Acknowledgements}

HGM is grateful for support from the UK Science and Technology Facilities Council through the grant
ST/I001204/1.
WJP is grateful for support from the UK Science and Technology Facilities Research Council through the grant
ST/I001204/1, and the European  Research Council through the grant ``Darksurvey". FSK acknowledges the support of the Karl-Schwarzschild Program from the Leibniz Society.
 FSK, SRT, CC, and FP acknowledge support from the Spanish MICINNs Consolider-Ingenio 2010 Programme under grant MultiDark CSD2009-00064, MINECO Centro de Excelencia Severo Ochoa Programme under grant SEV-2012-0249, and grant AYA2014-60641-C2- 1-P.  The massive production of all MultiDark Patchy BOSS DR12 mocks has been performed at the BSC Marenostrum supercomputer, the Hydra cluster at the Instituto de Fisica Teorica UAM/CSIC and NERSC at the Lawrence Berkeley National Laboratory.

We thank Shun Saito and Florian Beutler for comments and helpful discussions. 

Funding for SDSS-III has been provided by the Alfred P. Sloan
Foundation, the Participating Institutions, the National Science
Foundation, and the U.S. Department of Energy Office of Science. The
SDSS-III web site is http://www.sdss3.org/.

SDSS-III is managed by the Astrophysical Research Consortium for the
Participating Institutions of the SDSS-III Collaboration including the
University of Arizona,
the Brazilian Participation Group,
Brookhaven National Laboratory,
University of Cambridge,
Carnegie Mellon University,
University of Florida,
the French Participation Group,
the German Participation Group,
Harvard University,
the Instituto de Astrofisica de Canarias,
the Michigan State/Notre Dame/JINA Participation Group,
Johns Hopkins University,
Lawrence Berkeley National Laboratory,
Max Planck Institute for Astrophysics,
Max Planck Institute for Extraterrestrial Physics,
New Mexico State University,
New York University,
Ohio State University,
Pennsylvania State University,
University of Portsmouth,
Princeton University,
the Spanish Participation Group,
University of Tokyo,
University of Utah,
Vanderbilt University,
University of Virginia,
University of Washington,
and Yale University.
This research used resources of the National Energy Research Scientific
Computing Center, which is supported by the Office of Science of the
U.S. Department of Energy under Contract No. DE-AC02-05CH11231.

Numerical computations were done on the Sciama High Performance Compute (HPC) cluster which is supported by the ICG, SEPNet and the University of Portsmouth.

%
%
%


\def\jnl@style{\it}
\def\aaref@jnl#1{{\jnl@style#1}}

\def\aaref@jnl#1{{\jnl@style#1}}

\def\aj{\aaref@jnl{AJ}}                   
\def\araa{\aaref@jnl{ARA\&A}}             
\def\apj{\aaref@jnl{ApJ}}                 
\def\apjl{\aaref@jnl{ApJ}}                
\def\apjs{\aaref@jnl{ApJS}}               
\def\ao{\aaref@jnl{Appl.~Opt.}}           
\def\apss{\aaref@jnl{Ap\&SS}}             
\def\aap{\aaref@jnl{A\&A}}                
\def\aapr{\aaref@jnl{A\&A~Rev.}}          
\def\aaps{\aaref@jnl{A\&AS}}              
\def\azh{\aaref@jnl{AZh}}                 
\def\baas{\aaref@jnl{BAAS}}               
\def\jrasc{\aaref@jnl{JRASC}}             
\def\memras{\aaref@jnl{MmRAS}}            
\def\mnras{\aaref@jnl{MNRAS}}             
\def\pra{\aaref@jnl{Phys.~Rev.~A}}        
\def\prb{\aaref@jnl{Phys.~Rev.~B}}        
\def\prc{\aaref@jnl{Phys.~Rev.~C}}        
\def\prd{\aaref@jnl{Phys.~Rev.~D}}        
\def\pre{\aaref@jnl{Phys.~Rev.~E}}        
\def\prl{\aaref@jnl{Phys.~Rev.~Lett.}}    
\def\pasp{\aaref@jnl{PASP}}               
\def\pasj{\aaref@jnl{PASJ}}               
\def\qjras{\aaref@jnl{QJRAS}}             
\def\skytel{\aaref@jnl{S\&T}}             
\def\solphys{\aaref@jnl{Sol.~Phys.}}      
\def\sovast{\aaref@jnl{Soviet~Ast.}}      
\def\ssr{\aaref@jnl{Space~Sci.~Rev.}}     
\def\zap{\aaref@jnl{ZAp}}                 
\def\nat{\aaref@jnl{Nature}}              
\def\iaucirc{\aaref@jnl{IAU~Circ.}}       
\def\aplett{\aaref@jnl{Astrophys.~Lett.}} 
\def\apspr{\aaref@jnl{Astrophys.~Space~Phys.~Res.}}
\def\bain{\aaref@jnl{Bull.~Astron.~Inst.~Netherlands}} 
\def\fcp{\aaref@jnl{Fund.~Cosmic~Phys.}}  
\def\gca{\aaref@jnl{Geochim.~Cosmochim.~Acta}}   
\def\grl{\aaref@jnl{Geophys.~Res.~Lett.}} 
\def\jcp{\aaref@jnl{J.~Chem.~Phys.}}      
\def\jgr{\aaref@jnl{J.~Geophys.~Res.}}    
\def\jqsrt{\aaref@jnl{J.~Quant.~Spec.~Radiat.~Transf.}}
\def\memsai{\aaref@jnl{Mem.~Soc.~Astron.~Italiana}}
\def\nphysa{\aaref@jnl{Nucl.~Phys.~A}}   
\def\physrep{\aaref@jnl{Phys.~Rep.}}   
\def\physscr{\aaref@jnl{Phys.~Scr}}   
\def\planss{\aaref@jnl{Planet.~Space~Sci.}}   
\def\procspie{\aaref@jnl{Proc.~SPIE}}   
\def\jcap{\aaref@jnl{J. Cosmology Astropart. Phys.}}

\let\astap=\aap
\let\apjlett=\apjl
\let\apjsupp=\apjs
\let\applopt=\ao

\newcommand{\etal}{et al.\ }

\newcommand{\mpc}{\, {\rm Mpc}}
\newcommand{\kpc}{\, {\rm kpc}}
\newcommand{\hmpc}{\, h^{-1} \mpc}
\newcommand{\ihmpc}{\, h\, {\rm Mpc}^{-1}}
\newcommand{\ikms}{\, {\rm s\, km}^{-1}}
\newcommand{\kms}{\, {\rm km\, s}^{-1}}
\newcommand{\hkpc}{\, h^{-1} \kpc}
\newcommand{\lya}{Ly$\alpha$\ }
\newcommand{\lyb}{Lyman-$\beta$\ }
\newcommand{\lyaf}{Ly$\alpha$ forest}
\newcommand{\lr}{\lambda_{{\rm rest}}}
\newcommand{\bF}{\bar{F}}
\newcommand{\bS}{\bar{S}}
\newcommand{\bC}{\bar{C}}
\newcommand{\bB}{\bar{B}}
\newcommand{\vdF}{{\mathbf \delta_F}}
\newcommand{\vdS}{{\mathbf \delta_S}}
\newcommand{\vdf}{{\mathbf \delta_f}}
\newcommand{\vdn}{{\mathbf \delta_n}}
\newcommand{\vdC}{{\mathbf \delta_C}}
\newcommand{\vdX}{{\mathbf \delta_X}}
\newcommand{\xrei}{x_{rei}}
\newcommand{\lrmin}{\lambda_{{\rm rest, min}}}
\newcommand{\lrmax}{\lambda_{{\rm rest, max}}}
\newcommand{\lmin}{\lambda_{{\rm min}}}
\newcommand{\lmax}{\lambda_{{\rm max}}}
\newcommand{\hi}{\mbox{H\,{\scriptsize I}\ }}
\newcommand{\heii}{\mbox{He\,{\scriptsize II}\ }}
\newcommand{\vp}{\mathbf{p}}
\newcommand{\vq}{\mathbf{q}}
\newcommand{\vxperp}{\mathbf{x_\perp}}
\newcommand{\vkperp}{\mathbf{k_\perp}}
\newcommand{\vrperp}{\mathbf{r_\perp}}
\newcommand{\vx}{\mathbf{x}}
\newcommand{\vy}{\mathbf{y}}
\newcommand{\vk}{\mathbf{k}}
\newcommand{\vR}{\mathbf{r}}
\newcommand{\tdtwo}{\tilde{b}_{\delta^2}}
\newcommand{\tstwo}{\tilde{b}_{s^2}}
\newcommand{\tbthree}{\tilde{b}_3}
\newcommand{\tadtwo}{\tilde{a}_{\delta^2}}
\newcommand{\tastwo}{\tilde{a}_{s^2}}
\newcommand{\tabthree}{\tilde{a}_3}
\newcommand{\vnabla}{\mathbf{\nabla}}
\newcommand{\tpsi}{\tilde{\psi}}
\newcommand{\vv}{\mathbf{v}}
\newcommand{\fnl}{{f_{\rm NL}}}
\newcommand{\tfnl}{{\tilde{f}_{\rm NL}}}
\newcommand{\gnl}{g_{\rm NL}}
\newcommand{\orderfour}{\mathcal{O}\left(\delta_1^4\right)}
\newcommand{\SDSSPF}{\cite{2006ApJS..163...80M}}
\newcommand{\PF}{$P_F^{\rm 1D}(k_\parallel,z)$}
\newcommand\ion[2]{#1$\;${\small \uppercase\expandafter{\romannumeral #2}}}%
\newcommand\ionalt[2]{#1$\;${\scriptsize \uppercase\expandafter{\romannumeral #2}}}%
\newcommand{\vxone}{\mathbf{x_1}}
\newcommand{\vxtwo}{\mathbf{x_2}}
\newcommand{\vRot}{\mathbf{r_{12}}}
\newcommand{\cm}{\, {\rm cm}}

\bibliographystyle{mn2e}
\bibliography{dr12_rsd.bib}

\appendix

 \section{Effect of systematic weights at large scales}\label{appendix_b}
In this appendix we study the impact of the systematic weights in the power spectrum monopole and quadrupole of the CMASS sample. The systematic weights are designed to correct for fluctuations in the target density caused by changes in the observational efficiency \citep{Rossetalinprep}. The CMASS sample presents correlations between the galaxy density at large scales and the systematic weights are designed  to correct for these variations giving an isotropic weighted field. However, the accuracy of these weights is limited and we are interested in  their accuracy not only for the monopole, but also for the quadrupole, which we expect to be more sensitive to these corrections. 

In order to test  this effect, we have analysed the CMASS data power spectrum multipoles before (pre-systematic-weight correction power spectrum, $P_{\rm no-sys}$) and after  the application of the systematic weights (post-systematic-weight correction power spectrum, $P_{\rm sys}$).  We expect that these pre- and post-systematic-weight power spectrum multipoles converge to the same values at sufficiently small scales, where the fluctuations in the target density are not relevant. However, at larger scales, where the effects of the fluctuations in the target density are not negligible, they will predict different power spectrum amplitudes. By analyzing this difference, we will estimate the percentile correction of the systematic weights on the power spectrum multipoles. Setting a limit of $\sim5\%$ correction, we define a large scale cutoff for the power spectrum multipoles. 

\begin{figure}
\centering
\includegraphics[scale=0.3]{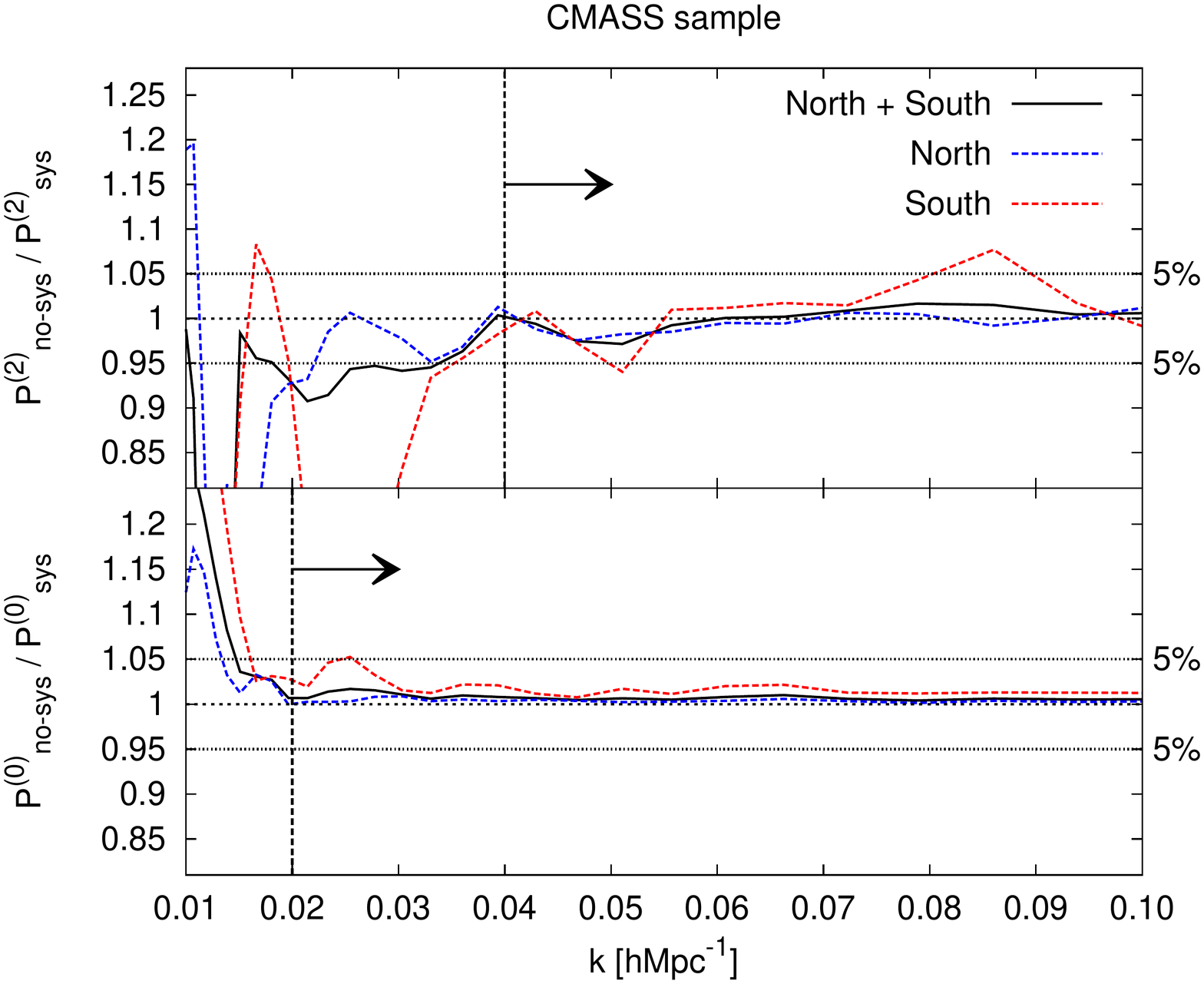}
\caption{Effect of systematic weights in the CMASS data sample for the NGC (blue dashed lines), SGC (red dashed lines) and NGC+SGC (black solid lines). The arrows indicate the large scale cuts applied in order to keep those scales where the correction represents less than $\sim 5\%$ of the signal. These $k$-cuts are $k=0.02\,h{\rm Mpc}^{-1}$ for the monopole (lower sub-panel) and $k=0.04\,h{\rm Mpc}^{-1}$ for the quadrupole (upper sub-panel). }
\label{plot:appendixb}
\end{figure}

  This is displayed in Fig. \ref{plot:appendixb}, where the ratio between the data power spectrum multipoles is shown for the monopole (lower sub-panel) and quadrupole (upper sub-panel). The different colour lines show this effect for the NGC, SGC, and NGC+SGC as labeled. 

As expected, the effect of the systematic weights is relevant only at large scales, where they suppress spurious correlations and has a higher impact on the quadrupole respect to the monopole. Setting an accuracy limit of $\sim5\%$ correction, we  discard those scales where the systematic weight correction on the power spectrum multipoles exceeds this limit. Furthermore, we can compute the $\chi^2$ for the difference between the weighted and unweighted $P^{(0)}$ and $P^{(2)}$. This is $\chi^2=DC^{-1}D^t$, where $D\equiv P_{\rm sys}-P_{\rm no-sys}$. We obtain that for $k_{\rm min}=0.02\,h{\rm Mpc}^{-1}$ for the monopole and $k_{\rm min}=0.04\,h{\rm Mpc}^{-1}$ for the quadrupole, (as indicated by the black arrows) and $k_{\rm max}=0.25\,h{\rm Mpc}^{-1}$ for both statistic, the reduced $\chi^2$ is 0.044. This means that by applying the weights we are maximally correcting by just $\sim0.2\sigma$ ($0.2\simeq\sqrt{0.044}$) in some measured parameter. This set a maximal impact of the known systematic without the need of defining any particular model. 

Therefore, in this paper we will only consider for our analyses  those scales smaller than $k=0.02\,h{\rm Mpc}^{-1}$ for the monopole and $k=0.04\,h{\rm Mpc}^{-1}$ in the quadrupole, as indicated by the black arrows.  

\section{Effect of fiber collisions weights on the best-fit parameter estimation}\label{sec:fc}

In this appendix we study  the effect of the fiber collision weights on the power spectrum multipoles, and more precisely on the $f\sigma_8$ measurement according to the model presented in \S\ref{sec:modelling}.
As described in \S\ref{sec:bossdata}, the fiber collision weights are included in order to account for those galaxy pairs that are too close to each other ($<62''$) to put two fibre detectors. The fraction of collided galaxies, $f_{cg}\equiv\sum_i [w_{\rm fc}({\bf x}_{i})-1)] / \sum_i w_{\rm fc}({\bf x}_{i})$, for the data is presented in the first column of Table \ref{table:collidedgalaxies}, for the CMASS and LOWZ samples. Since the number density of galaxies is higher in the CMASS sample, the fraction of collided galaxies is also higher in this sample.  The second and the third column of Table \ref{table:collidedgalaxies} present the fraction of collided galaxies found in the \textsc{qpm} and \textsc{MD-Patchy} mocks, respectively. For the CMASS sample, the \textsc{MD-Patchy} mocks present a smaller value of $f_{cg}$ respect to the data and \textsc{qpm} mocks, which both are in close agreement. This is due to a limitation in the resolution of substructure inside \textsc{MD-Patchy} haloes. This limitation will be solved in future versions of the \textsc{MD-Patchy} mocks.  In the LOWZ sample, both data and  \textsc{MD-Patchy} mocks agree well with the value of $f_{cg}$, because for the number density of galaxies of this sample, the resolution of substructure was not a limiting factor. However, we see that  \textsc{qpm} mocks present a higher value of $f_{cg}$ with respect to the data. This is due to the version of \textsc{qpm} mocks used in this paper matched an old LOWZ catalogue, where galaxies with previously known redshifts were subsample to match the BOSS close-pairs selection. This subsample has now been discontinued.  As for the \textsc{MD-Patchy} mocks, this will be fixed in future releases.
\begin{table}
\begin{center}
\begin{tabular}{c|c|c|c}
 & DR12 dataset & \textsc{qpm} & \textsc{MD-Patchy}\\
\hline
LOWZ & 0.0157 & 0.0418 & 0.0148\\
CMASS & 0.0528 & 0.0529 & 0.0349
\end{tabular}
\end{center}
\caption{Fraction of collided galaxies, $f_{cg}\equiv\sum_i [w_{\rm fc}({\bf x}_{i})-1] / \sum_i w_{\rm fc}({\bf x}_{i})$, for the data, for the \textsc{qpm} and \textsc{MD-Patchy} mocks, for the CMASS and LOWZ samples. }
\label{table:collidedgalaxies}
\end{table}%

In order to test the effect of the fiber collisions in the parameter estimation, we focus on the \textsc{qpm} mocks for the CMASS sample, which has the higher value of $f_{cg}$ amongst all the cases. We measure the monopole and quadrupole of 1000 mock realizations and take their average in order to gain signal to noise. We consider the two following selection of galaxies
\begin{enumerate}
\item We treat the galaxies as in a real survey. When two or more galaxies present an angular separation of $\leq62''$, we weight one of them by the number of galaxies within the $\leq62''$ angular radius, and remove the others. This mimics what it is done with the real dataset  
\item  We consider all the galaxies resolved in the mocks and weight them equally. This is the ideal case we would have if all the targeted galaxies in the survey were analysed spectroscopically. 
\end{enumerate}
\begin{figure*}
\centering
\includegraphics[scale=0.27]{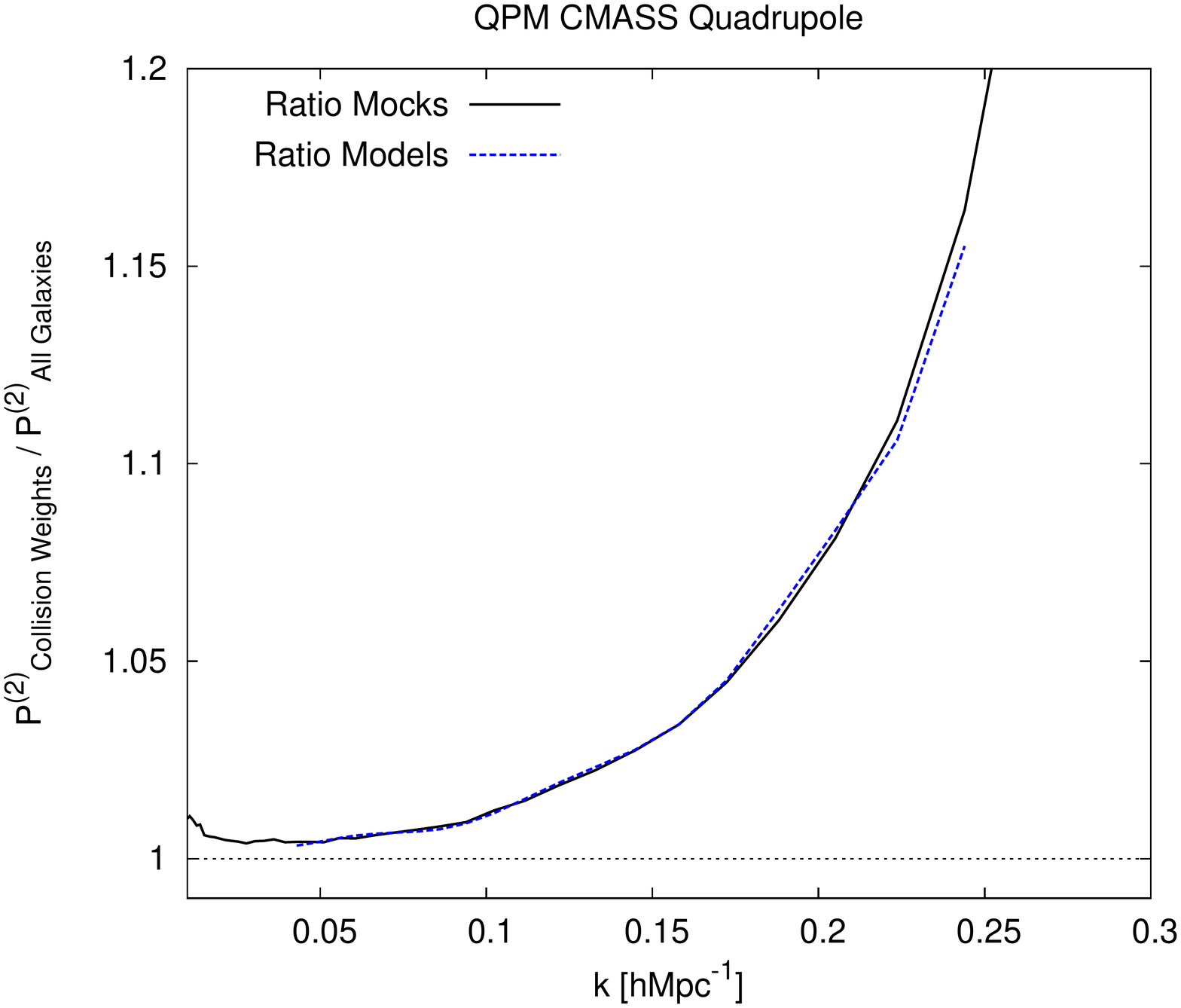}
\includegraphics[scale=0.27]{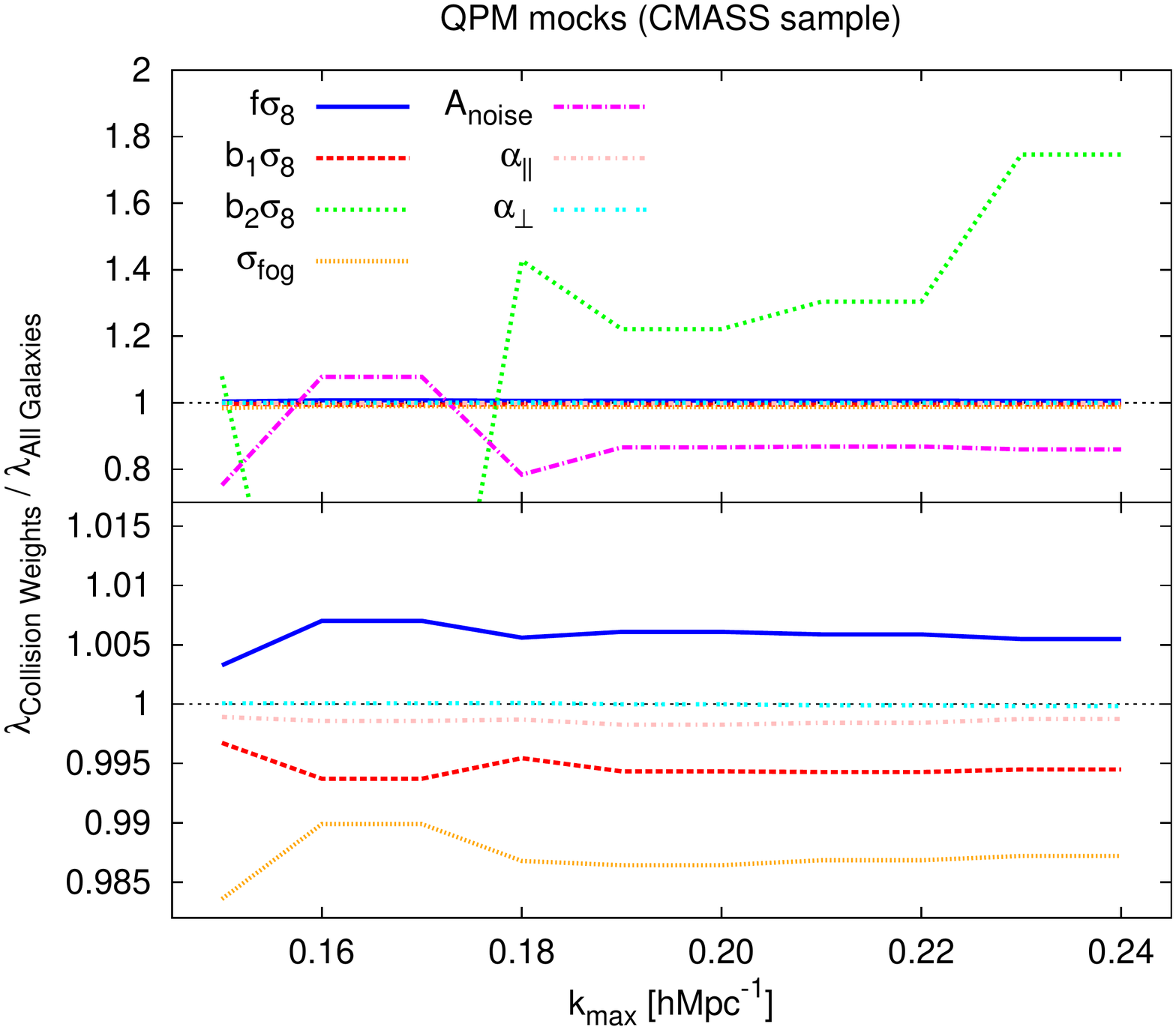}
\caption{Effect of fiber collision weights in the CMASS sample, where the fraction of collided galaxies is 0.0529. The left panel display the ratio between power spectrum quadrupole of the cases (i) and (ii) (see text),  where the fiber collisions are applied, ${P^{(2)}}_{\rm Collision Weights}$, and where all the galaxies are considered, ${P^{(2)}}_{\rm All Galaxies}$. The black solid line is the ratio of the measured quadrupole of 1000 realizations of the \textsc{qpm} mocks, and the blue dashed line the ratio between the best-fit models to the cases (i) and (ii) at $k_{\rm max}=0.24\,h{\rm Mpc}^{-1}$. The right panel displays the ratio between the best-fit parameters of the best-fit models of the cases (i) and (ii). Top and bottom panel display the ratio at different scales.      }
\label{plot:fiber_collisions}
\end{figure*}

The case (ii) has the correct clustering and the correct anisotropic signal. Comparing both will show how the collision weights affect the power spectrum multipoles and the estimation of parameters. 

When we compare the power spectrum monopole (and therefore the isotropic clustering) of cases (i) and (ii), we see that  the effects of the collision weights are degenerate with the amplitude of the shot noise parameter, $A_{\rm noise}$, which we treat as a free nuisance parameter. This  was already reported in \cite{hector_bispectrum1} for the DR11 sample using \textsc{PThalos} mocks \citep{Maneraetal:2013}. The effect of the fiber collision weights in the power spectrum quadrupole (i.e. in the anisotropic clustering) is more complex. We know that a fraction of the angular close pairs correspond to galaxies which  share the same dark matter host halo. Using the regions where we have a  superposition of plates, we have estimated that this fraction is about $\sim60\%$. On the other hand, $\sim40\%$ of the angular close pairs correspond to galaxies that happens to share a similar LOS, but that are actually reasonably uncorrelated. By applying the fiber collision weights to galaxies that are actual close pairs, we are removing signal in the direction of the LOS with respect to the signal in the transverse direction, which is not modified. Therefore, by applying the collision weights, the anisotropic power spectrum is affected, which potentially can alter the estimation of $f\sigma_8$. This is shown in the left panel of Fig. \ref{plot:fiber_collisions}, where the black solid line show the fractional change in the power spectrum {  quadrupole} of case (i) respect to case (ii), for \textsc{qpm} mocks. We see that this change is sub-percent at large scales, but rapidly grows as we go to smaller scales, reaching $\simeq10\%$ at $k\simeq0.2\,h{\rm Mpc}^{-1}$ and $\simeq20\%$ at $k\simeq0.25\,h{\rm Mpc}^{-1}$. In order to test the impact of this change in the fitted parameters, we fit the RSD bias model presented in \S\ref{sec:modelling} to the power spectrum monopole and quadrupole of cases (i) and (ii) and compare them, using the scales $0.02 \,h{\rm Mpc}^{-1}\leq k \leq k_{\rm max}$ for the monopole and  $0.04 \,h{\rm Mpc}^{-1}\leq k \leq k_{\rm max}$ for the quadrupole, as we did for the data, varying $k_{\rm max}$ from $0.15\,h{\rm Mpc}^{-1}$ to $0.24\,h{\rm Mpc}^{-1}$. The blue dashed line of Fig. \ref{plot:fiber_collisions} shows the ratio of the best-fit models to power spectrum multipoles of cases (i) and (ii). We see that the difference among the models describes well the differences in the power spectrum quadrupole produced by the fiber collision weights. In the right panel of Fig. \ref{plot:fiber_collisions} the ratio between the best-fit parameters, $\{b_1\sigma_8, b_2\sigma_8, f\sigma_8, \sigma_{\rm FoG}, A_{\rm noise}, \alpha_\parallel, \alpha_\perp  \}$ of these two models are shown. Both top and bottom panels display the same ratio of parameters, but at different ranges, for clarity. We observe that the change in the monopole and quadrupole due to the fiber collision weights is absorbed mainly by $A_{\rm noise}$ and $b_2\sigma_8$, which changes of order of $20\% - 100\%$ in the fiber collision case respect to case (i).  The FoG damping parameter, $\sigma_{\rm FoG}$ changes by about $2.5\%$ and does not present any significant change with the minimum scale of the fit. The AP parameters, $\alpha_\parallel$ and $\alpha_\perp$ do not present any significant change due to the effect of fiber collisions. Finally $f\sigma_8$ and $b_1\sigma_8$ are modified by about $\sim0.5\%$, and show no dependence with the minimum scale of the fit. In Table \ref{table_sys} we summarise these results and compare them with the statistical errors for the CMASS sample using $k_{\rm max}=0.24\,h{\rm Mpc}^{-1}$.

\begin{table}
\begin{center}
\begin{tabular}{|c|c|c}
Parameter & statistical error [\%] & systematic error due to fc [\%]\\
\hline
$f\sigma_8$ & 8.5\% & 0.5\%\\
$\alpha_\parallel$ & 3.1\% & 0.1\% \\
$\alpha_\perp$ & 1.6\% & $\ll0.1\%$ \\
\hline
$b_1\sigma_8$ & 1.7\% & 0.5\%\\
$b_2\sigma_8$ & 130\% & 70\%\\
$\sigma_{\rm FoG}$ & 9.5\% & 1.5\%\\
$A_{\rm noise}$ & 175\% & 10\%
\end{tabular}
\end{center}
\caption{Statistical and systematic errors caused by the effect of fiber collisions in the free parameters of the model, for the CMASS sample at $k_{\rm max}=0.24\,h{\rm Mpc}^{-1}$. For the cosmological parameters, $f\sigma_8$, $\alpha_\parallel$ and $\alpha_\perp$ the systematic errors caused by the fiber collisions are much smaller than the statistical errors of the data for the CMASS sample.}
\label{table_sys}
\end{table}%

We conclude that the effect of fiber collision are absorbed chiefly by $A_{\rm noise}$ and $b_2\sigma_8$. The impact of the fiber collision on  $f\sigma_8$ and $b_1\sigma_8$ is a sub-percent. In this case, $f\sigma_8$ tend to be overestimated with respect to the ideal case where all the galaxies were considered, whereas $b_1\sigma_8$ tend to be underestimated. Finally the AP parameters present changes of order $\simeq 0.1\%$ for $\alpha_\parallel$ and $\ll 0.1\%$ for $\alpha_\perp$.

Since the $1\sigma$ statistical errors of these parameters are about 10 times larger than this systematic shift, we do not  correct our result by this effect. Because the fraction of collided galaxies in the CMASS sample is higher than in the LOWZ sample, we expect that these changes are  also negligible in the LOWZ sample.

   \end{document}